%% file: main.tex
\documentclass[11pt]{article}

\usepackage[utf8]{inputenc}
\usepackage[T1]{fontenc}
\usepackage{mathpazo}

\usepackage[letterpaper,margin=1in]{geometry}
\usepackage{amsmath,amssymb,amsthm,mathtools}
\usepackage{microtype,booktabs,array,tabularx,longtable,enumitem,needspace}
\usepackage{float}
\usepackage{tikz}
\usetikzlibrary{arrows.meta}
\floatstyle{ruled}
\newfloat{algorithm}{tbp}{loa}
\floatname{algorithm}{Algorithm}

\newenvironment{algorithmsteps}{%
  \begin{enumerate}[label=\arabic*.,leftmargin=1.8em,labelsep=0.5em,
    itemsep=0.55em,parsep=0pt,topsep=0.35em,partopsep=0pt]
  \raggedright
}{\end{enumerate}}

\usepackage[dvipsnames]{xcolor}
\definecolor{ForestGreen}{rgb}{0.1333,0.5451,0.1333}
\definecolor{DarkRed}{rgb}{0.65,0,0}
\usepackage{hyperref}
\hypersetup{colorlinks=true,linkcolor=DarkRed,citecolor=ForestGreen,
  urlcolor=cyan,
  pdftitle={Fully Dynamic k-Median with Worst-Case Update Time and Recourse},
  pdfauthor={Chenglin Fan},
  pdfsubject={Research manuscript, 2026}}
\usepackage{aliascnt}
\usepackage[nameinlink,noabbrev,capitalise]{cleveref}
\crefname{invariant}{Invariant}{Invariants}
\crefname{algorithm}{Algorithm}{Algorithms}
\Crefname{algorithm}{Algorithm}{Algorithms}
\allowdisplaybreaks[2]
\numberwithin{equation}{section}
\newtheorem{theorem}{Theorem}[section]
\newaliascnt{lemma}{theorem}
\newtheorem{lemma}[lemma]{Lemma}
\aliascntresetthe{lemma}
\newaliascnt{proposition}{theorem}
\newtheorem{proposition}[proposition]{Proposition}
\aliascntresetthe{proposition}
\newaliascnt{corollary}{theorem}
\newtheorem{corollary}[corollary]{Corollary}
\aliascntresetthe{corollary}
\newaliascnt{definition}{theorem}
\newtheorem{definition}[definition]{Definition}
\aliascntresetthe{definition}
\newaliascnt{invariant}{theorem}
\newtheorem{invariant}[invariant]{Invariant}
\aliascntresetthe{invariant}
\newaliascnt{remark}{theorem}
\newtheorem{remark}[remark]{Remark}
\aliascntresetthe{remark}
\newcommand{\OPT}{\operatorname{OPT}}
\newcommand{\cost}{\operatorname{cost}}

\newcommand{\supp}{\operatorname{supp}}
\newcommand{\Ball}{\operatorname{Ball}}
\newcommand{\MakeRobustProc}{\textnormal{\textsc{MakeRobust}}}
\newcommand{\BlockRobustify}{\textnormal{\textsc{BlockRobustify}}}
\newcommand{\WindowCore}{\textnormal{\textsc{WindowCore}}}
\newcommand{\Refresh}{\textnormal{\textsc{Refresh}}}
\newcommand{\Develop}{\textnormal{\textsc{Develop}}}
\newcommand{\Remove}{\textnormal{\textsc{Remove}}}
\newcommand{\sym}{\mathbin{\triangle}}
\newcommand{\wtO}{\widetilde O}

\newcommand{\one}{\mathbf 1}
\newsavebox{\questioncontent}

\title{ Dynamic \(k\)-Median with
Worst-Case \\ Update Time and Recourse}
\author{Chenglin Fan}

\begin{document}
\hypersetup{pageanchor=false}
\pagenumbering{gobble}
\maketitle

\begin{abstract}
We study fully dynamic metric $k$-median under unit-weight
client insertions and deletions. After each update, the
algorithm must maintain at most $k$ centers chosen from
the current client locations. We give a randomized
constant-factor approximation algorithm with
$\widetilde O(k)$ worst-case update time and
$O((1+\log\Delta)^3)$ worst-case center recourse,
where $\Delta$ is the aspect ratio of the underlying
metric space. The approximation guarantee holds
simultaneously over any prescribed finite adaptive
update sequence with high probability, while the
update time and recourse bounds hold on every execution.
We also obtain the same asymptotic guarantees for
metric $k$-means.

\end{abstract}

\clearpage
\pagenumbering{roman}
\hypersetup{pageanchor=true}
\clearpage
\pagenumbering{arabic}

\part{Overview and Main Results}\label{part:overview}
\input{sections/introduction}
\input{sections/overview_sixstep}

\clearpage
\part{The Six-Step Construction}
\label{part:inner}

We now give the algorithms and proofs in the order of the six-step
overview. \Cref{sec:step1} fixes the weighted model and explains lazy
protection. \Cref{sec:step2} constructs protected cores and their
stability windows. \Cref{sec:step3} proves the local repair bounds
and uses them in an amortized rebuilding algorithm.
\Cref{sec:step4} turns a change of solutions into a budget-preserving
migration, and \Cref{sec:step5} schedules construction and migration
with worst-case bounds. Finally, \Cref{sec:step6} maintains the
weighted summary and proves \Cref{thm:main} by composition.

\input{sections/step1_model_lazy}

\input{sections/step2_cores}
\input{sections/step3_rebuilding}

\input{sections/step4_recourse}
\input{sections/step5_time}

\input{sections/step6_summary}

\clearpage
\part{Extensions}\label{part:extensions}
\input{sections/kmeans_and_conclusion}

\clearpage
\appendix
\input{sections/restricted_local_search}

\section*{AI Methodology Statement}

OpenAI ChatGPT was used during the development of this work to discuss selected proof approaches and to assist in developing and refining portions of proofs. It was also used to improve the exposition, organization, and LaTeX source of the manuscript.

\bibliographystyle{alpha}
{\normalsize\interlinepenalty=10000\bibliography{references/bibliography}}
\end{document}

%% file: sections/introduction.tex
\section{Introduction}\label{sec:intro}

In the metric $k$-median problem, we are given a set $P$
of clients in a metric space and an integer $k$.
The goal is to choose at most $k$ centers from $P$
to minimize the sum of distances from each client to
its nearest center. In the static setting, the client
set is fixed. In the fully dynamic setting, clients
may be inserted or deleted over time, and the algorithm
must maintain a good approximation after every update.
We require the centers to be chosen from the current
client locations.

We measure an algorithm by its approximation ratio, update
time, and center recourse, the number of centers added or
removed after an update. A single client change can make
the preferred centers change substantially. For example,
deleting an isolated client frees a center that can be
used elsewhere. A good dynamic algorithm must respond
without replacing many centers or recomputing the entire
solution.

Worst-case bounds impose an additional requirement: the
algorithm must continue serving the current input while
a replacement solution is being built. The old solution
must remain a good approximation until the replacement
is ready, and the transition must stay within the
$k$-center budget. Bounding the total work and total
recourse over a long sequence does not meet these
per-update requirements.
Previous work obtained constant-factor approximations
with $\widetilde O(k)$ amortized update time and
polylogarithmic amortized center
recourse~\cite{BCGLP24,BCF25}.
In this paper, we obtain worst-case guarantees for
both update time and center recourse while preserving
a constant-factor approximation.

\subsection{Model and Main Results}

Let \((X,d)\) be a fixed metric space. We assume that
the algorithm is given \(\Delta\ge1\) such that, after
normalizing the metric, every nonzero distance lies in
\([1,\Delta]\). The bound need not be tight. Distances between
known locations can be queried in constant time, even
when a location no longer contains a client. The algorithm
does not need to enumerate \(X\).

At any time, the input \(P\) is a finite multiset of
unit-weight clients, and \(\supp(P)\) denotes the set
of occupied locations. For a center set \(S\subseteq X\),
define
\[
 d(x,S)=\min_{s\in S}d(x,s),\qquad
 \cost_P(S)=\sum_{x\in P}d(x,S),
\]
and
\[
 \OPT_k(P)=
 \min_{\substack{S\subseteq\supp(P)\\|S|\le k}}
 \cost_P(S).
\]
We take \(d(x,\varnothing)=+\infty\) and assign cost zero
to the empty input. A center set is \emph{proper} if all
its centers are occupied client locations.

The recourse of update \(t\) is
\(\lvert S_t\triangle S_{t-1}\rvert\), so replacing one
center by another counts as two changes. We maintain
the centers themselves, not an explicit nearest-center
assignment for every client.
Let $n_0$ be the initial number of clients and
$T$ an upper bound on the number of updates.
Since each update inserts or deletes one client,
at most $n_0+T$ clients can be active at any time.
We set
\[
N=\max\{n_0+T,2k,2\}.
\]
We also fix a failure probability $\delta\in(0,1)$.
We write \(\widetilde O(\cdot)\) to suppress logarithmic
factors in \(N+2\), \(T+2\), \(\Delta+2\), and
\(1/\delta\). We work in the randomized real-RAM
model described in \Cref{sec:prelim}. We assume that
the ambient optima used in the analysis are attained,
as is automatic when \(X\) is finite.

\begin{theorem}[Fully dynamic metric \(k\)-median]
\label{thm:main}
There is a randomized algorithm for unit-weight client
insertions and deletions that maintains a proper set
\(S_t\) of at most \(k\) centers. For any adaptive
update sequence of length at most \(T\), with
probability at least \(1-\delta\),
\[
 \cost_{P_t}(S_t)=O(\OPT_k(P_t))
 \qquad
 \text{for every completed update }t.
\]
The hidden constant is absolute. Initialization on \(n_0\)
clients takes \(\widetilde O(k(n_0+k))\) time.
After initialization, each update takes
\(\widetilde O(k)\) time and changes
\(O((1+\log\Delta)^3)\) centers in the worst case.
Properness, cardinality, update time, and recourse
are guaranteed for every realization of the random
choices. Only the approximation guarantee can fail.
\end{theorem}

The adversary may choose each update after observing
all center sets published so far. The known distance
bound applies to the fixed metric space, including
locations occupied in the past, rather than only
to the current client support. These assumptions
are used in the snapshot and projection arguments.
Initialization is a one-time preprocessing cost. All later
rebuilding, including synchronous rebuilding in a short
window, is included in the stated per-update bound.
If desired, stable filler centers give exactly
\(\min\{k,|\supp(P_t)|\}\) distinct centers without
changing the asymptotic bounds.

\begin{table}[t]
\centering
\setlength{\tabcolsep}{4pt}
\begin{tabular}{@{}lcccc@{}}
\toprule
Reference & Approximation & Update time
          & Recourse & Resource bounds\\
\midrule
\cite{BCGLP24}
 & \(O(1/\varepsilon)\)
 & \(\widetilde O(k^{1+\varepsilon})\)
 & \(\widetilde O(k^\varepsilon)\)
 & Amortized\\
\cite{BCF25}
 & \(O(1)\)
 & \(\widetilde O(k)\)
 & \(O(\log^2\Delta)\)
 & Amortized\\
This paper
 & \(O(1)\)
 & \(\widetilde O(k)\)
 & \(O((1+\log\Delta)^3)\)
 & Worst case\\
\bottomrule
\end{tabular}
\caption{Selected simultaneous guarantees for fully dynamic
metric \(k\)-median. The final column applies to both
update time and recourse. The parameter
\(\varepsilon>0\) is fixed in the first row.}
\label{tab:median-comparison}
\end{table}

The algorithm also processes weighted records internally.
A record edit may change the weight at a single location
by an arbitrary amount, including to or from zero.
This is necessary because one client update can transfer
the entire weight of a cluster in the summary.
The original client updates still have unit weight,
while the generated records have integer weights
in \([1,2N]\).
Our approach also extends to metric $k$-means,
where connection costs are squared distances.
For a center set $S$, define
\[
 \Psi_P(S)=\sum_{x\in P}d(x,S)^2
\]
and
\[
 \OPT_{k,2}(P)
 =
 \min_{\substack{S\subseteq\supp(P)\\|S|\le k}}
 \Psi_P(S).
\]
As in the $k$-median problem, centers are restricted
to current client locations. The underlying distance
$d$ remains a metric, but $d^2$ need not satisfy the
triangle inequality. We assume that ambient
squared distance optima are attained, as is
automatic in finite metric spaces.

\begin{corollary}[Fully dynamic metric $k$-means]
\label{cor:kmeans-main}
Under the same update model, there is a randomized
algorithm that maintains at most $k$ centers
chosen from the current client locations.
For any adaptive update sequence of length at most
$T$, with probability at least $1-\delta$,
\[
 \Psi_{P_t}(S_t)
 =O\bigl(\OPT_{k,2}(P_t)\bigr)
 \qquad
 \text{for every completed update }t.
\]
The hidden constant is absolute. The worst-case
update time, recourse, and initialization bounds
are those of \Cref{thm:main}, with the same hidden
logarithmic parameters. The resource and feasibility
guarantees hold for every realization of the random
choices.
\end{corollary}
The model includes finite sets of Euclidean locations
with their induced metric. In this case, the proper
squared optimum is at most twice the optimum obtained
by allowing arbitrary Euclidean centroids.
Thus the corollary also gives a constant approximation
against the unrestricted centroid benchmark
(\Cref{km:euclidean}).

\subsection{Main Technical Ingredients}
\label{sec:technical-comparison}

\paragraph{Controlling local repairs.}
Previous robust-center methods~\cite{FLNS21,BCF25}
achieve low amortized recourse. The difficulty in
obtaining a worst-case bound is that moving one center
may trigger changes to many others.

We associate each center with an anchor, which serves
as a geometric reference point, and a proxy, which is
the location used to serve clients.
We also store the sequence of intermediate locations
used to construct each proxy. This sequence, called
a robust certificate, helps us determine when the
proxy needs to be recomputed after an update. Rather than discarding this information after
each update, we retain the relevant history.
This history helps in two ways. A strong-reset
condition limits the number of anchor relocations
triggered by a single change. A packing argument
shows that one record edit can directly affect only
$O(1+\log\Delta)$ existing certificates.
Consequently, the work and center changes of each
repair batch can be bounded in terms of that batch's
own edits and proposed center changes
(\Cref{block:theorem}).
The weak offset condition alone does not control this
particular rebase-to-proxy repair rule. The
\hyperref[geom:weak-only-obstruction]{example in
Section~\ref{geom:rebase-subsection}}
forces arbitrarily many relocations after one update,
even at constant aspect ratio, if only that weak condition
is maintained. The strong-reset invariant excludes this
configuration and yields the logarithmic closure bound.

\paragraph{Rebuilding solutions in the background.}
We build on the restricted-removal procedure and
double-sided stability argument of~\cite{BCF25}.
At each checkpoint, we retain a protected subset
of the current centers and reserve some center
slots for handling subsequent updates.
These tools allow us to determine a period during
which the protected centers, together with locations
affected by recent updates, remain a constant-factor
approximation. The length of this period can be
computed without knowing the optimum.

Our main challenge is to complete the necessary
work within this period while maintaining worst-case
bounds on every update.
We construct the replacement solution in the
background and gradually move from the old centers
to the new ones. A four-stage schedule distributes
reconstruction, weighted-record updates, and
center migration across the available updates.
It ensures that all work finishes before the
current checkpoint expires, that the algorithm
never maintains more than $k$ centers, and that
both update time and center recourse are bounded
in the worst case. When the available period is
too short for background rebuilding, we complete
the reconstruction within a single update
(\Cref{thm:window-core,thm:refresh,thm:inner};
\Cref{lem:migration}).

\paragraph{Maintaining a dynamic weighted summary.}
We reduce the original client set to a weighted summary
on $\widetilde O(k)$ locations using the fixed-assignment
representation of~\cite{BCLP23} and the layered controller
of~\cite{GS26}. The additional requirement is that each
original update changes only $O(1)$ weighted records,
including during a partition replacement.

We address this by moving from the old
representation to the new one gradually.
During a transition, a new weighted record
is activated with its full current weight
before the corresponding old records are removed.
Thus, every client remains represented throughout
the transition, although some clients may
temporarily be counted twice.
By carefully scheduling these changes, we ensure
that each original client update produces only
$O(1)$ weighted-record edits.
The resulting summary may contain clusters
from different snapshots, and these clusters
need not be nested. We therefore prove a
separate packing bound on the cost of representing
the original clients by the current weighted
summary. This allows us to preserve the
clustering approximation while maintaining
both small support and a constant number of
record edits per update
(\Cref{outer:event-consistency,outer:tail-packing};
\Cref{outer:summary}).
All weighted-record edits are processed by
the inner solver before the updated center
set is published.

\paragraph{Organization.}
\Cref{sec:stoc-overview} gives a six-step overview.
Part~II follows exactly that order: the weighted model and
lazy updates (\Cref{sec:step1}); protected cores and stability
windows (\Cref{sec:step2}); robust repair and amortized
rebuilding (\Cref{sec:step3}); worst-case recourse
(\Cref{sec:step4}); worst-case update time
(\Cref{sec:step5}); and weighted summaries and the final
composition (\Cref{sec:step6}). Each detailed section states
the interfaces it provides to the next step. Part~III proves
the squared-distance extension.

\subsection{Relation to Earlier Work}

The metric \(k\)-median problem has a long history,
beginning with facility location on graphs
~\cite{Hakimi64,Hakimi65}. Classical approximation
algorithms include local search~\cite{AryaGKMMP04},
primal-dual methods based on Lagrangian relaxation
~\cite{JainV01}, and successive sampling
~\cite{MP02,MP04}. Our algorithm uses static removal
and augmentation procedures from this literature.
The versions needed for weighted instances and
ambient benchmarks are stated in \Cref{sec:prelim}.

Lattanzi and Vassilvitskii~\cite{LattanziV17}
studied consistent clustering in insertion-only
streams. Subsequent work developed robust-center
methods for general metrics~\cite{FLNS21} and
fully dynamic consistent facility location
~\cite{CohenAddadHPSS19}. Dynamic coresets
~\cite{HenzingerK20} and clustering sparsifiers
~\cite{BCLP23} provide ways to maintain compressed
representations of changing inputs. For fully
dynamic metric \(k\)-median, \cite{BCGLP24,BCF25}
achieve the amortized guarantees summarized in
\Cref{tab:median-comparison}. In particular,
the robustification analysis in
\cite[Section~14.3]{BCF25} bounds the total cost
over an update sequence. We strengthen this part
of the analysis by bounding the cost of each
repair invocation, which is needed to obtain
worst-case guarantees.

Worst-case recourse has also been studied for
other clustering objectives. For \(k\)-center,
\cite{BateniEFHJMW23} consider adaptive and
oblivious adversaries, while \cite{LackiHGRJ24}
give a deterministic constant approximation
with constant worst-case recourse. More recently,
Grilnberger and Skarlatos~\cite{GS26} developed
an adaptive dynamic bicriteria construction
combined with a worst-case inner solver.
They also discuss an extension of this approach
to \((k,z)\)-clustering in Section~1.2.1.
Our construction uses their structural layer
schedule. Applying this approach to metric
\(k\)-median requires a weighted inner solver
with worst-case guarantees, together with
bounds on weighted record changes and
transportation between successive summaries.
We also establish that these guarantees hold
at the same update boundaries.
The correspondence with~\cite{GS26} is given
in \Cref{tab:outer-correspondence}.

For Euclidean clustering, Bhattacharya et
al.~\cite{BCFJJL26},
give a \(\mathrm{poly}(1/\varepsilon)\)-approximation
for \(k\)-means with
\(\widetilde O(k^\varepsilon)\) amortized update
time and \(\widetilde O(1)\) amortized recourse
for fixed \(\varepsilon\in(0,1)\). They also
obtain a tradeoff for Euclidean \(k\)-median.
Their running time has polynomial dependence
on the dimension and \(1/\varepsilon\), and
their final guarantee assumes an oblivious
adversary. In contrast, our algorithm applies
to general metrics and guarantees worst-case
update time and recourse on every execution.
For any prescribed finite update horizon,
its approximation guarantee holds simultaneously
with high probability, including when updates
are chosen adaptively based on previously
published center sets.

%% file: sections/overview_sixstep.tex
\section{Algorithm Overview}\label{sec:stoc-overview}

We develop the algorithm in six steps, each removing a limitation of
the preceding one. The first five give a solver for an explicit
weighted instance with \(n\) occupied locations and
\(\widetilde O(n+k)\) worst-case update time. The sixth maintains
a weighted summary with \(\widetilde O(k)\) occupied locations
and only \(O(1)\) record edits per original client update. Combining
the two gives the desired \(\widetilde O(k)\) bound. We explain the
idea, obstacle, and main technical tool at each step; Part~II gives
the corresponding algorithms and proofs in the same order.

A weighted record edit changes the weight at one location by any
amount. We use \(\Phi_P=\Phi_{P,1}\) for weighted median cost;
on unit-weight inputs this agrees with \(\cost_P\). Squared cost is
\(\Phi_{P,2}\), also denoted \(\Psi_P\) for unit-weight clients.
The quantity \(F_j(P)\) is the optimum using at most \(j\) facilities
anywhere in the ambient metric, whereas \(\OPT_{k,p}\) refers to the
proper optimum on original clients. In Steps~1--5, \(P,Q\) may denote
weighted snapshots; in Step~6, they denote the original clients and
their weighted summary. Original-client update time \(t\) and inner
weighted-record time \(u\) are different clocks.
Intermediate facilities need not be occupied locations; the published
centers are their proper projection. The precise model and the
projection guarantee are given in \Cref{sec:step1}.
Approximation statements below are conditioned on the success of
the relevant randomized calls. The final algorithm enforces its
feasibility and resource bounds on every execution.

\begin{figure}[ht]
\centering
\begin{tikzpicture}[
  >=Latex,
  every node/.style={font=\normalsize},
  stage/.style={
    draw=black!65, line width=0.55pt, rounded corners=2pt,
    text width=3.85cm, minimum height=1.65cm,
    inner sep=4pt, align=center, fill=black!3
  },
  final/.style={stage,fill=black!9},
  flow/.style={-Latex,line width=0.85pt,draw=black!75}
]
  \node[stage] (s1) at (0,0)
    {\textbf{Step 1}\\lazy updates\\\(2k\) facilities};
  \node[stage] (s2) at (4.55,0)
    {\textbf{Step 2}\\protected cores\\\(k\)-facility budget};
  \node[stage] (s3) at (9.10,0)
    {\textbf{Step 3}\\robust rebuilding\\amortized bounds};
  \node[stage] (s4) at (9.10,-2.20)
    {\textbf{Step 4}\\gradual migration\\worst-case recourse};
  \node[stage] (s5) at (4.55,-2.20)
    {\textbf{Step 5}\\background work\\worst-case time};
  \node[final] (s6) at (0,-2.20)
    {\textbf{Step 6}\\weighted summaries\\\(\widetilde O(k)\) time};
  \draw[flow] (s1.east) -- (s2.west);
  \draw[flow] (s2.east) -- (s3.west);
  \draw[flow] (s3.south) -- (s4.north);
  \draw[flow] (s4.west) -- (s5.east);
  \draw[flow] (s5.west) -- (s6.east);
\end{tikzpicture}
\caption{The six-step construction. Steps 1--5 operate on an
explicit weighted instance. Step 6 supplies a small instance with
a bounded number of record changes after each original update.}
\label{fig:sixstep-roadmap}
\end{figure}
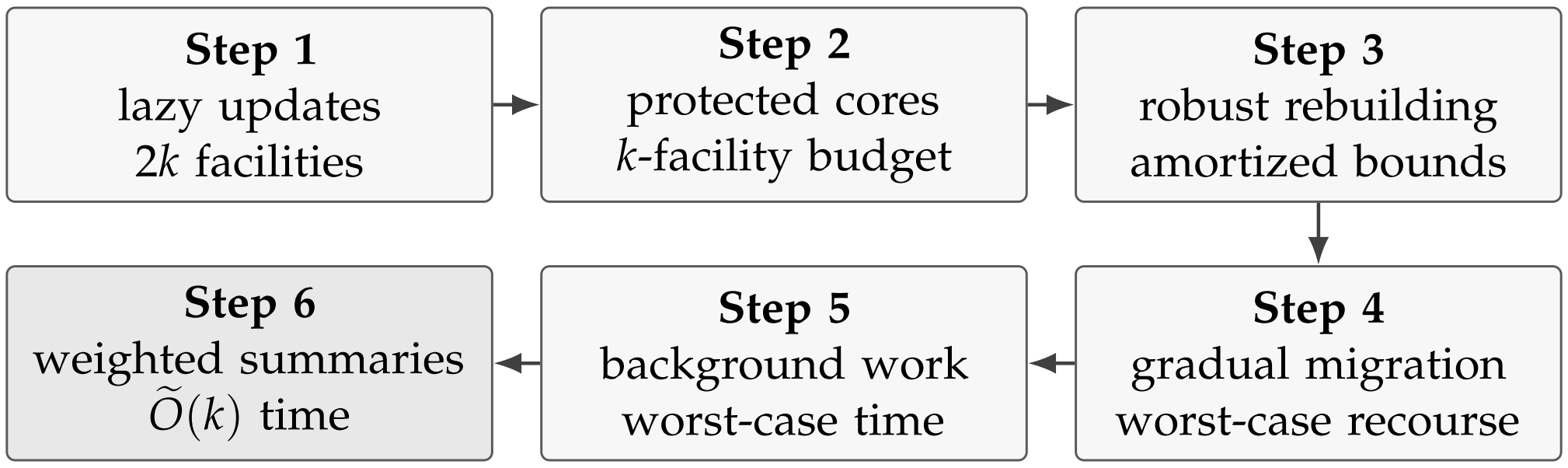

\subsection{Step 1: Lazy Updates and Periodic Rebuilding}
\label{sec:overview-preliminaries}

We begin with a simple dynamic algorithm that maintains
a constant-factor approximation using at most \(2k\)
facilities. For now, facilities may lie anywhere in the
ambient metric space, and we allow periodic rebuilding
without requiring worst-case update-time guarantees.

\paragraph{Basic observation.}
Suppose two instances \(P\) and \(Q\) differ only at
a set \(I\) of at most \(h\) locations. For any facility
set \(S\), opening facilities at these locations gives
\(\Phi_Q(S\cup I)\le \Phi_P(S)\).
Indeed, edited locations incur zero connection cost,
while the cost at all other locations cannot increase.
Applying this observation to an optimal solution for
\(Q\) also gives \(F_{j+h}(P)\le F_j(Q)\).
The second inequality is particularly useful for
deletions: it allows us to compare an earlier instance
with the current optimum even when the latter has
decreased substantially.

\paragraph{Algorithm and guarantees.}
Let \(h=\lfloor k/2\rfloor\), for \(k\ge2\).
At the beginning of each epoch, we compute a static
solution \(A_p\) containing at most \(k+h\) facilities
and satisfying
\(\Phi_{P_p}(A_p)\le\beta F_{k+h}(P_p)\)
for a constant static factor \(\beta\). During the next
\(h\) updates, we keep \(A_p\) unchanged and collect all
edited locations in a set \(I\), including locations
whose clients have been deleted. We maintain
\(S_t=A_p\cup I\) and compute a new static solution
at the end of the epoch.

The two inequalities above imply
\(\Phi_{P_t}(S_t)\le\Phi_{P_p}(A_p)
\le\beta F_{k+h}(P_p)\le\beta F_k(P_t)\).
Moreover, \(|S_t|\le k+2h\le2k\).
Each ordinary update adds at most one facility.
Rebuilding may change \(O(k)\) facilities, but occurs
only once every \(\Theta(k)\) updates. Thus, the
algorithm achieves a constant-factor approximation
with at most \(2k\) facilities and \(O(1)\)
amortized center recourse.

\paragraph{Towards a feasible dynamic algorithm.}
The warm-up still exceeds the \(k\)-facility budget,
may open historical locations, and performs an
expensive rebuild at each epoch boundary. Step~2
addresses the first issue by choosing how many
positions to reserve from the structure of the instance.
The formal treatment in \Cref{sec:step1} also develops
three static tools: restricted removal decreases the
facility count at controlled cost; augmentation adds
a few facilities to improve an existing solution; and
proper projection converts internal facilities into
current client centers with at most twice the cost
and constant recourse per source or target edit.

\subsection{Step 2: Protected Cores and Stability Windows}
\label{overview:subsec-core}

Instead of opening extra checkpoint facilities as
in Step~1, we remove some and reserve those slots
for future updates.
Let \(p\) denote a checkpoint time and \(P_p\)
the instance at that time. If we can safely
remove \(H\) facilities from a \(k\)-facility
reference solution, the remaining \(k-H\)
facilities form a \emph{protected core} \(K_p\).
During the next \(H\) updates, we open a
facility at each edited location, as in Step~1.
The total number of facilities never exceeds \(k\).

\paragraph{The main difficulty.}
Removing facilities increases cost, while deletions
may sharply decrease the new optimum. We therefore need
\(\Phi_{P_p}(K_p)\le C_0F_{k+H}(P_p)\),
where \(C_0\) is an absolute constant.
By the inequalities established in Step~1,
the maintained solution \(S_t\) at any time
\(t\le p+H\) satisfies
\(\Phi_{P_t}(S_t)\le C_0F_k(P_t)\).
Thus, the protected core provides a constant-factor
approximation throughout the update window.

The removal budget must depend on the instance:
when \(k\) well-separated clients have optimum
zero, removing any necessary facility makes
the cost positive.

\paragraph{Computing the core.}
Our procedure \(\WindowCore\) determines how many
facilities can be removed without substantially
increasing the connection cost. Starting from
a constant-factor approximate reference solution,
it tests geometrically increasing removal budgets
and measures the costs of the resulting solutions.
Let \(R\) be the last accepted budget: set \(R=0\) if none accepts
and \(R=k-1\) if all tests accept. For a sufficiently large fixed
constant \(D\), choose \(H=\lfloor R/D\rfloor\) and \(B=H+1\).

The main technical ingredient is an LP
interpolation argument. It shows that if removing
a relatively large number of facilities causes
only a bounded increase in cost, then removing
a smaller number yields the stronger comparison
with \(F_{k+H}(P_p)\) required above. This allows
us to construct the protected core without
computing the optimal clustering cost.

\paragraph{Stability windows.}
The reserved capacity also determines how long
the core remains valid. We define \(B=H+1\)
as the \emph{stability window}, so the core
can absorb the next \(B-1\) updates.
To support later rebuilding, we also need
window lengths to remain comparable across
nearby checkpoints. We establish this through
an instance-dependent measure \(b(P)\) of removable
slack, which changes by at most one under
a single record edit, even when the edited
weight changes arbitrarily. The precise comparison is
\(B\le b(P)+1\le CB\) for an absolute constant \(C\).
The additive one covers \(b(P)=0\) and ensures comparable
window lengths for nearby instances.

Step~2 therefore maintains at most \(k\)
facilities with a constant-factor approximation
throughout a certified update window.
Given the maintained distance rows, the protected
core and its window can be computed in
\(\widetilde O((n+k)B)\) time.
However, replacing an expired core may still
change many facilities at once. Step~3
addresses this problem by developing a
rebuilding procedure with small amortized
recourse. The formal guarantees appear in
\Cref{sec:step2}.

\subsection{Step 3: Robust Rebuilding and Amortized Recourse}
\label{overview:subsec-blocks}
\label{overview:subsec-state}

Step~2 maintains a constant-factor approximation
using at most \(k\) facilities. However, when a
protected core expires, computing a new solution
from scratch may replace almost all \(k\) facilities,
even if only a few clients have changed. Our goal
is to construct a new reference solution whose
number of center changes is proportional to the
length of the preceding stability window.

\paragraph{Robust reference solutions.}
At each checkpoint \(p\), we maintain a reference
solution \(U_p\) of \(k\) facilities with geometric
annotations. Each facility has a geometric reference
location, called its \emph{anchor}, and a serving
location, called its \emph{proxy}. A \emph{robust
certificate} records local improvements from the anchor
at its generation time to the serving proxy across
decreasing scales. A later rebase may move the current
anchor to its proxy, while the full certificate sequence,
original top radius, and generation time are retained
until the certificate is replaced. For label \(i\), we
write \(a_i\) for its current anchor, \(v_i\) for its proxy,
and \(q_i\) for its certificate's historical top;
the strong-reset time \(\tau_i\) and certificate generation
time \(\sigma_i\) record different events.

\paragraph{Rebuilding with few center changes.}
For a stability window of length \(B\), the
procedure \(\Refresh\) constructs a new reference
after at most \(B\) edits. Its first challenge is
to obtain a good solution while adding few facilities.
Our small-augmentation argument matches well-separated
facilities of a comparison solution to existing
robust proxies; removable slack bounds the unmatched
facilities. Only \(O(B)\) additions are needed.
The procedure then opens all edited locations and
uses restricted removal to restore the \(k\)-facility
budget. The resulting candidate \(W\) is a
constant-factor approximation differing from \(U_p\)
at only \(O(B)\) facilities.

\paragraph{Local robustification.}
Although \(W\) has low cost, its geometric certificates
may be invalid. We repair anchors, proxies, and
certificates locally. The obstacle is that one
anchor insertion may cause a cascade of relocations,
and one client edit may affect many old certificates.
A \emph{strong-reset condition} controls the cascade;
a \emph{chronological packing argument} bounds
certificate overlap by scale and creation time.
Together they bound each repair invocation by its
own edits and proposed center changes, while a
fixed client partition controls the cost increase.

Consequently, \(\Refresh\) produces a new
robust reference \(U'\) satisfying
\(\Phi_{P_t}(U')\le\alpha F_k(P_t)\),
where the approximation constant \(\alpha\)
remains unchanged across rebuilding operations.
Moreover, the number of changed facilities is
\(|U'\triangle U_p|=O(B(1+\log\Delta)^3)\),
and the rebuilding work is
\(\widetilde O((n+k)B)\).

\paragraph{Amortized guarantees.}
We combine \(\Refresh\) with \(\WindowCore\),
replacing checkpoints after a fixed fraction of
their windows and rebuilding immediately for
short windows. Comparable neighboring windows
allow the rebuilding costs to be charged to
the intervening edits. We obtain
\(\widetilde O(n+k)\) amortized update time
and \(O((1+\log\Delta)^3)\) amortized
center recourse, while maintaining at most
\(k\) facilities and a constant-factor
approximation.

These bounds remain amortized because a rebuild
may occur in a single update. Step~4 spreads
center changes across edits; Step~5 spreads
the computation. See \Cref{sec:step3} for the
formal construction.

\subsection{Step 4: Worst-Case Center Recourse}
\label{overview:subsec-wc-recourse}

Step~3 reduces the total number of center changes
during a window of length \(B\) to
\(O(B(1+\log\Delta)^3)\), but they could all
occur in one update. We distribute them over
the window to obtain worst-case recourse.

\paragraph{The main difficulty.}
Suppose we have an old solution and a new
constant-factor approximate solution.
Replacing facilities in an arbitrary order
may exceed the \(k\)-facility budget or
produce poor intermediate solutions even
when both endpoints are good.

\paragraph{Budget-preserving migration.}
For every departing facility, choose a nearby
target facility. Remove an old facility when
its replacement is already open; otherwise
swap the two atomically. Insert remaining
targets only after making enough room.

This gives a \emph{budget-preserving migration}
from an old facility set \(A\) to a target
set \(D\). Every intermediate set \(M\)
contains at most the larger of the two
endpoint sizes. More importantly, for
every client location \(x\), it satisfies
\(d(x,M)\le2d(x,A)+d(x,D)\).

Thus intermediate connection costs are bounded
by a constant combination of the endpoint costs.

\paragraph{Handling updates during migration.}
The input keeps changing during migration.
We freeze the old serving set, compute the new
protected core, and keep later edited locations
in a common buffer. If the transition ends within
both certified windows, the migration inequality
applied with that buffer preserves the budget
and approximation throughout.

\paragraph{Worst-case recourse.}
The refresh bound from Step~3 implies that
the old and new solutions differ
in only \(O(B(1+\log\Delta)^3)\) facilities.
Our migration order uses only a constant
number of operations per changed facility.
Executing \(O((1+\log\Delta)^3)\) transitions per
update over \(\Theta(B)\) edits therefore gives
the same worst-case center recourse bound;
proper projection preserves it for published
client centers. However, computing the target
and preparing the path may still take excessive
time at one update. Step~5 resolves that issue;
the migration proof is in \Cref{sec:step4}.

\subsection{Step 5: Worst-Case Update Time}
\label{overview:subsec-obstacles}

Step~4 guarantees that only a small number
of facilities change per update, but building
the target may still require a large computation
at once. We distribute that work across updates
and finish before the old core expires.

\paragraph{Background rebuilding.}
We freeze the instance \(P_t\), build the next
checkpoint in the background, and continue
serving clients with the old solution while
recording subsequent edits in a buffer.

For sufficiently large stability windows,
we divide the rebuilding process into
four stages:

\begin{enumerate}
    \item \textbf{Refresh.}
    Construct a new robust reference solution
    for the frozen instance using the
    procedure from Step~3.

    \item \textbf{Protected core.}
    Apply \(\WindowCore\) to compute the
    new protected core and its stability
    window.

    \item \textbf{Migration preparation.}
    Construct the budget-preserving migration
    sequence from Step~4 and update the
    supporting distance data structures.

    \item \textbf{Gradual migration.}
    Execute the migration in bounded batches,
    keeping all locations edited since the
    frozen snapshot in the common buffer.
\end{enumerate}

Only Stage~4 changes the migration base \(M_u\). The edit buffer
\(J_u\) and proper projection are maintained throughout all four
stages, so the serving set \(Z_u=M_u\cup J_u\) can change earlier.
After migration, the new checkpoint is adopted
with the edits made since its snapshot.

\paragraph{Meeting the rebuilding deadlines.}
The main challenge is to meet every stage's
deadline, not just bound its total work.
We start while the old core is young. Window
stability from Step~2 gives comparable old
and new lifetimes, and Step~3 bounds the
work to be scheduled. Fixed quotas of
\(\widetilde O(n+k)\) operations and
\(O((1+\log\Delta)^3)\) center changes per edit
complete reconstruction and migration within
both windows. Very short windows are rebuilt
synchronously; their total work already fits
the worst-case budget.

\paragraph{Maintaining consistent snapshots.}
The target is computed on a frozen instance,
but current demand keeps changing. Persistent
distance structures keep the frozen and live
versions separate. On completion, incremental
reconciliation updates only the edited locations,
and generation checks prevent stale background
work from overwriting newer edits.

\paragraph{Worst-case guarantees.}
Every stage has a fixed work budget; randomized
calls have predetermined iteration caps. If a feasibility
guard fails, a job exhausts its total work allowance, or
a deadline is missed, optional rebuilding stops. Exhausting
one work slice merely pauses that job. The foreground still
processes edits and maintains proper projection within the
resource budget. Feasibility, time, and recourse therefore
hold unconditionally. Approximation requires a common
randomized success event, which the algorithm does not test.

We obtain a fully dynamic algorithm for
weighted instances with
\(\widetilde O(n+k)\) worst-case update time
and \(O((1+\log\Delta)^3)\) worst-case
center recourse. The approximation guarantee
holds simultaneously over any prescribed
finite adaptive update sequence with high
probability.

The dependence on the number \(n\) of occupied
locations remains. Step~6 removes it using a
small weighted summary; see \Cref{sec:step5}
for the formal schedule.

\subsection{Step 6: Dynamic Weighted Summaries}
\label{overview:subsec-summary}

Step~5 achieves worst-case update time
\(\widetilde O(n+k)\) for a weighted instance,
where \(n\) is the number of occupied locations.
To obtain our final \(\widetilde O(k)\) bound,
we replace the original client set \(P\)
with a much smaller weighted instance \(Q\).

The idea is to group nearby clients into
clusters and represent each cluster by a
single current client location, weighted
by the number of clients it represents.
We aim to maintain only \(\widetilde O(k)\)
such weighted locations.

However, a small summary alone is not enough.
Its clustering cost must remain comparable
to that of the original instance, and each
original client update must cause only
a constant number of weighted-record edits.
Otherwise, processing changes to the summary
could still take too much time.

\paragraph{Maintaining a small summary.}
We construct the summary through multiple
sampling layers. Each layer selects
representative locations and assigns
a portion of the clients to clusters.
Within a layer generation, these assignments
remain fixed, while cluster weights and
representatives are updated as clients
are inserted or deleted.

The layers are maintained by background
rebuilding. Reconstructing one layer may
invalidate the layers that follow it,
so our controller rebuilds an entire
suffix of layers and publishes it only
after the suffix is complete. This keeps
the number of active clusters small
and avoids exposing partially constructed
partitions.

\paragraph{Constant weighted-record changes.}
The main challenge is that replacing a
partition may change many cluster records
simultaneously. Directly updating the
weighted summary would therefore produce
a large number of record edits at one update.

We avoid this by separating the maintained
cluster partition from the weighted records
visible to the algorithm in Step~5.
Each cluster is represented by an active
\emph{token} containing its current
representative and total client weight.
When switching to a new partition, we
gradually activate its tokens with their
full current weights before retiring
the old tokens.

During this transition, each client belongs
to one or two active tokens. Thus, the
summary may temporarily count some clients
twice, but it never loses their contribution.
Moreover, an insertion or deletion affects
at most two active tokens. If a representative
is deleted, another live client from its
cluster can be selected without modifying
the individual records of all its members.

By performing a fixed number of token
activations and retirements per update,
we ensure that each original client update
produces only \(O(1)\) weighted-record edits.
The summary maintains \(\widetilde O(k)\)
occupied locations with
\(\widetilde O(k)\) worst-case maintenance time.

\paragraph{Controlling the approximation error.}
We must also show that replacing clients
by their representatives does not
significantly increase the clustering cost.
Let \(E\) denote the total transportation
cost of moving clients to their active
representatives, counting both copies
when a client is temporarily represented
twice. By the triangle inequality,
the clustering costs on the original
instance and the weighted summary
can be compared using \(E\).

The main difficulty is bounding this
transportation cost over all active layers.
Different layers may come from different
snapshots, so their current client sets
need not be nested. Consequently,
analyzing each layer separately could
accumulate an excessive error.

Our key technical tool is a
\emph{geometric tail-packing argument}.
It uses the geometric decrease in
layer sizes to charge the transportation
costs of different layers to the
original clustering optimum, even
without nested client sets.
We obtain \(E=O(\OPT_k(P))\), where \(\OPT_k(P)\)
is the proper optimum on the original clients. This
ensures that a constant-factor solution
for the weighted summary also gives
a constant-factor solution for the
original instance.

\paragraph{Combining the two algorithms.}
We now run the weighted algorithm from
Step~5 on the dynamic summary \(Q\).
Since \(Q\) has only \(\widetilde O(k)\)
occupied locations and each original
update produces \(O(1)\) weighted-record
edits, the resulting worst-case update
time is \(\widetilde O(k)\).

All weighted-record edits generated by
one original update are processed before
a new solution is published. This ensures
that the published centers correspond
to the completed summary update, rather
than an intermediate transition state.
Because every summary representative is
a current client location, the resulting
centers are also valid for the original
instance.

Combining the summary guarantees with
those of Step~5 yields a fully dynamic
constant-factor approximation with
\(\widetilde O(k)\) worst-case update time
and \(O((1+\log\Delta)^3)\) worst-case
center recourse. These resource bounds
hold on every execution, while the
approximation guarantee holds
simultaneously over any prescribed
finite adaptive update sequence with
high probability.

The complete summary construction and
analysis appear in \Cref{sec:step6}.
The extension to squared connection
costs is discussed in \Cref{sec:kmeans}.

%% file: sections/step1_model_lazy.tex
\section{Step 1: Weighted Model and Lazy Updates}
\label{sec:step1}\label{sec:prelim}\label{sec:framework}

We begin with the simplest way to accommodate a changed input:
keep the old facilities and open a facility at every edited
location. Starting with just \(k\) facilities protects the old
cost but need not approximate the new optimum. Allowing extra
facilities at a checkpoint gives a simple \(2k\)-facility algorithm
with a current-optimum guarantee. Step~2 shows how to recover the
exact \(k\)-facility budget.

The first five steps work with an explicit weighted instance.
Their virtual facilities may occupy historical or auxiliary
locations, even when those locations carry no current demand.
Only the published solution must be proper. After the lazy-update
argument, we state the static removal, augmentation, and projection
interfaces needed by the later steps. This section also fixes the
computational model and the conventions for empty and zero-cost
instances.

\subsection{Weighted Instances, Time, and Snapshots}

Let \((X,d)\) be the fixed metric space, and let a weighted instance
\(P\) be represented by a function
\(w_P:X\to\mathbb R_{\ge0}\) with finite positive support.
For a facility set \(S\subseteq X\), let
\(d(x,S)=\min_{s\in S}d(x,s)\), with
\(d(x,\varnothing)=+\infty\). Define
\[
 \Phi_P(S)=\sum_{x\in\supp(P)}w_P(x)d(x,S),\qquad
 F_j(P)=\min_{\substack{S\subseteq X\\|S|\le j}}\Phi_P(S).
\]
Thus \(F_j(P)\) is the optimum with at most \(j\) facilities
anywhere in \(X\), whereas a proper solution must place its
facilities in \(\supp(P)\). As stated in the introduction,
the algorithm is given \(\Delta\ge1\) and works with the
normalized metric satisfying \(1\le d(x,y)\le\Delta\) for
distinct real locations. This bound determines the geometric
scales below; its value need not be tight. Distances
between known locations can be queried in constant time, even
after those locations leave the support. The algorithm does not
enumerate \(X\). We assume that the ambient optima appearing
in the analysis are attained. The same assumption is made for
the squared-distance optima in \Cref{sec:kmeans}; both assumptions
hold automatically when \(X\) is finite.

A \emph{record edit} changes the weight at one location by an
arbitrary amount, including to or from zero. Multiple clients at
the same location are represented by one weight. We write
\(n=|\supp(P)|\); for a procedure involving two snapshots,
\(n\) denotes the larger support size unless stated otherwise.
We use \(n_{\rm init}=|\supp(P_0)|\) for the initial support
size of an explicit weighted instance. This is distinct from the
number \(n_0\) of original unit-weight clients in \Cref{thm:main}.
The ratio of the largest to the smallest positive weight may
appear in hidden logarithmic factors. On an empty input,
\(F_j(P)=0\) for all \(j\ge0\). On a nonempty input,
\(F_0(P)=+\infty\). We only apply LP interpolation to
positive facility budgets.

We use a randomized real-RAM model. Arithmetic and comparisons on
distances and weights, as well as an independent uniform draw
from \([0,1)\), take constant time. A balanced cumulative-weight
tree supports weighted sampling and weight changes in
\(O(\log(n+2))\) time. Our bounds count arithmetic operations;
they do not give bit-complexity bounds for arbitrary real inputs.

The inner algorithm measures time in record edits. Write \(P_u\)
for its input after the \(u\)-th edit, and let
\[
 I(a,b)=\{x\in X:\text{a record at }x\text{ is edited at a time in }(a,b]\}.
\]
This set includes edited locations whose final weight is zero.
Its cardinality is at most \(b-a\), even if a location is edited
repeatedly. A checkpoint is an earlier record time \(p\) at which
we retain an input and a reference solution. An inner snapshot
freezes both the weighted dictionary and the roots of its auxiliary
structures; later edits do not alter that frozen input.

The outer algorithm in Step~6 instead processes individual client
insertions and deletions, indexed by \(\tau\). Each inserted client
receives a distinct birth identifier, and reinsertion at a previously
occupied location creates a new identifier. A client birth and a
weighted location are therefore different objects. The outer
snapshots freeze cluster ownership, not the masses or representatives
of those clusters. These remain live until their records are retired.

\subsection{Lazy Protection and the \texorpdfstring{\(2k\)}{2k}-Facility Warm-Up}

\begin{lemma}[Lazy inequalities]\label{lem:lazy}
Suppose \(s\) record edits change \(P\) into \(Q\), and let \(I\)
be the distinct edited locations, including those of final weight zero.
For every facility set \(S\) and integer \(j\ge0\),
\begin{equation}\label{overview:lazy}
 \Phi_Q(S\cup I)\le\Phi_P(S),\qquad
 F_{j+s}(Q)\le F_j(P),\qquad F_{j+s}(P)\le F_j(Q).
\end{equation}
\end{lemma}
\begin{proof}
Outside \(I\), the weights in \(P\) and \(Q\) agree. Opening a
facility at each location in \(I\) makes the contribution of
every edited location zero, while adding facilities cannot
increase the cost at the other locations. This proves the first
inequality. Since \(|I|\le s\), applying it to an optimal
\(j\)-facility solution, first for \(P\) and then for \(Q\),
gives the two budget inequalities. The argument uses only
nonnegative connection costs and zero cost at an opened location,
so it also applies to squared distances.
\end{proof}

\begin{lemma}[The \(2k\)-facility warm-up]\label{sixstep:warmup}
Fix \(k\ge2\) and \(h=\lfloor k/2\rfloor\). Divide the record edits
into epochs of \(h\) edits. At the beginning of each epoch, at time
\(p\), compute a static \(\beta\)-approximate set \(A_p\),
using the static guarantee of \Cref{lem:develop}, with
\(|A_p|\le k+h\) and
\(\Phi_{P_p}(A_p)\le\beta F_{k+h}(P_p)\).
Until the next rebuilding, maintain
\(Z_u=A_p\cup I(p,u)\). Then throughout every epoch,
\[
 |Z_u|\le2k,\qquad \Phi_{P_u}(Z_u)\le\beta F_k(P_u).
\]
Replacing the reference after every \(h\) edits incurs
\(O(1)\) amortized center recourse. No worst-case update-time
bound is asserted for this procedure.
\end{lemma}
\begin{proof}
Within an epoch, \(u-p\le h\), so at most \(h\) distinct
locations have been edited. Hence
\(|Z_u|\le(k+h)+h\le2k\). The lazy inequalities give
\[
 \Phi_{P_u}(Z_u)\le\Phi_{P_p}(A_p)
 \le\beta F_{k+h}(P_p)\le\beta F_k(P_u).
\]
An ordinary edit adds at most one location to the virtual set.
At an epoch boundary, replacing the old set by a new static
reference changes at most \(O(k)\) facilities. Since
\(h=\Theta(k)\) for \(k\ge2\), charging these changes to the
preceding epoch gives constant amortized recourse.
\end{proof}

The extra checkpoint facilities are essential for comparison with
the current optimum. For instance, with \(k=2\) and clients at
\(0,1,D\) on a line, where \(D>2\), the optimal two-facility
solution \(\{1,D\}\) has cost one. Deleting the client at \(D\)
reduces the two-facility optimum to zero, while the old solution
still costs one. Using a \((k+h)\)-facility reference avoids this
obstruction but leaves more than \(k\) facilities open. Step~2
instead reserves a variable number of slots while preserving the
comparison with a larger-budget optimum.

\subsection{LP Interpolation on a Common Facility Domain}

We will compare solutions with different numbers of facilities.
For this purpose, consider a finite set \(\mathcal F\subseteq X\)
containing an ambient optimal solution for every integer budget
in the comparison. Such a set exists by the attainment assumption:
one may take the union of finitely many optimal solutions.
It is used only in the analysis and is never constructed by the
algorithm. For real \(z\ge1\), let \(L_z(P)\) be the optimum of
the standard assignment LP on \(\mathcal F\). The variable
\(y_c\) denotes the extent to which facility \(c\) is opened,
and \(x_{pc}\) is the fraction of the demand at \(p\) assigned
to \(c\):
\[
 \min\sum_{p\in\supp(P)}\sum_{c\in\mathcal F}w_P(p)d(p,c)x_{pc}
 \quad\text{subject to}\quad
 \sum_cx_{pc}=1,\quad 0\le x_{pc}\le y_c,\quad
 \sum_cy_c\le z,\quad y_c\ge0.
\]
By construction, the integral optimum on \(\mathcal F\) equals
\(F_j(P)\) at each integer budget \(j\) being compared.
The integrality-gap bound of~\cite{ARS03}, also used
in~\cite[Section~7.1]{BCF25}, then gives
\begin{equation}\label{eq:lp-gap}
 L_j(P)\le F_j(P)\le3L_j(P).
\end{equation}
The weighted value bound follows for rational weights by
scaling and replicating demand at the same locations. Continuity
of the finite integral and LP optimum values extends it to real
weights. This is an argument about values and does not expand
the algorithmic input.
The choice of \(\mathcal F\) matters here. On an arbitrary
restricted domain, the LP gap is relative to that domain's
integral optimum, which need not equal the ambient value \(F_j(P)\).

\begin{lemma}[Convexity in the opening budget]\label{lem:lp-convex}
For \(a,b\ge1\) and \(\lambda\in[0,1]\),
\[
 L_{\lambda a+(1-\lambda)b}(P)
 \le\lambda L_a(P)+(1-\lambda)L_b(P).
\]
\end{lemma}
\begin{proof}
Take optimal LP solutions for budgets \(a\) and \(b\).
Since both use the same facility domain, their convex combination
is feasible at budget \(\lambda a+(1-\lambda)b\) and has the
stated objective value.
\end{proof}

\subsection{Restricted Removal and Augmentation}

We use two static procedures. Restricted removal reduces the number
of facilities while controlling the increase in cost. Augmentation
keeps a given facility set open and adds a bounded number of new
locations. Both procedures use a prescribed number of randomized
iterations or samples, with a failure allowance fixed before the call.
Their size and running-time bounds hold on every execution; only
their approximation guarantees may fail.

\begin{lemma}[Restricted removal]
\label{lem:remove}
Let \(q\ge1\), \(r\ge0\), and \(|X_0|=q+r\). Given ordered
client-to-candidate distance trees, \(\Remove(P,X_0,r)\) returns a set
\(Y\subseteq X_0\) of size \(q\). Except with its assigned failure
probability,
\[
 \Phi_P(Y)\le2\Phi_P(X_0)+12F_q(P).
\]
Its work is \(\widetilde O((n+|X_0|)(r+1))\). For \(r=0\), it
returns \(X_0\).
\end{lemma}
\begin{proof}
Apply \Cref{lem:restricted-local-search} with \(p=1\).
That lemma proves the approximation and fixed work bound for
weighted demand, allowing both candidates and comparison
facilities anywhere in the ambient metric.
\end{proof}

The procedure is the restricted randomized local search
of~\cite{BCGLP24}, used in~\cite[Lemma~11.1]{BCF25}.
Starting from a \(q\)-subset of \(X_0\), it samples a
candidate \(z\) that is not currently open and deletes the
best facility from \(S\cup\{z\}\). Deleting \(z\) leaves
\(S\) unchanged, so costs never increase. The
\(\widetilde O(r)\) iteration cap is fixed before the call.
The proof in \Cref{app:restricted-local-search} groups comparison
facilities sharing a nearest candidate, and handles zero cost
and the positive-weight range explicitly.

\begin{lemma}[Augmentation {\cite[Lemma~11.3]{BCF25}}]\label{lem:develop}
For an absolute \(\beta\ge1\), a facility set \(U\), and an integer
\(a\ge0\), \(\Develop(P,U,a)\) returns \(U\cup F\), where
\(F\subseteq\supp(P)\) and \(|F|\le a\). Except with its assigned
failure probability,
\[
 \Phi_P(U\cup F)\le
 \beta\min_{\substack{H\subseteq\supp(P)\\|H|\le a}}
                  \Phi_P(U\cup H).
\]
Given distances to \(U\), the work is
\(\widetilde O((n+|U|)(a+1))\).
For each \(q\ge1\), a static solution of at most \(q\) facilities
and cost at most \(\beta F_q(P)\) can also be found in
\(\widetilde O((n+1)(q+1))\) work, enlarging \(\beta\) if necessary.
\end{lemma}

We briefly describe the reduction that implements augmentation
using a static solver. If \(a=0\) or \(\Phi_P(U)=0\), return
\(U\). If \(|\supp(P)\setminus U|\le a\), open all remaining
positive-weight locations. When \(U=\varnothing\) and \(a\ge1\),
run the static algorithm directly. In the remaining case,
contract all facilities of \(U\) to a single location \(u_*\),
transfer their client mass there, and define
\[
 d'(x,u_*)=d(x,U),\qquad
 d'(x,y)=\min\{d(x,y),d(x,U)+d(y,U)\}
\]
on \((\supp(P)\setminus U)\cup\{u_*\}\).
The static solver chooses facilities only from this finite
contracted set.
The distance between two remaining locations may use a shortcut
through \(u_*\). As long as \(u_*\) is open, however, such a
shortcut cannot improve on connecting directly to \(U\).
Consequently, any solution containing \(u_*\) has exactly the
same objective after its other facilities are lifted back to
the original metric.

We must ensure that the static solution contains \(u_*\).
Let \(C_U=\Phi_P(U)>0\),
\(\delta_U=\min_{x\in\supp(P)\setminus U}d(x,U)>0\), and let
\(\beta_{\rm st}\) denote the static approximation constant.
The following finite weight is sufficient:
\[
 w_* =\sum_{x\in U}w_P(x)+(\beta_{\rm st}+1)C_U/\delta_U.
\]
Opening \(u_*\) alone costs \(C_U\). Any solution omitting it
incurs more than \(\beta_{\rm st}C_U\) from the mass at
\(u_*\), so every successful static solution with at most
\(a+1\) facilities must include it. Lifting such a solution
opens at most \(a\) new positive-support facilities in addition
to \(U\). If a capped call omits \(u_*\), we return \(U\)
without retrying. The facility budget therefore holds even
when the static approximation fails.

The weighted static running time follows from Mettu and
Plaxton~\cite{MP02,MP04}. To compare with the ambient optimum,
project each ambient optimal facility to a nearest positive-support
location and discard duplicates. This increases client distances
by at most a factor of two, so we may choose
\(\beta\ge2\beta_{\rm st}\). The forcing weight affects only
the logarithmic weight-range factors. In the padded metric
introduced below, it is at most
\(nw_{\max}[1+2(\beta_{\rm st}+1)\Delta]\) and at least
\((\beta_{\rm st}+1)w_{\min}\). Here the extrema range over original
positive weights, and empty support is handled directly.

\subsection{Separated Pairs and Deletion}

The local repair analysis compares two facility sets through pairs
that are close to each other but far from competing facilities.
When few such pairs exist, the next lemma shows that one of the
sets contains facilities that can be removed at limited cost.

\begin{definition}[Well-separated pairs]\label{def:separated}
Fix a sufficiently large absolute \(\Gamma\). A pair
\((u,v)\in U\times V\) is well separated if
\[
 d(u,U\setminus\{u\})\ge\Gamma d(u,v),\qquad
 d(v,V\setminus\{v\})\ge\Gamma d(u,v).
\]
Ties are broken consistently. We fix \(\Gamma\) large enough for
the robust replacement arguments for both \(k\)-median and
\(k\)-means.
\end{definition}

For \(\Gamma>2\), no facility can belong to two such pairs,
so the well-separated pairs form a matching. The following
result converts a shortage of pairs into a bound on how many
centers can be deleted.

\begin{lemma}[Deletion from few separated pairs {\cite[Lemma~10.14]{BCF25}}]
\label{lem:separated-delete}
Let \(|U|=k\), \(|V|=k+s\), and suppose there are exactly \(k-m\)
well-separated pairs. If \(m\ge s\), some subset of \(U\) has at
most \(k-\lfloor(m-s)/4\rfloor\) facilities and cost at most
\[
 6\Gamma\bigl(\Phi_P(U)+\Phi_P(V)\bigr).
\]
\end{lemma}

We use the weighted formulation in the full version of~\cite{BCF25},
where facilities may lie outside the current demand support.
Its fractional openings lie in \(U\), and its assignment bounds
use distances and client weights; historical and dummy facilities
therefore satisfy the same hypotheses. The fractional opening
budget is at most \(k-(m-s)/4\). It is consequently feasible
for the integer budget \(k-\lfloor(m-s)/4\rfloor\), at which
the integrality-gap bound is applied on the finite domain \(U\).

\subsection{Padding and Proper Projection}\label{subsec:padding}

The inner algorithm keeps exactly \(k\) shadow facilities, even
when fewer than \(k\) client locations are occupied. To make
this possible, we add \(2k\) distinct zero-weight dummy locations
to the fixed metric space. Set the distance between any dummy
and any real location, or between two distinct dummies, to
\(2\Delta\). These distances preserve the triangle inequality.
The resulting metric has nonzero distances in
\([1,\Delta_d]\), where \(\Delta_d=2\Delta\).
For the resource bounds, write
\(L_\Delta=1+\lceil\log_2(2\Delta)\rceil\).
The multiscale constructions may use another fixed logarithm
base, which changes these bounds only by an absolute factor.
If a dummy serves positive demand, replacing it by a real
location cannot increase any connection distance. Thus padding
does not decrease \(F_j\) for \(j\ge1\).

The padded reference contains \(m=n+2k\) locations. We also
keep distance rows for at most \(k\) historical shadow facilities,
so the number of working row keys remains \(\Theta(n+k)\).
Zero-weight locations contribute no demand constraints.
Dummies may occupy internal facility slots, but they are never
included in the published proper solution.

\begin{lemma}[Dynamic distinct projection {\cite[Lemma~9.1 and Section~9.2]{BCGLP24}}]
\label{lem:projection}
For a facility set \(A\) and a dynamic target support \(B\), one can
maintain \(\pi(A,B)\subseteq B\) with
\[
 |\pi(A,B)|=\min\{|A|,|B|\},\qquad
 d(x,\pi(A,B))\le2d(x,A)\quad(x\in B).
\]
A source or target insertion or deletion changes the projected set by
at most two elements and takes \(\widetilde O(|A|+|B|)\) time.
If \(|B|<|A|\), the projected set is \(B\).
\end{lemma}

Here \(\pi(A,B)\) denotes a set of distinct projected facilities;
it is not the nearest-facility map \(\pi_A(x)\) used elsewhere.
The projection processes source facilities in a fixed order and
selects a nearest target that has not already been selected.
Once all targets have been selected, every remaining prefix
projection is \(B\); its available-target tree is empty.
This convention handles support sizes crossing \(|A|\), including
empty support, without rebuilding the projection. The recourse
bound applies to one source or target insertion or deletion;
an operation changing both sets is processed as two atomic events.
Summing the pointwise inequality over the current client weights
gives the corresponding cost bound.

\paragraph{Persistence and allocation.}
Snapshots share persistent roots. We charge each newly allocated
node to the operation that creates it. Retiring a root does not
trigger recursive reclamation during an update, so the time
bound counts only the work actually performed. Total allocation
is bounded by preprocessing plus the executed work over the
prescribed horizon. We do not claim a space bound depending
only on the current input size.

We now have the ingredients for selecting a checkpoint core:
lazy protection controls edited locations, restricted removal
creates free facility slots, and LP interpolation compares the
optima at nearby budgets. The next section combines them into a
window that can be computed from the reference solution itself.

%% file: sections/step2_cores.tex
\section{Step 2: Protected Cores and Stability Windows}
\label{sec:step2}\label{sec:windows}

The warm-up already achieves a constant-factor approximation to the
current optimum, but may use up to \(2k\) facilities. We recover the
exact budget by retaining fewer checkpoint facilities and reserving
the remaining positions for edits. The main question is how many
positions can be reserved while keeping the approximation guarantee.

Throughout Steps~2--5, assume \(k\ge2\). Let \(\beta\ge1\) be the
fixed approximation constant for the static initialization and
augmentation interfaces from Step~1, and set
\begin{equation}\label{eq:alpha-fixed}
 \alpha=160\beta.
\end{equation}
Every checkpoint reference has exactly \(k\) facilities; on the
common success event, its cost is at most \(\alpha F_k\).
The constants \(\alpha\) and \(\Gamma\)
remain fixed throughout the execution. The geometric annotations needed
to refresh a reference are constructed in Step~3; they are not needed
for the window computation itself.

\subsection{The Protected-Core Invariant}
At checkpoint time \(p\), let \(U_p\) be the current reference.
For an integer \(0\le H\le k-1\), restricted removal returns
\(K_p=\Remove(P_p,U_p,H)\subseteq U_p\) with \(k-H\) facilities;
when \(H=0\), we simply keep \(K_p=U_p\). Here \(\Remove\) is the
static procedure of \Cref{lem:remove}, not an optimization oracle.
We seek a choice of \(H\) for which
\begin{equation}\label{overview:core}
 |K_p|=k-H,\qquad
 \Phi_{P_p}(K_p)\le C_0F_{k+H}(P_p),
\end{equation}
where \(C_0\) is an absolute constant. Comparing with the optimum
using \(k+H\) facilities accounts for decreases in the optimum over
the next \(H\) edits. Restricted removal alone does not give this
comparison; the remainder of the section certifies an appropriate
removal budget.

\begin{lemma}[Protection during a bounded age]
\label{lem:core-protection}\label{overview:lemma-protection}
If \eqref{overview:core} holds, then for every record time \(u\) with
\(u-p\le H\), the set \(K_p\cup I(p,u)\) has at most \(k\)
facilities and cost at most \(C_0F_k(P_u)\).
\end{lemma}
\begin{proof}
The buffer uses at most one facility per record edit, so the
cardinality is at most \(k-H+(u-p)\). For cost, lazy updating
and monotonicity in the center budget give
\begin{equation}\label{overview:protected}
 \Phi_{P_u}(K_p\cup I(p,u))
 \le\Phi_{P_p}(K_p)
 \le C_0F_{k+H}(P_p)
 \le C_0F_{k+(u-p)}(P_p)
 \le C_0F_k(P_u).
\end{equation}
\end{proof}

Set \(B=H+1\). The value \(B\) is the checkpoint's stability window:
the core remains protected for the next \(B-1\) edits. Every record
edit increases its age, including repeated edits at the same location.
When \(B=1\), no time or spare position is available for a buffer;
Step~5 handles the next edit synchronously.

\subsection{Stability Parameters and Interpolation}
An edit can change one weight by an arbitrary amount. We therefore
measure stability through the number of facilities, rather than a
bound on the change in cost. For \(M\ge1\), define the
\emph{removable slack}
\[
 r_M(P)=\max\{r\in\{0,\ldots,k-1\}:
                   F_{k-r}(P)\le M F_k(P)\}.
\]
This measures how many facilities can be given up at a loss of at
most a factor \(M\). To compare the old core with an optimum that
uses additional facilities, define also
\begin{equation}\label{eq:universal-slack}
 b(P)=\max\{b\in\{0,\ldots,k-1\}:
                   F_{k-b}(P)\le16F_{k+b}(P)\}.
\end{equation}
Zero is feasible in both definitions. Neither parameter is computed
by the algorithm; they are used to analyze the removal tests below.

\begin{lemma}[Lipschitz stability]\label{lem:lipschitz}
One weighted record edit changes \(b(P)\) by at most one.
\end{lemma}
\begin{proof}
Let \(P,Q\) differ by one edit. For \(b=b(P)\ge1\), the lazy
inequalities give
\[
 F_{k-b+1}(Q)\le F_{k-b}(P)
 \le16F_{k+b}(P)\le16F_{k+b-1}(Q).
\]
Hence \(b(Q)\ge b(P)-1\). This inequality is automatic if
\(b(P)=0\). Interchanging \(P,Q\) proves the claim.
\end{proof}

Suppose a solution can lose \(R\) facilities while keeping its cost
within a factor \(M\). We need a similar comparison at a smaller loss
\(H\), and also against a solution with \(k+H\) facilities. The LP
interpolation from \Cref{sec:prelim} provides both bounds. We state
the argument in a form that also applies to squared connection costs.

\begin{lemma}[Two-sided interpolation from an LP gap]
\label{lem:gap-interpolation}
Suppose the integral optimum values \(G_j\) and convex LP values \(L_z\)
are defined on a common facility domain, with
\(L_j\le G_j\le gL_j\) at all integer budgets under comparison,
where \(g\ge1\). If \(0\le R\le k-1\) is integral,
\(M\ge1\), and \(G_{k-R}\le MG_k\), then every integer
\(0\le H\le R/(4gM)\) satisfies
\[
 G_{k-H}\le(g+\tfrac14)G_k,
 \qquad G_k\le\tfrac{4g}{3}G_{k+H}.
\]
\end{lemma}
\begin{proof}
For \(H=0\) both claims are immediate. Otherwise put \(\lambda=H/R\).
Convexity and the gap bound imply
\[
 G_{k-H}\le g\bigl(\lambda L_{k-R}+(1-\lambda)L_k\bigr)
 \le g(\lambda M+1-\lambda)G_k
 \le(g+\tfrac14)G_k.
\]
Next set \(\mu=H/(R+H)\), so
\(k=\mu(k-R)+(1-\mu)(k+H)\). Then
\[
 G_k\le g\mu MG_k+g(1-\mu)G_{k+H}.
\]
Since \(g\mu M\le1/4\), rearrangement proves the second inequality.
No optimum value is used as a denominator, so the argument also covers
zero optimum.
\end{proof}

\begin{corollary}[Median slack comparison]
\label{lem:two-interpolations}\label{cor:slack-equivalence}
If \(M\ge1\), \(R\in\mathbb Z\) with \(0\le R\le k-1\), and
\(F_{k-R}\le MF_k\), then every integer \(0\le H\le R/(12M)\)
satisfies
\[
 F_{k-H}\le\tfrac{13}{4}F_k,
 \qquad F_k\le4F_{k+H},
 \qquad H\le b(P).
\]
For \(M\ge16\), we also have
\(b(P)\le r_M(P)<12M(b(P)+1)\).
\end{corollary}
\begin{proof}
Apply \Cref{lem:gap-interpolation} with \(g=3\). The product of its
factors is \(13<16\), which proves \(H\le b(P)\).
Furthermore, \(F_{k-b}\le16F_{k+b}\le MF_k\) gives \(b\le r_M\).
For the reverse comparison, apply the first conclusion with
\(R=r_M\) and \(H=\lfloor R/(12M)\rfloor\).
\end{proof}

\subsection{Computing the Window and Core}
Starting from the known reference \(U\), we try increasingly large
removal budgets and stop when the computed cost becomes too large.
A fixed fraction of the last accepted budget is reserved for edits.
This gives an explicit window without evaluating any optimum value.

Fix the constants
\begin{equation}\label{eq:window-constants}
 \begin{split}
 \eta&\ge\max\{16,6\Gamma(\alpha+2)\},\qquad A_{\rm test}=14\eta,\\
 M_0&=A_{\rm test}\alpha,\qquad
 D=\lceil12M_0\rceil,\qquad C=2D.
 \end{split}
\end{equation}
The procedure \(\WindowCore(P,U)\) takes a weighted input \(P\)
and an \(\alpha\)-approximate \(k\)-facility reference \(U\).
It tests removal budgets in the ordered list \(1,2,4,\ldots\),
stopping at \(k-1\) and appending \(k-1\) when necessary.
A test at \(r\) accepts if the measured cost of
\(\Remove(P,U,r)\) is at most \(A_{\rm test}\Phi_P(U)\).
Stop at the first rejection, and let \(R\) be the last accepted budget;
put \(R=0\) if the first test rejects. Finally return
\begin{equation}\label{eq:window-choice}
 H=\lfloor R/D\rfloor,\qquad B=H+1,
 \qquad K=\Remove(P,U,H).
\end{equation}
When \(H=0\), keep \(K=U\) without calling the removal procedure.
Every test uses a computed solution and its measured cost, not the
parameters \(b(P)\) or \(r_M(P)\).

\begin{theorem}[Computable window and core]
\label{thm:window-core}\label{overview:lemma-window}
Let \(k\ge2\), \(|U|=k\), and
\(\Phi_P(U)\le\alpha F_k(P)\). If the static calls succeed,
\(\WindowCore(P,U)\) returns \(B,H,K\) such that
\begin{subequations}\label{overview:window}
\begin{align}
 H&=B-1,\quad K\subseteq U,\quad |K|=k-H,
       \label{eq:window-size}\\
 B&\le b(P)+1\le CB,\qquad r_\eta(P)<2DB,
       \label{eq:window-scale}\\
 \Phi_P(K)&\le C_0F_{k+H}(P),\qquad C_0=8\alpha+156.
       \label{eq:window-protection}
\end{align}
\end{subequations}
Given the distance-tree data structure, the work is
\(\wtO((n+k)B)\).
\end{theorem}
\begin{proof}
\emph{Accepted prefix.} The static removal guarantee shows that
every tested budget \(r\le r_\eta\) is accepted:
\[
 \Phi_P(\Remove(P,U,r))
 \le2\Phi_P(U)+12\eta F_k(P)
 \le(2+12\eta)\Phi_P(U)
 \le A_{\rm test}\Phi_P(U).
\]
Here \(F_k(P)\le\Phi_P(U)\) because \(U\) has \(k\) facilities.
If the first rejected budget is \(r_f\), then \(r_\eta<r_f\).
When \(R>0\), the doubling schedule gives \(r_f\le2R\).
For \(R=0\), rejection occurs at one and \(r_\eta=0\).
If every test accepts, then \(R=k-1\ge r_\eta\).
Thus \(r_\eta\le2R\) in every case. The argument relies only on
the guaranteed accepted prefix; tests beyond that prefix need not
behave monotonically.

\emph{Window scale.} Acceptance at \(R\) gives
\(F_{k-R}\le M_0F_k\), also valid when \(R=0\).
By \Cref{lem:two-interpolations}, \(H\le b(P)\). Since
\(\eta\ge16\) and \(R<D(H+1)\),
\[
 b+1\le r_\eta+1\le2R+1\le2D(H+1)=CB,
 \qquad r_\eta<2DB.
\]

\emph{Core cost.} The removal guarantee and the same interpolation give
\[
 \Phi_P(K)\le2\alpha F_k+12F_{k-H}
 \le(2\alpha+39)F_k
 \le(8\alpha+156)F_{k+H}.
\]
For \(H=0\), \(K=U\) satisfies this bound directly.

\emph{Work.} The sum of tested budgets, including the first rejected
one, is \(O(R+1)\). The final budget is at most \(R\), and
\(R+1\le D(H+1)=DB\). Summing the static work bounds proves the
claim.
\end{proof}

The distance-tree implementation is given in \Cref{subsec:trees}.
The theorem needs only the static interface from Step~1: deleting
\(r\) candidates costs \(\wtO((n+k)(r+1))\) on a private version
of the stored rows. Step~3 uses the resulting core and window to
construct a nearby replacement reference.

Different snapshots may yield different accepted removal budgets, so
the computed window need not change by at most one. It is enough that
nearby snapshots yield windows of comparable lengths. The following
bound follows from the stability of \(b\).

\begin{lemma}[Neighboring computed windows]\label{lem:window-comparison}
Let \(B,B'\) satisfy \eqref{eq:window-scale} for instances separated
by \(s\le aB\) record edits, with \(0\le a<1\). Then
\[
 \frac{(1-a)B}{C}\le B'\le(C+a)B.
\]
\end{lemma}
\begin{proof}
Lipschitz stability gives
\(b(Q)+1\ge b(P)+1-s\ge(1-a)B\) and
\(b(Q)+1\le b(P)+1+s\le(C+a)B\).
Now use \(b(Q)+1\le CB'\) and \(B'\le b(Q)+1\).
\end{proof}

%% file: sections/step3_rebuilding.tex
\section{Step 3: Robustification and Amortized Rebuilding}
\label{sec:step3}

The protected core from Step~2 remains useful for a certified number
of record edits. To continue beyond that period, we need a new
reference solution and a new core. Recomputing an arbitrary static
solution would not suffice: it could replace all \(k\) facilities,
even after only a few edits. Our goal is to make the number of changes
and the rebuilding work proportional to the old window.

We first prove the local statement needed for this goal. Given an
annotated reference and a proposed \(k\)-facility set, batch
robustification repairs the geometric annotations and robust
certificates at a cost controlled by that batch's edits and proposed
facility changes. We then initialize these annotations, construct a
new checkpoint with \Refresh, and combine it with \WindowCore\ to
obtain an amortized baseline. All subroutines used by that baseline
are thus specified before the algorithm.

\subsection{Stable Geometric Certificates}
\label{geom:section}\label{sec:local-repair}

Each center label carries two locations. Its \emph{anchor} determines
the geometric scale at which we construct a certificate; its
\emph{proxy} is the facility that serves clients. A certificate records
the multiscale sequence used to obtain that proxy. Its precise
weighted definition appears in \Cref{block:robust-sequence}.

We must control two kinds of changes. A new anchor can shorten other
anchors' nearest-neighbor distances and cause a cascade of relocations.
A client edit can also invalidate certificates inherited from earlier
snapshots. We retain two kinds of history to handle these effects:
strong-reset times for anchor offsets and generation times for full
certificates. The bounds below are deterministic and do not depend on
the sizes of the client weights.

We work in the padded metric of \Cref{subsec:padding}, whose nonzero
distances lie in \([1,\Delta_d]\), where \(\Delta_d=2\Delta\). Write
\[
 L_\Delta=1+\lceil\log_2\Delta_d\rceil.
\]
For a finite set \(A\) of distinct anchors, write
\[
 r_A(a)=d(a,A\setminus\{a\})\qquad(a\in A),
\]
with \(r_{\{a\}}(a)=+\infty\). We interpret conditions indexed by an
empty set as vacuous. Since the metric space is fixed, distances to a
location remain available after its client or anchor has been deleted.

\subsubsection{Maintaining Approximate Nearest-Neighbor Scales}
\label{geom:witness-subsection}

Maintaining exact nearest-neighbor distances would be expensive: one
new anchor can become the nearest neighbor of many old anchors. We
instead maintain approximate scales that may be smaller than the true
distances but change at only a few locations per update.

To describe the update history, we distinguish a center label from an
\emph{anchor incarnation}. Each insertion creates a new incarnation,
even if its coordinate has appeared before. That coordinate remains
fixed until deletion. Relocating an anchor ends one incarnation and
starts another, without changing the center label. The distinction
will let us compare witness edges by their creation times.

\begin{lemma}[Stable nearest-neighbor scales]
\label{geom:stable-labels}
A deterministic data structure maintains labels \(\rho(a)\) for a dynamic
set \(A\) of distinct anchors. Whenever \(|A|\ge2\),
\[
 \frac{r_A(a)}{64}\le\rho(a)\le r_A(a)\qquad(a\in A).
\]
One insertion or deletion changes the labels of \(O(L_\Delta)\)
surviving incarnations, in addition to the inserted or deleted label.
For \(m=\max\{2,|A_{\rm before}|,|A_{\rm after}|\}\), the worst-case
work is \(O(mL_\Delta)\). The structure uses \(O(m)\) records and a
shared table of \(O(L_\Delta)\) dyadic scales. Building this table once
takes \(O(L_\Delta)\) work; thereafter, initialization on \(m\)
anchors takes \(O(m^2+m\log(L_\Delta+1))\) work.
\end{lemma}

\begin{proof}
Assume \(|A|\ge2\). We assign each anchor an outgoing witness edge.
During an update an anchor may temporarily have no witness; all such
anchors are repaired before the update finishes. The witnesses have
two possible forms.
\begin{enumerate}
\item An \emph{unpaired} anchor \(v\) points to a distinct anchor \(u\),
with length \(\ell(v)=d(v,u)\) satisfying
\[
 \ell(v)\le4r_A(v).
 \tag{G1}\label{geom:unpaired-invariant}
\]
For each target \(u\), at most one incoming unpaired edge lies in each
dyadic length interval \([2^b,2^{b+1})\).
\item A \emph{paired} anchor belongs to a disjoint pair \(\{v,w\}\).
The endpoints point to each other and satisfy
\[
 \ell(v)=\ell(w)=d(v,w),\qquad
 \ell(v)\le64r_A(v),\quad \ell(w)\le64r_A(w).
 \tag{G2}\label{geom:paired-invariant}
\]
Paired edges do not appear in the dictionaries of incoming unpaired edges.
\end{enumerate}
An anchor whose witness has been removed is called \emph{free}; it
remains in \(A\). For each witness we record when it was last reset.
An unpaired edge is exact at its reset time:
\[
 \ell(v)=r_A(v),
 \tag{G3}\label{geom:unpaired-reset}
\]
and a newly formed pair obeys the stronger bounds
\[
 \ell(v)\le12r_A(v),\qquad \ell(w)\le12r_A(w).
 \tag{G4}\label{geom:paired-reset}
\]
The endpoints of a newly formed pair have the same reset time. All
other resets receive distinct times.

\paragraph{Repair rule.}
To repair a free anchor \(v\), scan \(A\) for an exact nearest
neighbor \(u\), and put \(h=d(v,u)\). If \(u\) has no incoming
unpaired edge in the dyadic bucket containing \(h\), install
\(v\to u\). Otherwise, let \(w\to u\) be the edge already in that
bucket, with length \(h_w\). Remove it and pair \(v\) with \(w\).
The two lengths lie in the same bucket, so
\(h_w<2h\) and \(h<2h_w\); hence
\[
 d(v,w)<3h=3r_A(v),\qquad
 d(v,w)<3h_w\le12r_A(w).
\]
The pair therefore satisfies \eqref{geom:paired-reset}. Repairing
\(v\) changes at most one existing witness and leaves no additional
anchor free. It does not matter whether the target \(u\) is itself
paired or unpaired.

\paragraph{Insertion.}
Insert \(x\). Its appearance can only decrease nearest-neighbor
distances, so we remove witnesses whose lengths may now be too large.
Before repairing anything, mark every unpaired anchor
\(v\) with \(d(v,x)<\ell(v)/4\), and every paired anchor \(v\) with
\(d(v,x)<\ell(v)/64\). Call these anchors \emph{triggers}. Remove
all triggered unpaired edges and every pair containing a trigger. Make
\(x\) free, and repair all free anchors on the new, fixed set \(A\).
Every witness that survives the erasures satisfies its required bound:
both its previous nearest-neighbor distance and its distance to \(x\)
are at least \(\ell(v)/c\), where \(c=4\) or \(64\).

To count triggers, take two of the same type, \(u,v\), with \(v\)
reset later. The current incarnation of \(u\) was present when \(v\)
was reset. If \(a\) is the reset constant and \(c\) the trigger constant,
then
\[
 d(u,v)\ge\ell(v)/a,
 \qquad
 d(u,v)<(\ell(u)+\ell(v))/c.
\]
Consequently,
\[
 \ell(u)>\left(\frac ca-1\right)\ell(v).
 \tag{G5}\label{geom:trigger-growth}
\]
For unpaired triggers, \((a,c)=(1,4)\), giving a factor greater than
\(3\). For paired triggers, \((a,c)=(12,64)\), giving a factor greater
than \(13/3\). Equal reset times occur only within one current pair;
both endpoints cannot trigger, since that would imply
\[
 d(u,v)\le d(u,x)+d(x,v)<2d(u,v)/64.
\]
Witness lengths lie in \([1,\Delta_d]\). Thus each type has
\(O(L_\Delta)\) triggers, and the erasures free only
\(O(L_\Delta)\) anchors. Consequently, only \(O(L_\Delta)\)
repairs and witness changes are needed.

\paragraph{Deletion.}
When \(x\) is deleted, remove its outgoing witness, all incoming unpaired
edges, and its pair if present. The incoming dictionary contains at most
one edge per dyadic scale, so \(O(L_\Delta)\) anchors become free.
Deleting a point can only increase the surviving nearest-neighbor
distances; all other witnesses remain valid. Repair the free anchors
on the new set. Again there are \(O(L_\Delta)\) repairs and changes.

\paragraph{Labels and work.}
Set \(\rho(v)=\ell(v)/64\). Because a witness is a distinct current
anchor, \(r_A(v)\le\ell(v)\le64r_A(v)\), which proves the label bounds.
Store each witness, its optional mate, and balanced incoming-edge
dictionaries indexed by dyadic bucket. These use one record per edge.
An insertion scans all anchors for triggers; a deletion enumerates its
incoming dictionary. At most \(\min\{m,O(L_\Delta)\}\) repairs each
perform an \(O(m)\) nearest-neighbor scan. Bucket lookup and dictionary
updates fit within \(O(mL_\Delta)\), also when \(m<L_\Delta\).
Initialization repairs each free anchor once. Empty and singleton sets
are handled directly, with label \(+\infty\) for a singleton.
Transitions among sizes zero, one, and two change only constantly
many labels.
\end{proof}

\subsubsection{Keeping Anchors Close to Their Proxies}
\label{geom:rebase-subsection}

The approximate scales alone do not control relocations. Moving an
anchor to its proxy may shorten another anchor's nearest-neighbor
distance and force a second move. To prevent a long chain of such
relocations, we impose a stronger offset condition at selected reset
times and remember those times thereafter.

Let center label \(i\) have anchor \(a_i\), proxy \(v_i\), and offset
\(\delta_i=d(a_i,v_i)\). The proxies stay fixed during the repair
considered here. We maintain the weak offset condition
\[
 \delta_i\le r_A(a_i)/10,
 \qquad A=\{a_i\},
 \tag{G6}\label{geom:weak-offset}
\]
together with the following history. At its latest \emph{strong reset},
label \(i\) satisfied
\[
 \delta_i\le r_A(a_i)/40.
 \tag{G7}\label{geom:strong-offset}
\]
Its anchor and proxy have remained fixed since that reset, and the
label has remained present. Strong-reset times \(\tau_i\) are distinct.
Creating a label with \(a_i=v_i\), or rebasing its anchor to its proxy,
provides a strong reset with offset zero.

The reset history is part of the maintained state, not merely an
assumption made for the analysis. In particular, if
\(\tau_i>\tau_j\), then \(a_j\) was present at its current coordinate
when \(i\) was reset, and therefore
\[
 40\delta_i\le d(a_i,a_j).
 \tag{G8}\label{geom:chronological-offset}
\]
This separation between newer offsets and older anchors will bound
entire rebase cascades.

\paragraph{Why the weak condition does not suffice.}
\label{geom:weak-only-obstruction}
The limitation of the weak offset condition can be seen in a metric of constant
aspect ratio. Consider a tree with hub \(x\), edges \(xa_i\) of length \(6\), and
edges \(a_iv_i\) of length \(1\), for \(1\le i\le m\). Initially
\(r_A(a_i)=12\) and \(\delta_i=1\), so
\eqref{geom:weak-offset} holds. Inserting \(x\) as an anchor with
proxy \(x\) makes \(r_A(a_i)\le6<10\delta_i\) for every old label.
The violation persists until that label is rebased, regardless of which
other labels have moved. Thus the rebase-to-proxy rule requires \(m\)
moves, although the unpadded metric has aspect ratio \(14\).
This configuration does not admit the prescribed strong-reset history:
\eqref{geom:chronological-offset} would require \(40\le12\) for two
labels with distinct reset times.

The next lemma shows why keeping the strong-reset history is enough.
Its bound includes every relocation in a cascade, not just those
triggered directly by the original insertion.

\begin{lemma}[Rebase closure]
\label{geom:rebase-closure}
Suppose distinct anchors and distinct proxies satisfy
\eqref{geom:weak-offset} and the strong-reset history. Insert an anchor
\(x\) with proxy \(x\), where \(x\) is neither an existing anchor nor
an existing proxy. Keeping all proxies fixed, one can restore the weak
condition by rebasing \(O(L_\Delta)\) old anchors to their proxies.
Each label is rebased at most once.

For \(m\) labels after insertion, the closure requires
\(O(mL_\Delta)\) auxiliary work and \(O(L_\Delta)\) primitive anchor
insertions and deletions. Processing these operations with
\Cref{geom:stable-labels} gives combined worst-case work
\(O(mL_\Delta^2)\) and \(O(L_\Delta^2)\) stable-label changes.
\end{lemma}

\begin{proof}
For the counting argument, freeze the values of
\(a_i,v_i,\delta_i,\tau_i\) immediately before the insertion. We
continue to use these values even after a label has been rebased.
Labels with zero offset need no rebase.

We process violated labels with a queue. Initially, enqueue every
label \(i\) such that
\(d(a_i,x)<10\delta_i\). On processing \(i\), delete \(a_i\), insert
\(v_i\) as its anchor, and record a zero-offset strong reset. Scan all
labels not yet activated and enqueue each \(j\) satisfying
\(d(a_j,v_i)<10\delta_j\). A label enters the queue only once.
Once activated, it is processed even if the anchor that activated
it subsequently moves.

The anchors remain distinct. The initial weak condition prevents
\(v_i=a_j\) for \(i\ne j\), since this would give
\(d(a_i,a_j)=\delta_i\le d(a_i,a_j)/10\). Distinct proxies and the
choice of \(x\) also prevent collisions with \(x\) or with previously
inserted proxy anchors.

The activation history forms a forest. Labels activated directly by
\(x\) are its roots. If inserting \(v_a\) first activates label \(b\),
make \(a\) the parent of \(b\). Parents are processed before their
children, and every parent edge satisfies
\[
 d(a_a,a_b)<\delta_a+10\delta_b.
 \tag{G9}\label{geom:causal-distance}
\]
Write \(d_{ab}=d(a_a,a_b)\). If \(\tau_b>\tau_a\), chronological
separation gives \(\delta_b\le d_{ab}/40\), while the initial weak
condition gives \(\delta_a\le d_{ab}/10\). Then
\eqref{geom:causal-distance} would imply
\(d_{ab}<d_{ab}/10+10d_{ab}/40=7d_{ab}/20\), a contradiction.
Thus reset times strictly decrease along every parent edge:
\(\tau_a>\tau_b\). Applying chronological separation in this order
to \eqref{geom:causal-distance} gives
\[
 \delta_a<\frac{10}{39}\delta_b.
 \tag{G10}\label{geom:causal-offset-growth}
\]

We next show that every activated anchor is close to the original
insertion point, at a scale proportional to its original offset.
Put \(K_0=400/29\). Every activated label \(b\) satisfies
\[
 d(a_b,x)<K_0\delta_b.
 \tag{G11}\label{geom:activation-ball}
\]
The root case follows from the activation rule. For a child of \(a\),
induction gives
\[
 \begin{aligned}
 d(a_b,x)
 &\le d(a_b,a_a)+d(a_a,x)\\
 &<10\delta_b+(K_0+1)\delta_a\\
 &<\left(10+(K_0+1)\frac{10}{39}\right)\delta_b
 =K_0\delta_b.
 \end{aligned}
\]
This bound lets us count the entire forest, not just a single branch.
Order all activated labels by decreasing original reset time. For
two labels \(a,b\) in this order,
\eqref{geom:chronological-offset} and \eqref{geom:activation-ball} imply
\[
 40\delta_a\le d(a_a,a_b)<K_0(\delta_a+\delta_b).
\]
Hence
\[
 \delta_b>\left(\frac{40}{K_0}-1\right)\delta_a
 =\frac{19}{10}\delta_a.
 \tag{G12}\label{geom:all-activation-growth}
\]
Every activated offset is a nonzero distance in \([1,\Delta_d]\).
There are at most \(1+\lceil\log_{19/10}\Delta_d\rceil\) activated
labels.

It remains to check that the queue restores the invariant. Every
processed label has offset zero. For an unprocessed label \(b\),
all unchanged old anchors remain at distance at least
\(10\delta_b\) from \(a_b\). Any other final anchor is \(x\) or an
inserted proxy; if it were closer, its insertion would have activated
\(b\). Thus the weak condition holds for every label. Each rebase also
records a new valid strong reset, preserving the history for future
closures.

The initial scan and each rebase scan cost \(O(m)\). Each of the
\(O(L_\Delta)\) rebases performs two primitive anchor operations.
This proves the auxiliary bound; invoking \Cref{geom:stable-labels}
for every primitive operation proves the combined bounds.
\end{proof}

\begin{corollary}[Deletions, forced rebases, and coordinate coincidences]
\label{geom:forced-rebase}
Deleting a label preserves the weak offset condition for survivors.
A forced rebase of one anchor to its fixed proxy, followed by closure,
uses \(O(L_\Delta)\) primitive anchor operations. A new proxy distinct
from all current proxies can be installed with the same bound even
when it coincides with a current nominal anchor.
\end{corollary}

\begin{proof}
Deleting an anchor can only increase the nearest-neighbor distances
of those that survive, so no closure is needed for deletion alone.
For a forced rebase of label \(i\), delete \(a_i\) and insert \(v_i\)
with zero offset. If \(a_i=v_i\), no operation is needed. Otherwise,
the weak condition ensures that \(v_i\) is not another anchor, and
proxy distinctness ensures that it is not another proxy. Apply
\Cref{geom:rebase-closure} to the remaining labels.

The remaining case is a new proxy \(x\) that coincides with an
existing nominal anchor \(a_i\). Since proxies are distinct, it
differs from \(v_i\).
First rebase \(a_i\) to \(v_i\) and complete its closure. Every
rebase destination is an old proxy, so \(x\) is then unoccupied.
Insert \(x\) with zero offset and complete a second closure. The two
closures change the bound by only a constant factor.
\end{proof}

Applying the corollary successively, \(b\) requested insertions,
deletions, or forced rebases, each with its closure, incur
\(O(bL_\Delta)\) primitive anchor
operations, \(O(bL_\Delta^2)\) stable-label changes, and
\(O(mbL_\Delta^2)\) work. Additional strong resets satisfying
\eqref{geom:strong-offset} may occur between closures.

\subsubsection{Bounding the Overlap of Historical Certificates}
\label{geom:top-subsection}

The preceding bounds concern changes to anchors. Client edits create
a different problem: one edited location may lie in the regions used
to construct several old robust certificates. We need to know how many
certificates can be affected before deciding which ones to rebuild.

Let \(B_X(q,R)=\{x\in X:d(x,q)\le R\}\). The client-dependent
certificates defined in \Cref{block:robust-sequence} are built on
snapshots and may survive several anchor rebases. We now state the
lifetime conditions sufficient to bound the overlap of their full top
balls. The argument concerns center labels, which survive rebases,
rather than anchor incarnations, which do not.

For every current positive-height certificate \(i\), retain its
generation time \(\sigma_i\), top point \(q_i\), output proxy \(v_i\),
and full radius \(R_i=10^{\ell_i}\), where \(\ell_i\ge1\). Require:
\begin{enumerate}
\item At generation, \(q_i\) is the label's anchor, there are at least
two anchors, and
\[
 R_i<r_{A(\sigma_i)}(q_i)/20.
 \tag{G13}\label{geom:top-generation}
\]
\item The certificate's output obeys
\[
 d(q_i,v_i)\le R_i/2.
 \tag{G14}\label{geom:top-output-distance}
\]
\item Until replacement or retirement, the center label remains
present, its certificate data remain unchanged, and its anchor belongs
to \(\{q_i,v_i\}\).
\end{enumerate}
Generation times are distinct. We assign an order to certificates
produced in the same batch while the anchors stay fixed. A rebase changes a strong-reset
time \(\tau_i\), but leaves \(\sigma_i\) unchanged. Deleting a center
label retires its certificate, and later reintroducing that center creates
a fresh label. Rebuilding a certificate may instead change the proxy of
a surviving label; this replaces the old certificate and starts a new
certificate lifetime.

\begin{lemma}[Historical overlap and contamination]
\label{geom:top-ply}\label{geom:pollution}
Under the generation and lifetime conditions above, the family
\[
 \mathcal B=\{B_X(q_i,R_i):i\text{ is a current positive-height certificate}\}
\]
has ply \(O(L_\Delta)\): every point of \(X\) belongs to at most
\(O(L_\Delta)\) of its balls. If a certificate's defining client balls
are all contained in its full top ball, \(s\) weighted-record edits
directly contaminate at most \(O(sL_\Delta)\) certificates. Every
unmarked full certificate retains its previous validity.
\end{lemma}

\begin{proof}
Fix \(x\in X\) and consider two current certificates whose full top
balls contain \(x\). Suppose \(i\) was generated later than \(j\). At time
\(\sigma_i\), label \(j\) already carried its current certificate:
otherwise that certificate's generation time would be later than
\(\sigma_i\). By its lifetime condition,
\(a_j(\sigma_i)\in\{q_j,v_j\}\), so
\(d(q_j,a_j(\sigma_i))\le R_j/2\). Both \(q_i\) and
\(a_j(\sigma_i)\) were anchors. Their separation gives
\[
 20R_i<r_{A(\sigma_i)}(q_i)
 \le d(q_i,a_j(\sigma_i))
 \le R_i+\frac32R_j.
\]
Therefore
\[
 R_j>\frac{38}{3}R_i.
 \tag{G15}\label{geom:top-chronological-growth}
\]
Thus, when the balls containing \(x\) are listed from newest to
oldest, their radii grow by a factor greater than \(38/3\) at each
step. Each radius is at least \(1\) and,
by \eqref{geom:top-generation}, less than \(\Delta_d/20\). There are
at most \(1+\lceil\log_{38/3}\Delta_d\rceil\) such balls. No comparison
between an old top radius and a current nearest-neighbor scale is used.

To find affected certificates, mark every top ball containing an edited
coordinate. The ply bound gives \(O(sL_\Delta)\) marked certificates. Every defining ball
of an unmarked certificate has the same clients and weights as before,
so all of its conditions remain valid, independently of the magnitude
of each weight change.
\end{proof}

Height-zero certificates have no client-dependent conditions and need
no marking. When a required scale decreases, we keep the complete
old positive-height certificate, including its radius and generation
time. Without those data, the chronological comparison above and the
restart argument in \Cref{block:restart} would no longer apply.

\subsection{Batch Robustification}
\label{block:section}\label{sec:block}

Suppose the weighted instance changes from \(P\) to \(Q\), and a new
set of facilities \(W\) is proposed. We want to turn \(W\) into a
robust solution without paying for the entire history of earlier
updates. The work and the number of changed facilities must instead
depend on this batch of edits and on the change from the old facility
set to \(W\).

We handle geometry and certificate construction separately. We first
install the proposed facilities and repair the anchors, without moving
the proxies of retained labels. Only after the anchors have settled
do we rebuild the necessary certificates. The geometric lemmas of
\Cref{sec:local-repair} bound the first part. For the second, we
compare costs on the Voronoi cells of the original proposal \(W\),
which remain fixed throughout reconstruction.

Throughout the batch analysis, \(k\ge2\), nonzero distances lie in
\([1,\Delta_d]\), and
\[
 L=1+\lceil\log_{10}\Delta_d\rceil=\Theta(L_\Delta).
\]
The case \(k=1\) is handled separately in \Cref{sec:weighted-inner}.
Client scans visit only positive-weight locations. Zero-weight
padding and historical geometric data are kept separately, using
\(O(k)\) additional records. As in the geometric analysis above, optima here allow facilities
anywhere in the ambient space \(X\).

\subsubsection{Robust Certificates and the Maintained State}
\label{block:sequences}

We begin with the weighted version of a robust sequence. Starting
from an anchor, the sequence examines successively smaller balls.
At a given scale it either keeps the current point or moves to a
point whose connection cost on the current ball is small. The rule
below makes this choice precise, including when a ball has zero mass.

For a weighted client subset \(B\) of an instance \(P\), write
\[
 w(B)=\sum_{x\in B}w_P(x),\qquad
 \cost(z,B)=\sum_{x\in B}w_P(x)d(x,z),\qquad
 F_1(B)=\min_{z\in X}\cost(z,B).
\]
When \(w(B)>0\), let
\(\operatorname{av}(z,B)=\cost(z,B)/w(B)\).
The notation \(\Ball_R^P(z)\) denotes the clients of \(P\) within
closed distance \(R\) of \(z\), with their current weights. Ambient
balls are distinguished explicitly when needed.

\begin{definition}[Weighted robust sequence]
\label{block:robust-sequence}
A sequence \((p_0,\ldots,p_\ell)\) is \(\ell\)-robust for \(P\) if the
following rule holds for every \(i=\ell,\ell-1,\ldots,1\), with
\(B_i=\Ball_{10^i}^P(p_i)\).
\begin{enumerate}
\item If \(w(B_i)=0\), set \(p_{i-1}=p_i\).
\item If \(w(B_i)>0\) and
\(\operatorname{av}(p_i,B_i)\ge10^i/5\), set \(p_{i-1}=p_i\).
\item Otherwise choose \(p_{i-1}\in B_i\cup\{p_i\}\) such that
\begin{equation}\label{block:median-condition}
 \cost(p_{i-1},B_i)
 \le\min\{\cost(p_i,B_i),\,3F_1(B_i)\}.
\end{equation}
\end{enumerate}
A \(0\)-robust sequence has one point and no client-dependent condition.
A certificate stores the full sequence, its top \(p_\ell\), and its
generation time.
\end{definition}

This is the weighted version of~\cite[Definition 10.2]{BCF25}.
The procedure
\(\MakeRobustProc(P,a,\ell)\) starts at \(p_\ell=a\) and processes the
levels downward. Each invocation has a failure allowance
\(\varepsilon\in(0,1)\), omitted from the notation. At a positive-mass,
low-average level, it draws weighted-independent samples from \(B_i\),
evaluates their costs, and chooses the best candidate among the samples
and \(p_i\).

To see why sampling suffices, let \(Y\) be one weighted sample from
\(B_i\), and let \(z^*\) be an ambient one-median. The triangle
inequality gives
\[
 \mathbb E[\cost(Y,B_i)]\le2\cost(z^*,B_i).
\]
When the optimum is positive, Markov's inequality gives probability at least
\(1/3\) of cost at most \(3F_1(B_i)\). When the optimum is zero, the same
expectation bound implies that every sample has zero cost with
probability one. Taking \(O(1+\log((\ell+1)/\varepsilon))\) samples at
each nontrivial level therefore suffices for all median inequalities with
probability at least \(1-\varepsilon\). Conditional on the previously
constructed higher levels, the same argument applies at the current
ball. Empty balls need no samples, and \(\ell=0\) returns \(a\).
For \(n\) positive-support locations the prescribed work is
\begin{equation}\label{block:make-work}
 O\bigl((n+1)(\ell+1)[1+\log((\ell+1)/\varepsilon)]\bigr).
\end{equation}

It is useful to distinguish what holds for every sample outcome from
what holds only when the sampling succeeds. A \emph{cost-nonincreasing}
certificate obeys the zero-mass and high-average rules, and in the
remaining case chooses a point in \(B_i\cup\{p_i\}\) whose cost is no
larger than that of \(p_i\). Every execution has this property because
\(p_i\) is among the candidates. A \emph{robust-valid} certificate
also satisfies the \(3F_1(B_i)\) bound. Deterministic work and recourse
use only the first property. For approximation, we additionally require
the common success event for the certificates that remain in use.

\begin{lemma}[Sequence inequalities {\cite[Lemmas 10.3--10.4]{BCF25}}]
\label{block:sequence-bounds}
Every cost-nonincreasing certificate \((p_0,\ldots,p_\ell)\) satisfies,
for \(1\le i\le \ell\),
\begin{align}
 d(p_{i-1},p_i)&\le\frac25\,10^i,
 \label{block:step-displacement}\\
 d(p_0,p_i)&\le\frac12\,10^i.
 \label{block:total-displacement}
\end{align}
The ambient ball of radius \(10^{i-1}\) about \(p_{i-1}\) is contained
in the ambient ball of radius \(10^i\) about \(p_i\). Set
\(B_0=\Ball_1^P(p_0)\). For every \(0\le i\le \ell\) and weighted client
subset \(B_i\subseteq C\subseteq P\),
\begin{equation}\label{block:sequence-cost}
 \cost(p_0,C)\le\frac32\cost(p_i,C).
\end{equation}
These conclusions do not require robust validity.
\end{lemma}

\begin{proof}
The first bound controls how far the sequence moves at one level.
Whenever a move occurs, cost nonincrease and the triangle inequality
imply
\[
 d(p_{i-1},p_i)\le2\operatorname{av}(p_i,B_i)<\frac25\,10^i.
\]
Summing the geometric series proves \eqref{block:total-displacement}.
Moreover, \(10^{i-1}+d(p_{i-1},p_i)<10^i\), proving ambient containment
and hence \(B_0\subseteq B_1\subseteq\cdots\subseteq B_\ell\).

For the cost comparison, use induction on \(i\). The base case is
immediate. Since \(B_{i-1}\subseteq B_i\), the induction
hypothesis and cost nonincrease give
\[
 \cost(p_0,B_i)\le\tfrac32\cost(p_{i-1},B_i)
                 \le\tfrac32\cost(p_i,B_i).
\]
For \(x\in C\setminus B_i\), we have \(d(x,p_i)>10^i\), so
\eqref{block:total-displacement} implies
\(d(x,p_0)\le(3/2)d(x,p_i)\). Sum the inside and outside bounds.
The factor \(3/2\) is applied once to each client, rather than being
multiplied at every level. At a zero-mass level the sequence does not move.
\end{proof}

The algorithm cannot discard a certificate whenever its anchor moves:
doing so could create many reconstructions in one update. We instead
keep the certificate attached to the center label and record when it
was created, what its original top radius was, and whether its anchor
has since moved to the proxy. The following invariant lists the data
carried from one batch to the next.

\begin{invariant}[Deterministic state of a \(k\)-set]
\label{block:state}
For \(k\ge2\), the state on \(P\) contains \(k\) center labels, distinct
anchors \(A=\{a_i\}\), proxies \(V=\{v_i\}\), and these annotations.
\begin{enumerate}
\item With \(\delta_i=d(a_i,v_i)\), the offsets satisfy
\begin{equation}\label{block:weak-offset}
 \delta_i\le r_A(a_i)/10.
\end{equation}
Each label has a distinct strong-reset time \(\tau_i\), at which its
offset was at most \(r_{A(\tau_i)}(a_i)/40\). Its anchor and proxy have
remained fixed, and the label present, since then. Creation, rebase,
and certificate reconstruction record fresh reset times, including
when the offset is zero.
\item The data structure of \Cref{geom:stable-labels} maintains
\(\rho_i:=\rho(a_i)\), with
\begin{equation}\label{block:rho-range}
 r_A(a_i)/64\le\rho_i\le r_A(a_i).
\end{equation}
The required certificate level is
\begin{equation}\label{block:required-level}
 \ell_i^{\rm req}
 =\max\!\left\{0,\left\lceil\log_{10}(\rho_i/200)\right\rceil\right\}.
\end{equation}
\item Label \(i\) stores a full cost-nonincreasing certificate
\((p_{i,0},\ldots,p_{i,\ell_i})\) for \(P\), with \(p_{i,0}=v_i\) and
\(\ell_i\ge \ell_i^{\rm req}\). Writing \(q_i=p_{i,\ell_i}\) and
\(R_i=10^{\ell_i}\), its anchor satisfies
\begin{equation}\label{block:top-or-proxy}
 a_i\in\{q_i,v_i\}.
\end{equation}
Throughout this certificate's lifetime, its label remains present, its
full sequence and generation data remain fixed, and its anchor belongs
to \(\{q_i,v_i\}\). A rebase replaces the anchor incarnation, not the
center label or certificate.
\item A positive-height certificate was generated at time \(\sigma_i\)
with its top at the then-current anchor, with at least two anchors
present, and with
\begin{equation}\label{block:generation-radius}
 R_i<r_{A(\sigma_i)}(q_i)/20.
\end{equation}
Its generation time is retained until replacement. Generation times are
distinct; certificates built in the same batch are ordered while the
anchors remain fixed. Retired labels never recover old certificates.
\end{enumerate}
\end{invariant}

A \emph{full robust state}, also called an \emph{annotated robust
solution}, additionally has robust-valid certificates at every label.
This is a condition on the random choices, not a test carried out by
the algorithm. In particular, the algorithm does not retry indefinitely
after an unsuccessful sample.

If a required level decreases, we keep the old certificate in full.
We do not truncate the sequence or restart it from its old top: doing
either would lose the history needed for the overlap bound. Positive
levels are at most \(L\).
The deterministic invariant and \eqref{block:total-displacement}
therefore imply all hypotheses of \Cref{geom:top-ply}, including the
historical full-top ply bound.

\begin{lemma}[Comparison of anchor and proxy geometry]
\label{block:proxy-geometry}
Under \eqref{block:weak-offset}, the proxies are distinct and
\[
 \frac45d(a_i,a_j)\le d(v_i,v_j)\le\frac65d(a_i,a_j)
 \qquad(i\ne j).
\]
Thus \((4/5)r_A(a_i)\le r_V(v_i)\le(6/5)r_A(a_i)\), and every
certificate in \Cref{block:state} satisfies
\begin{equation}\label{block:proxy-radius}
 10^{\ell_i}\ge r_V(v_i)/15360.
\end{equation}
\end{lemma}
\begin{proof}
Each offset is at most one tenth of the corresponding nearest-anchor
distance, so the two offsets sum to at most \(d(a_i,a_j)/5\).
Apply the triangle inequality and minimize over \(j\ne i\). For the radius,
\[
 10^{\ell_i}\ge10^{\ell_i^{\rm req}}\ge\rho_i/200
 \ge r_A(a_i)/12800\ge r_V(v_i)/15360.
\]
The calculation also covers a required level of zero.
\end{proof}

\subsubsection{Cost Comparisons for Certificate Replacement}
\label{block:replacement-subsection}

We need two cost comparisons before describing the batch algorithm.
The first is the usual robust replacement statement: when a proxy and
a comparison facility are close to each other but well separated from
their respective competitors, the proxy serves the comparison
facility's cluster at comparable cost. The second comparison is used
when a certificate must be rebuilt. It shows that starting the new
sequence near the old proxy does not substantially increase the cost
on the relevant cluster. We prove both statements with the
anchor-based choice of certificate radius used here.

\begin{lemma}[Robust replacement]
\label{block:replacement}
Let \(V\) be a full robust state for \(P\) satisfying
\Cref{block:state}, and let \(Z\) be any comparison center set.
For \(\Gamma\ge400000\), if \((v,z)\in V\times Z\) is
\(\Gamma\)-well separated and \(C_z\) is the Voronoi cell of \(z\)
in \(Z\), then
\begin{equation}\label{block:replacement-cost}
 \cost(v,C_z)\le5\cost(z,C_z).
\end{equation}
\end{lemma}

\begin{proof}
There is nothing to prove when \(v=z\). Otherwise put \(d_0=d(v,z)\)
and choose the smallest integer \(i\) with
\(20d_0\le R=10^i<200d_0\). For the certificate
\((p_0,\ldots,p_\ell)\) of \(v=p_0\),
\[
 10^\ell\ge r_V(v)/15360\ge(\Gamma/15360)d_0\ge20d_0.
\]
Since \(d_0\ge1\), this implies \(1\le i\le \ell\).
For \(0\le j\le i\) and \(x\in B_j\),
\[
 d(x,z)\le10^j+d(p_j,p_0)+d_0
 \le\tfrac32R+R/20<2R<400d_0.
\]
Every other center in \(Z\) is at distance at least \(\Gamma d_0\)
from \(z\). Thus \(z\) is strictly closest to every client in these
balls, giving
\begin{equation}\label{block:replacement-ball-cell}
 B_j\subseteq C_z\qquad(0\le j\le i).
\end{equation}

We now distinguish the two possibilities in the robust-sequence rule.
If \(B_i\) has positive mass and average cost at least \(R/5\), then
\(p_i=p_{i-1}\). The lower-level displacement bound gives
\[
 d(z,p_i)\le d_0+10^{i-1}/2\le R/10.
\]
It follows that
\[
 \cost(z,B_i)\ge\cost(p_i,B_i)-w(B_i)R/10
                 \ge\tfrac12\cost(p_i,B_i).
\]
For \(x\notin B_i\), the same displacement gives
\(d(x,p_i)\le2d(x,z)\). Thus
\(\cost(p_i,C_z)\le2\cost(z,C_z)\), and
\eqref{block:sequence-cost} gives factor \(3\) for \(p_0\).

In the remaining cases, put \(b=p_{i-1}\). Robust validity, or zero
mass, gives
\[
 \cost(b,B_i)\le3\cost(z,B_i),\qquad
 d(z,b)\le d_0+10^{i-1}/2\le R/10.
\]
For \(x\notin B_i\), the step bound gives \(d(x,b)\ge R/2\).
Since \(d(z,b)\le R/10\), the reverse triangle inequality yields
\(d(x,z)\ge d(x,b)-R/10\ge(4/5)d(x,b)\), and hence
\(d(x,b)\le(5/4)d(x,z)\). Combining the inside and outside
bounds and applying \eqref{block:sequence-cost} at level \(i-1\),
\[
 \cost(v,C_z)\le\tfrac32\cost(b,C_z)
              \le\tfrac92\cost(z,C_z)\le5\cost(z,C_z).
\]
\end{proof}

\begin{lemma}[Near-anchor restart]
\label{block:restart}
Let \(\ell\ge1\), \(R=10^\ell\), and \(d(a,w)\le R/20\). For the
current instance \(Q\), set \(\mathcal B=\Ball_R^Q(a)\), and let \(v'\) be
returned by \(\MakeRobustProc(Q,a,\ell)\). If its sampled one-medians
satisfy \eqref{block:median-condition}, then every weighted client
subset \(C\) with \(\mathcal B\subseteq C\subseteq Q\) obeys
\begin{equation}\label{block:restart-cost}
 \cost(v',C)\le\frac92\cost(w,C).
\end{equation}
\end{lemma}

\begin{proof}
Write the new sequence as \((p_0,\ldots,p_\ell)\), with \(p_0=v'\)
and \(p_\ell=a\). The argument compares costs inside the top ball
with costs outside it. Outside \(\mathcal B\),
\(d(x,w)\ge(19/20)d(x,a)\).
If \(w(\mathcal B)=0\), nestedness makes every lower-level ball empty; the
procedure stays at \(a\), giving factor \(20/19\).

If the average cost on \(\mathcal B\) is at least \(R/5\), then
\[
 \cost(w,\mathcal B)\ge\cost(a,\mathcal B)-w(\mathcal B)R/20\ge\tfrac34\cost(a,\mathcal B).
\]
Together with the outside bound this gives
\(\cost(a,C)\le(4/3)\cost(w,C)\).
Equation~\eqref{block:sequence-cost} then gives factor \(2\) for \(v'\).

In the remaining case, \(b=p_{\ell-1}\) satisfies
\(\cost(b,\mathcal B)\le3F_1(\mathcal B)\le3\cost(w,\mathcal B)\) and
\(d(a,b)\le R/2\). For \(x\notin\mathcal B\),
\[
 d(x,w)\ge19R/20,\quad d(w,b)\le11R/20,\quad
 d(x,b)\le(30/19)d(x,w)<2d(x,w).
\]
Thus \(\cost(b,C)\le3\cost(w,C)\). The level-\((\ell-1)\) ball is
contained in \(\mathcal B\subseteq C\), so \eqref{block:sequence-cost} gives
\(\cost(v',C)\le(3/2)\cost(b,C)\le(9/2)\cost(w,C)\).
\end{proof}

\subsubsection{The Batch Algorithm and Its Analysis}
\label{block:algorithm-subsection}

The input to \BlockRobustify\ is an annotated state \(V\) satisfying
the deterministic \Cref{block:state} for \(P\), an updated instance
\(Q\), a log of \(s\) record edits from \(P\) to \(Q\), and a proposed
set \(W\) of exactly \(k\) distinct facilities. Write
\(h=|V\sym W|\). We do not assume that \(W\) is robust. Facilities
in \(V\cap W\) keep their labels, while facilities in
\(W\setminus V\) receive new labels. Write \(w_i\) for the location
of the proposed facility with label \(i\). The four phases below separate
the effects of edited weights, changed facilities, and changed
geometric scales.

\paragraph{Phase 1: mark contamination.}
We first identify certificates whose defining client balls might
have changed. Mark an old positive-height certificate if its full top ball contains
an edited coordinate. Discard marks on labels not retained in \(W\).
For every unmarked certificate, all defining balls keep exactly the
same clients and weights. Its deterministic local conditions, and its
robust validity when present, therefore survive. By
\Cref{geom:top-ply}, the number \(d\) of marked retained labels is
\begin{equation}\label{block:dirty-count}
 d=O(sL).
\end{equation}
Testing the edited coordinates against all old top balls costs
\(O(ks)\) work.

\paragraph{Phase 2: install the proposed labels.}
We now change the facility set to the proposal, without reconstructing
the surviving certificates. Retire every label with proxy outside
\(W\), removing its anchor and certificate. Insert each new proxy
\(w\) with anchor \(w\), offset
zero, and a certificate to be built in Phase 4. After each insertion,
complete the rebase closure. If \(w\) coincides with a retained
nominal anchor, first apply the forced-rebase rule of
\Cref{geom:forced-rebase}. Pass every primitive anchor operation to
the stable-label structure and record the changed \(\rho\) values.
The geometric invariants hold throughout these operations, although
new certificates are not yet available. Intermediate sets of size zero
or one use the boundary rules of \Cref{geom:stable-labels}; required
levels are evaluated only after all \(k\) labels have been installed.

\paragraph{Phase 3: prepare marked labels.}
The remaining marked certificates cannot be reused, so their anchors
are reset before new certificates are constructed. Force each marked
retained anchor to its proxy and complete its closure.
New labels already have zero offset. Once a proxy has been installed,
its location does not change during Phases 2 and 3. Retained
certificates keep their complete sequences and generation data.

\paragraph{Phase 4: reconstruct certificates.}
All geometric changes are now finished. Fix the final anchors and
stable labels, and compute the required levels
from \eqref{block:required-level}. Process each label once:
\begin{enumerate}
\item For a marked or new label, invoke
\(\MakeRobustProc(Q,a_i,\ell_i^{\rm req})\). Here \(a_i=w_i\).
\item For an unmarked old label with \(\ell_i^{\rm req}>\ell_i\), invoke
the same procedure from its current anchor.
\item Otherwise retain the proxy and its entire old certificate.
\end{enumerate}
Every reconstructed certificate receives fresh generation and
strong-reset times. Anchors do not change in this phase, so
reconstruction cannot trigger another geometric repair.

\begin{lemma}[No geometric feedback]
\label{block:no-feedback}
On every sampling outcome, Phase 4 restores the deterministic
\Cref{block:state}, without another anchor update or a second
reconstruction at any label. If all inherited certificates kept by the
call were robust-valid for \(P\), and all new samples satisfy
\eqref{block:median-condition}, the output is a full robust state for
\(Q\).
\end{lemma}

\begin{proof}
We first check the offset after a reconstruction. For a positive
required level \(\ell\),
\[
 R=10^\ell<\rho_i/20\le r_A(a_i)/20.
\]
By \eqref{block:total-displacement}, the new proxy's offset is less
than \(r_A(a_i)/40\), so it satisfies the strong-reset condition.
At level zero the proxy equals the starting anchor. Unchanged proxies
keep the weak offset condition. Since the anchors are fixed, these
proxy changes affect neither \(\rho_i\) nor the required levels;
\Cref{block:proxy-geometry} also guarantees distinct final proxies.

An unmarked certificate keeps all its local conditions for \(Q\), and is
retained only if its full height meets the new requirement. Every
reconstructed certificate is cost-nonincreasing on every outcome and
robust-valid on the stated sampling event. Its top is the fixed anchor,
and its positive top radius satisfies \eqref{block:generation-radius}.

It remains to check the historical annotations. For an inherited
certificate, the earlier phases either leave its anchor unchanged
or move it from the top to the fixed proxy. They do not retire
its center label or alter its certificate. Thus its lifetime condition
continues to hold. In Phase 4 a proxy changes only when its certificate
is replaced, starting a new lifetime at the fixed anchor with fresh
generation and reset data. These observations verify every part of the
invariant, including the history needed by \Cref{geom:top-ply} in the
next batch. The robust-validity conclusion uses both the success of
inherited certificates and that of the new samples.
\end{proof}

\begin{lemma}[Localization in the original proposed cells]
\label{block:localization}
At the start of Phase 4, let \(C_i\) be the Voronoi cell of \(w_i\)
in \(W\) for \(Q\). A new positive-height call at label \(i\), with
top radius \(R<r_A(a_i)/20\), satisfies
\[
 \Ball_R^Q(a_i)\subseteq C_i.
\]
\end{lemma}

\begin{proof}
The localization uses only the weak offset condition; it does not
depend on the new sampled certificate. Set \(r_i=r_A(a_i)\).
For \(x\in\Ball_R^Q(a_i)\), the weak offset
condition gives \(d(x,w_i)<3r_i/20\). For \(j\ne i\),
\[
 \begin{aligned}
 d(x,w_j)
 &\ge d(a_i,a_j)-d(a_j,w_j)-d(x,a_i)\\
 &\ge\tfrac9{10}d(a_i,a_j)-R>\tfrac{17}{20}r_i.
 \end{aligned}
\]
Thus \(w_i\) is strictly closest. The cells are defined using the
original proposal \(W\) and do not change when proxies are rebuilt.
\end{proof}

\begin{theorem}[Iterable batch robustification]
\label{block:theorem}\label{overview:lemma-batch}
Let \(V,P\) satisfy \Cref{block:state}, including its deterministic
geometry and certificate lifetimes. For the batch \(Q,W,s,h\) above,
let \(n=\max\{|\supp(P)|,|\supp(Q)|\}\). On every sampling outcome,
\BlockRobustify\ returns exactly \(k\) distinct facilities \(V'\)
satisfying the same deterministic invariant for \(Q\). If all retained
inherited certificates were robust-valid for \(P\), and all new
samples satisfy \eqref{block:median-condition}, its output is a full
robust state and
\begin{equation}\label{block:batch-cost}
 \Phi_Q(V')\le5\Phi_Q(W).
\end{equation}
On every outcome, the recourse and worst-case work satisfy
\begin{subequations}\label{overview:block-guarantee}
\begin{align}
 |V\sym V'|&=O((h+s)L^3),\label{block:batch-recourse}\\
 \text{work}&=\wtO((n+k)(h+s+1)).\label{block:batch-work}
\end{align}
\end{subequations}
The call performs \(O(hL+sL^2)\) primitive anchor operations and at
most \(O((h+s)L^3)\) calls to \MakeRobustProc.
The suppressed factors include powers of \(L\), dictionary costs,
and the prescribed confidence factors. These are bounds for each
individual call.
\end{theorem}

\begin{proof}
State preservation follows from \Cref{block:no-feedback}. To prove
the cost bound, condition on the stated success event. Fix the cells
\(C_i=C_{w_i}(W,Q)\) before reconstructing any proxy, and keep this
partition unchanged throughout the argument.

A marked or new label starts at \(a_i=w_i\). At positive height,
\Cref{block:localization,block:sequence-bounds} give
\(\cost(v_i',C_i)\le(3/2)\cost(w_i,C_i)\); at height zero the proxy
does not move. For an unmarked label that is reconstructed, let
\(R_i=10^{\ell_i}\) be its old full radius and \(R_i'\) its new radius.
The level increases, so
\[
 R_i'\ge10R_i,\qquad
 d(a_i,w_i)\le R_i/2\le R_i'/20.
\]
The offset bound uses \(a_i\in\{q_i,w_i\}\) and the complete old
certificate. The new top ball lies in \(C_i\), so
\Cref{block:restart} gives
\(\cost(v_i',C_i)\le(9/2)\cost(w_i,C_i)\).
An unchanged proxy contributes exactly its previous cost on its cell.
Summing over the fixed partition yields
\[
 \Phi_Q(V')\le\sum_i\cost(v_i',C_i)
 \le5\sum_i\cost(w_i,C_i)=5\Phi_Q(W).
\]
The approximation loss is incurred once per cell, regardless of how
many labels have been reconstructed.

For the resource bounds, we use only the deterministic invariant.
Phase 2
requests \(O(h)\) insertions, deletions, and forced rebases, including
coordinate coincidences. Phase 3 requests at most
\(d=O(sL)\) further rebases. By \Cref{geom:rebase-closure} and
\Cref{geom:forced-rebase},
the total number of primitive anchor operations is
\begin{equation}\label{block:anchor-edit-count}
 E=O((h+d)L)=O(hL+sL^2).
\end{equation}
Each primitive operation changes \(O(L)\) stable labels, counting a
fresh incarnation as an additional event. Thus the number of stable-label
change events is
\begin{equation}\label{block:rho-change-count}
 Z=O(EL)=O(hL^2+sL^3).
\end{equation}
An unmarked old label needs reconstruction only if its stable value
changed, since its old full height already met the previous
requirement. Therefore
the number \(M\) of calls to \MakeRobustProc\ satisfies
\[
 M\le d+h+Z=O((h+s)L^3).
\]
Each label is reconstructed at most once. Anchor changes do not move
proxies, and every proxy replacement changes the facility set by at
most two elements. Hence \(|V\sym V'|\le h+2M\).

Finally, we account for the running time. Marking costs \(O(ks)\),
and set comparisons and bookkeeping cost
\(\wtO(k(h+s+1))\). The geometric scans and stable-label maintenance
cost \(O(kEL)\). Every generation height is at most \(L\), so
\eqref{block:make-work} bounds certificate work by
\(\wtO((n+1)ML)\). Substitution gives, before confidence and dictionary
factors,
\[
 O\bigl(k(h+s+1)L^3+(n+1)(h+s)L^4\bigr),
\]
which proves \eqref{block:batch-work}. The zero-weight records occur
only in the \(O(k)\) geometric bookkeeping. If sorted client-to-proxy
rows are maintained, a proxy change updates one entry per client in
\(\wtO(n+k)\) work, within the same bound.
\end{proof}

The deterministic annotations returned by the procedure can be used
by the next batch even if some samples were unsuccessful. On the
common success event, robust validity is passed from inherited
certificates to the new state. We next construct the first such state
and then use the per-call bound to rebuild a checkpoint. Steps~4
and~5 will distribute the resulting center changes and work over
individual record edits.

\subsection{Initializing the Annotated State}
\label{subsec:annotation-initialization}

Batch repair assumes that the old reference already has the geometric
annotations and certificate history in \Cref{block:state}. We therefore
cannot use it to create the first checkpoint. Instead, we construct
the initial certificates directly, keeping all anchors fixed.

\begin{lemma}[Initializing the geometric annotations]
\label{lem:annotation-initialization}
Given any set \(W\) of exactly \(k\ge2\) distinct facilities and a
weighted instance \(Q\), one can construct a state satisfying the
deterministic Invariant~\ref{block:state} with proxies \(U\) such that
\[
 \Phi_Q(U)\le\tfrac32\Phi_Q(W).
\]
The work is \(\wtO(k(n+k))\), where \(n=|\supp(Q)|\), including
all initial distance rows. On the common success event of its certificate
calls, the state is also robust-valid.
\end{lemma}
\begin{proof}
Create one label per \(a\in W\), with anchor and initial proxy both
at \(a\). Initialize the stable-label structure on \(A=W\), and
compute each required height from \eqref{block:required-level}. Keep
all anchors fixed and construct one certificate
\(\MakeRobustProc(Q,a,\ell_a^{\rm req})\) per label. Record fresh,
distinct generation and strong-reset times as these calls complete.

Because the anchors stay fixed throughout initialization, each
certificate can be checked against the same set \(W\). At positive
height, its radius satisfies
\(R_a<r_W(a)/20\), and its proxy is within \(R_a/2<r_W(a)/40\)
of \(a\). At height zero the proxy stays at \(a\). Thus the strong
reset condition holds, the proxies are distinct, and each certificate
has the required height and generation radius. Every anchor remains
at its certificate's top. These observations establish all geometry
and lifetime conditions directly, without invoking batch repair on an
uninitialized state. Each constructed certificate is cost-nonincreasing
on every outcome; robust validity additionally uses the sampling-success
event.

For the cost bound, fix the Voronoi cells \(C_a\) of \(W\).
Every positive-height top ball is contained in its corresponding
cell, since its radius is less than
\(r_W(a)/20\). Equation~\eqref{block:sequence-cost} gives
\(\cost(v_a,C_a)\le(3/2)\cost(a,C_a)\); a height-zero proxy gives
the same inequality. Sum over the fixed cells. The stable-label
initialization, \(k\) certificate calls, and rows against the resulting
\(k\)-set together take \(\wtO(k(n+k))\) work.
\end{proof}

For \(k\ge2\), we first install \(k\) dummy virtual facilities and
initialize their proper projection onto the live support. This costs
\(\wtO(k(n_{\rm init}+k))\) and makes the deterministic failure rule of
\Cref{subsec:failure-rule} available before any randomized preprocessing.
Use \Cref{lem:develop} to compute a static approximate \(k\)-set,
pad as needed, apply \Cref{lem:annotation-initialization}, and call
\(\WindowCore\) on the resulting annotated state and distance rows.
The total work remains \(\wtO(k(n_{\rm init}+k))\). The initial cost is at most
\((3/2)\beta F_k\le\alpha F_k\),
so the checkpoint has the required approximation factor.

\subsection{Refreshing the Robust Solution}
\label{subsec:refresh}

After a bounded number of edits, we must replace the old checkpoint
solution by one that is again a constant-factor approximation for the
current weighted instance. The replacement should use only \(O(B)\)
changes before robustification, where \(B\) is the old checkpoint's
window. We obtain it by augmenting the old solution, adding the edited
locations, removing excess facilities, and finally repairing the robust
certificates.

We use the already fixed constants \(\beta\) and
\(\alpha=160\beta\) from \eqref{eq:alpha-fixed}. Recall that an
annotated robust solution satisfies \Cref{block:state} and has
robust-valid certificates, including every historical certificate
still retained. The refresh routine takes the old weighted snapshot,
its annotated reference, the certified window, and a target weighted
snapshot; it also uses the edit log between the two snapshots.

The following lemma is the reason augmentation needs only a small
number of additional facilities. A comparison facility that belongs
to a well-separated pair can be served through the corresponding
robust proxy. Only the remaining facilities need to be added.

\begin{lemma}[Small augmentation exists]\label{lem:small-augmentation}
Let \(U\) be an annotated robust \(k\)-set for \(P\), with
\(\Phi_P(U)\le\alpha F_k(P)\). For every
\(s\in\{0,\ldots,k\}\), a set \(F\subseteq\supp(P)\) of size at
most \(4r_\eta(P)+2s+3\) satisfies
\[
 \Phi_P(U\cup F)\le10F_{k+s}(P).
\]
\end{lemma}
\begin{proof}
The empty instance is immediate. Otherwise, project an ambient
\((k+s)\)-center optimum onto the positive support, discard duplicates,
and pad with distinct dummies to obtain a set \(V\) of exactly
\(k+s\) facilities. Then
\(\Phi_P(V)\le2F_{k+s}(P)\le2F_k(P)\). Padding may overlap \(U\).

Because \(\Gamma>2\), well-separated pairs form a matching. Write
\(k-m\) for their number. If \(m>s+3\), put
\(r=\lfloor(m-s)/4\rfloor\). The separated-pair deletion lemma gives
\[
 F_{k-r}(P)\le6\Gamma(\alpha+2)F_k(P)\le\eta F_k(P),
\]
so \(r\le r_\eta(P)\). The resulting inequality
\(m\le4r_\eta(P)+s+3\) also holds when \(m\le s+3\).

It remains to keep only the facilities not covered by this matching.
Keep the \(m+s\) unpaired facilities of \(V\). For each paired
facility, the robust replacement bound
(Lemma~\ref{block:replacement}) transfers its clients to its partner
in \(U\) at a factor of at most five. Therefore the union costs at
most \(5\Phi_P(V)\le10F_{k+s}(P)\) and adds at most
\(4r_\eta(P)+2s+3\) facilities. The padding dummies serve no positive
clients and can be omitted from the augmentation.
\end{proof}

\begin{theorem}[Fixed-factor refresh]\label{thm:refresh}
Let \(U\) be an annotated robust \(k\)-set for \(P\) satisfying
\(\Phi_P(U)\le\alpha F_k(P)\), and let \(B\) satisfy
\eqref{eq:window-scale}. Suppose \(Q\) is obtained from \(P\) by
\(s\le B\) record edits. On the common success event, the procedure
\(\Refresh(P,U,B,Q)\) returns an annotated robust \(k\)-set \(U'\)
with
\begin{equation}\label{overview:refresh}
 \begin{aligned}
 \Phi_Q(U')&\le\alpha F_k(Q),\\
 |U'\sym U|&=O(BL_\Delta^3),\\
 \text{work}&=\wtO((n+k)B).
 \end{aligned}
\end{equation}
Here \(n=\max\{|\supp(P)|,|\supp(Q)|\}\); the work bound assumes
the edit log and persistent distance rows for the old snapshot,
shadow facilities, and dummies. The implementation uses a prescribed
work cap.
\end{theorem}
\begin{proof}
Let \(I\) contain the distinct edited locations, including deletions
and positive-to-positive weight changes; thus \(|I|\le s\). Set
\(a=\lceil8DB+2s+3\rceil\) and perform the following three steps.

\emph{Augment on the old snapshot.} We first enlarge the old solution
while its distance data are still available. Since \(r_\eta(P)<2DB\),
\Cref{lem:small-augmentation} and the augmentation guarantee imply
\[
 \widetilde U=\Develop(P,U,a),\qquad
 \Phi_P(\widetilde U)\le10\beta F_{k+s}(P).
\]

\emph{Absorb edits and restore the budget.} Opening every edited
location makes the old augmentation usable for \(Q\), even when an
edit changes a weight by an arbitrary amount. Put
\(X_0=\widetilde U\cup I\). The lazy inequalities give
\[
 \Phi_Q(X_0)\le10\beta F_{k+s}(P)\le10\beta F_k(Q).
\]
The set \(X_0\) contains \(U\) and only \(f=O(B)\) additional
facilities. Removal returns a \(k\)-subset \(W\subseteq X_0\) with
\[
 \Phi_Q(W)\le(20\beta+12)F_k(Q)\le32\beta F_k(Q),
 \qquad |W\sym U|\le2f.
\]

\emph{Repair the annotations.} The proposal \(W\) need not carry
valid robust certificates. Apply Theorem~\ref{block:theorem} to
the old annotated state, new snapshot \(Q\), and proposal \(W\).
Its cost bound yields
\[
 \Phi_Q(U')\le5\Phi_Q(W)\le160\beta F_k(Q)=\alpha F_k(Q),
\]
and its recourse bound gives the stated number of changes.

For the running time, retain the old distance rows while adding the \(f\)
new facility columns and preparing rows at edited locations. This costs
\[
 \wtO\bigl((n+k)f+f^2+s(k+f)\bigr)
       =\wtO((n+k)B),
\]
because \(f=O(B)\), \(s\le B\), and \(B\le k\).
Augmentation, removal, and batch repair have the same work bound.
Persistent versions preserve the old snapshot; updates arriving during
this computation are reconciled separately in \Cref{subsec:trees}.
\end{proof}

Each refresh returns to the same approximation factor \(\alpha\).
The factor does not compound across checkpoints; the window constants
affect how much work is scheduled, not the approximation factor.

\subsection{Periodic Rebuilding and Amortized Recourse}
\label{subsec:amortized-rebuilding}

We now combine the preceding procedures. Let \(u\) count weighted
record edits and let \(p\) be the current checkpoint time.
The checkpoint stores its frozen input \(P_p\), annotated reference
\(U_p\), and the output \((B,H,K_p)\) of \(\WindowCore(P_p,U_p)\).
The set \(I\) contains the distinct locations edited since \(p\).
A location remains in \(I\) even if its current weight is zero.

The call \(\Refresh(P_p,U_p,B,P_u)\) constructs an annotated reference
for the current input using the checkpoint, its certified window, and
the intervening edit log. Applying \WindowCore\ to that reference
then gives the next core. In this first algorithm both calls are
completed at a single update boundary; no work is performed in the
background. We rebuild after a fixed fraction of the window, except
that a window shorter than \(16\) is rebuilt after one edit.
Initialization is performed once; each subsequent record edit executes
the Edit, Rebuild, and Serve steps in that order.

\begin{algorithm}[H]
\caption{Lazy Updates with Periodic Rebuilding: the amortized baseline}
\label{alg:lazy-rebuilding}
\normalsize
\begin{algorithmsteps}
\item \textbf{Initialize.} Construct \(U_0\) as in
  \Cref{subsec:annotation-initialization}, compute
  \((B,H,K_0)\gets\WindowCore(P_0,U_0)\), and set
  \(p=u=0\), \(I=\varnothing\), and \(Z=K_0\).
\item \textbf{Edit.} Apply the record edit at \(x\), increment \(u\),
  and add \(x\) to \(I\), even if its weight becomes zero.
\item \textbf{Rebuild.} If \(B<16\), rebuild in this handler.
  Otherwise, rebuild when \(u-p=\lfloor B/16\rfloor\).
  Compute \(U'\gets\Refresh(P_p,U_p,B,P_u)\) and
  \((B',H',K')\gets\WindowCore(P_u,U')\).
  Adopt
  \((p,P_p,U_p,B,H,K_p)\gets(u,P_u,U',B',H',K')\)
  and clear \(I\).
\item \textbf{Serve.} Set \(Z=K_p\cup I\) and publish its proper
  projection as defined in \Cref{lem:projection}.
\end{algorithmsteps}
\end{algorithm}

All bounds in the following baseline analysis are conditioned on the
common success event of the static and certificate calls. The
deterministic caps and failure rules needed for the final algorithm
are treated in Step~5. In particular, if \(B=1\), the old core has no
spare slot: the next update is rebuilt synchronously before a new
solution is published.

Let an epoch contain \(s\) record edits. For \(B\ge16\),
\(s=\lfloor B/16\rfloor=\Theta(B)\), and
\Cref{lem:window-comparison} gives \(B'=O(B)\).
For \(B<16\), the epoch has one edit. The same upper bound follows
directly from \Cref{lem:lipschitz} and \eqref{eq:window-scale}:
\[
 B'\le b(P_u)+1\le b(P_p)+2\le CB+1=O(B).
\]
This argument also covers \(B=1\), without applying a
neighboring-window bound that requires \(s<B\).

At a rebuild, each core omits only \(O(B)\) facilities from its
reference, and the old edit buffer has \(O(B)\) locations. Let
\(Z^-\) be the virtual set published just before the rebuild update.
Since \(\Refresh\) changes \(O(BL_\Delta^3)\) reference facilities,
\begin{equation}\label{overview:warmup-recourse}
 |Z^-\sym K'|
 \le |U_p\sym U'|+O(B)=O(BL_\Delta^3).
\end{equation}
Edits that do not rebuild add at most one location to the virtual set.

\begin{lemma}[Amortized epoch bound]
\label{overview:lemma-epoch}
On the common success event, \Cref{alg:lazy-rebuilding} maintains a
virtual constant-factor solution with at most \(k\) facilities.
After initialization, its amortized update time is
\(\wtO(n+k)\), and its amortized virtual recourse is
\(O(L_\Delta^3)\), where \(n\) bounds the weighted support size
during the execution.
\end{lemma}

\begin{proof}
Consider a completed update before its epoch's rebuild. If \(B\ge16\),
the checkpoint age is at most \(\lfloor B/16\rfloor-1\le B-1=H\).
\Cref{lem:core-protection} therefore gives cardinality at most
\(k\) and constant-factor approximation. For \(B<16\), every edit
rebuilds before publishing. At a rebuild, \Cref{thm:refresh} restores
the reference guarantee and \Cref{thm:window-core} supplies a new core
at age zero. Thus feasibility and approximation hold at every
completed update.

The refresh uses \(\wtO((n+k)B)\) work. Since \(B'=O(B)\), computing
the new window and core has the same bound. The distance rows required
by these calls can be updated eagerly: an ordinary record edit costs
\(\wtO(n+k)\), and the rows for changed facilities are included in the
rebuild bound. Equation~\eqref{overview:warmup-recourse} bounds the
boundary's virtual changes by \(O(BL_\Delta^3)\).

Charge these rebuilding costs to the edits of the epoch. When
\(B\ge16\), there are \(\Theta(B)\) such edits. When \(B<16\),
there is one edit and \(B\) is bounded by an absolute constant.
The resulting amortized bounds are \(\wtO(n+k)\) work and
\(O(L_\Delta^3)\) virtual recourse per record edit.
\end{proof}

The baseline shows why a window of length \(B\) is enough to pay for
the next checkpoint. It does not yet give worst-case guarantees:
all \(O(BL_\Delta^3)\) center changes and \(\wtO((n+k)B)\) work
can occur at one boundary. Step~4 addresses the changes by migrating
between cores while respecting the center budget. Step~5 schedules
the construction and migration so that each record edit receives
only \(\wtO(n+k)\) work and \(O(L_\Delta^3)\) center changes.

%% file: sections/step4_recourse.tex
\section{Step 4: Worst-Case Center Recourse}
\label{sec:step4}

Step~3 constructs a complete replacement reference using
\(O(BL_\Delta^3)\) changes over an epoch. Installing all these
changes at once gives only an amortized recourse bound. Opening all
target facilities before closing the old ones could also exceed
\(k\). We instead order the changes so that intermediate sets remain
small and have controlled cost, then execute a bounded number of
changes per edit. Constructing a target in the background and
installing that target are separate operations.

\subsection{A Budget-Preserving Migration Order}
\begin{lemma}[Budget-preserving migration]
\label{lem:migration}\label{overview:lemma-migration}
For nonempty facility sets \(A,B\), there is an ordered sequence of
\(O(|A\sym B|)\) insertions, deletions, and atomic swaps from \(A\)
to \(B\) such that every intermediate set \(S\) obeys
\begin{equation}\label{overview:migration}
 |S|\le\max\{|A|,|B|\},\qquad
 d(x,S)\le2d(x,A)+d(x,B)\quad(x\in X).
\end{equation}
If both endpoint sizes are at most \(k\), the sequence can be prepared
in \(O(k|A\sym B|+k)\) work. If a common set \(J\) remains open,
then also
\[
 d(x,S\cup J)\le2d(x,A\cup J)+d(x,B\cup J).
\]
\end{lemma}
\begin{proof}
Keep \(A\cap B\) open throughout. For each \(a\in A\setminus B\),
choose a nearest facility \(f(a)\in B\). Facilities assigned to an
already open target can be removed immediately; the others are removed
when their targets open. More precisely, use the following order:
\begin{enumerate}[label=\arabic*.,leftmargin=2em,itemsep=0.3ex]
\item Delete every \(a\) whose assigned target lies in \(A\cap B\).
\item For each \(b\in B\setminus A\) with a nonempty preimage, swap
      \(b\) for one facility in that preimage, then delete its other
      facilities.
\item Insert the remaining targets, whose preimages are empty.
\end{enumerate}
Swaps preserve cardinality and deletions reduce it. Before the last
step, all remaining facilities belong to \(B\). Inserting the
remaining targets therefore never exceeds the larger endpoint size.
For a nearest \(a\in A\) to \(x\), either \(a\) is still open or
its target \(f(a)\) is open. In the latter case, for a nearest
\(b_x\in B\) to \(x\),
\[
 d(x,f(a))\le d(x,a)+d(a,f(a))
 \le d(x,a)+d(a,b_x)\le2d(x,A)+d(x,B).
\]
This gives the distance bound. If \(J\) attains either endpoint minimum,
use \(d(x,S\cup J)\le d(x,J)\); otherwise apply the bound just
proved. Finding all assignments takes \(O(k|A\setminus B|)\) queries,
and set differences and the operation list take the remaining linear
work.
\end{proof}

We regard a swap as one transition between published center sets.
Internally it may require a deletion and an insertion within a record
handler. Before starting the transition, we reserve all work for its at most
two virtual membership changes and their projection updates. No pause
or failure check occurs in the middle of that commit. Separate membership
flags for the base and common set ensure that removing a buffered
location from the base does not close it in their union.

\subsection{Checkpoint State and the Recourse Quota}
At a phase start \(t\), a checkpoint at \(p\le t\) stores the
snapshot \(P_p\), an annotated robust \(\alpha\)-approximate
\(k\)-set \(U_p\), its window and core \((B,H,K_p)\), and the
persistent data-structure roots. The invariant is
\[
 t-p\le B/16,\qquad Z_t=K_p\cup I(p,t),
\]
where \(Z_t\) is the virtual serving set. The published set is its
proper projection onto the current support. Initialization has
\(p=t=0\). The edited-location set \(I(p,t)\) is the one defined
in Step~1; in particular, it retains edited locations of zero current
weight and satisfies \(I(p,u)=I(p,t)\cup I(t,u)\).

Freeze the initial base \(A=K_p\cup I(p,t)\). A replacement
checkpoint for \(P_t\) supplies a core \(K_t\) and reserve
\(H_t=B_t-1\). During migration, let \(M_u\) be the current
base on the \(A\)-to-\(K_t\) path and keep the common buffer
\(J_u=I(t,u)\) open. Thus \(Z_u=M_u\cup J_u\). The buffer may
grow while the path is executed, but the lemma applies to its current
value at every update boundary.

\begin{proposition}[Per-update migration quota]
\label{sixstep:recourse-quota}
Suppose both cores satisfy the protected-core invariant for their
respective snapshots and
\(|A\sym K_t|=O(BL_\Delta^3)\). Assume a migration interval
contains at least a fixed positive constant times \(B\) record
edits, the path has been prepared before that interval, and throughout
the interval
\[
 u-p\le H,\qquad u-t\le H_t.
\]
Then a quota of \(O(L_\Delta^3)\) atomic transitions per record
edit completes the migration within the interval. Every resulting
virtual set has at most \(k\) facilities, costs at most
\(3C_0F_k(P_u)\), and changes in \(O(L_\Delta^3)\) locations
per edit.
\end{proposition}
\begin{proof}
The path has \(O(BL_\Delta^3)\) transitions, so the stated quota
suffices over the prescribed interval. Put \(s=t-p\) and
\(j=u-t\). The migration size bound gives
\[
 |M_u\cup J_u|
 \le\max\{k-H+s,k-H_t\}+j\le k,
\]
since \(s+j\le H\) and \(j\le H_t\). Both buffered endpoints
\(A\cup J_u=K_p\cup I(p,u)\) and
\(K_t\cup J_u=K_t\cup I(t,u)\) have cost at most
\(C_0F_k(P_u)\) by \Cref{lem:core-protection}. Summing the
buffered distance bound of \Cref{lem:migration} with current client
weights proves the cost claim. Each atomic swap changes at most two
facilities, and the current edit adds at most one location to the
buffer, proving the per-edit recourse bound.
\end{proof}

The proposition separates the migration argument from the timing
conditions it needs. Step~5 supplies those conditions: it constructs
the target, prepares the path, and reserves enough of both protected
windows to finish migration. Proper projection preserves the recourse
order by \Cref{lem:projection}. Path preparation and the data-structure
work of each transition are accounted for separately in that step.

%% file: sections/step5_time.tex
\section{Step 5: Worst-Case Update Time}
\label{sec:step5}\label{sec:weighted-scheduling}
\label{sec:inner}\label{sec:weighted-inner}

The migration order controls recourse once a target and its path have
been prepared. We must also construct that target within a bounded
amount of work per edit, keep its data structures current, and finish
before either protected core expires. This step gives the complete
weighted-record handler. The input after \(u\) record edits is \(P_u\);
write \(n_u=|\supp(P_u)|\). The initial annotated checkpoint is the
one constructed in Step~3.

\subsection{Scheduling Large and Small Windows}
The computed window determines how much background work we can spread
over future edits. When \(B\) is large, we refresh and install the
next checkpoint in four stages, each with a fixed amount of work and
center movement per update. When \(B\) is small, the whole refresh
fits within a single update. We first describe the large-window case.

\paragraph{Large windows.}
Suppose \(B\ge1024C\). At phase start \(t\), freeze \(P_t\)
and set
\[
 A=K_p\cup I(p,t),\qquad q=\left\lfloor B/(256C)\right\rfloor.
\]
The phase lasts exactly \(4q\) record updates. Every location edited
since \(t\) enters the common buffer \(J_u=I(t,u)\); opening this
buffer lets the old and new solutions serve those edits without
immediate reconstruction. The base set \(M_u\) starts at \(A\), and
the virtual output is \(Z_u=M_u\cup J_u\). Use four stages of
\(q\) updates each:
\begin{enumerate}[label=\textbf{Stage \arabic*.},leftmargin=5.8em,labelsep=0.7em,
                  itemsep=0.5ex]
\item Refresh the old annotated solution on the frozen input:
      \(U_t=\Refresh(P_p,U_p,B,P_t)\).
\item Choose the next protected core and its window:
      \((B_t,H_t,K_t)=\WindowCore(P_t,U_t)\).
\item Prepare the migration from \(A\) to \(K_t\), and bring the
      new shadow's distance rows up to date as described below.
\item Carry out the migration in fixed-size batches, keeping
      \(J_u\) open throughout.
\end{enumerate}
The base remains \(A\) through Stage~3, so no partially constructed
target is published. At the end of Stage~4, adopt
\((P_t,U_t,B_t,H_t,K_t)\) as the checkpoint, with timestamp \(p'=t\).
Retain \(J_u\), since its edits occurred after that snapshot.

\begin{lemma}[Feasibility and budgets of a large-window phase]
\label{lem:stage-horizons}\label{lem:phase-quality}\label{lem:phase-budgets}
On the common success event, the target window satisfies
\[
 \frac{15B}{16C}\le B_t\le(C+1/16)B.
\]
At every record time \(u\) during the phase,
\[
 u-p\le5B/64<B-1,\qquad
 u-t\le B/(64C)\le B_t/60<B_t-1.
\]
The virtual output has at most \(k\) facilities and cost at most
\(3C_0F_k(P_u)\). Fixed quotas of \(\wtO(n_u+k)\) work and
\(O(L_\Delta^3)\) recourse per update finish every stage on time,
and the new checkpoint satisfies the age invariant.
\end{lemma}
\begin{proof}
\emph{Ages.} At phase start, the old checkpoint is at most \(B/16\)
edits old. This keeps the new window comparable to the old one.
Apply \Cref{lem:window-comparison} with \(a=1/16\) to obtain
the bounds on \(B_t\). Since \(4q\le B/(64C)\),
\[
 u-p\le B/16+B/(64C)\le5B/64,
 \qquad u-t\le B/(64C)\le B_t/60.
\]
The strict bounds follow from \(B\ge1024C\) and \(B_t\ge960\).
At adoption, \(u-p'=4q\le B_t/60<B_t/16\).

\emph{Cost and capacity.} Both migration endpoints remain protected
throughout the phase. The old and new cores each have enough unused
slots for their respective edit buffers. Including the common buffer,
the endpoints are
\[
 A\cup J_u=K_p\cup I(p,u),\qquad
 K_t\cup J_u=K_t\cup I(t,u).
\]
Both cost at most \(C_0F_k(P_u)\) by
\Cref{lem:core-protection}. \Cref{lem:migration} therefore bounds every
intermediate cost by \(3C_0F_k(P_u)\). Write \(s=t-p\) and
\(j=u-t\). The migration size bound gives
\[
 |M_u\cup J_u|\le\max\{k-H+s,k-H_t\}+j\le k,
\]
because \(s+j\le H\) and \(j\le H_t\).

\emph{Deadlines.} Each stage has \(q\ge B/(512C)\) updates available.
Refresh and the next
window require \(\wtO((n_t+k)B)\) work each, using the known bound
\(B_t\le(C+1/16)B\). Also
\[
 \begin{split}
 |A\sym K_t|
 &\le |U_p\sym U_t|+|U_p\setminus K_p|
       +|U_t\setminus K_t|+|I(p,t)|\\
 &=O(BL_\Delta^3).
 \end{split}
\]
Migration preparation and row reconciliation take \(\wtO(kB)\)
work. Each executed center change, including projection, costs
\(\wtO(n_u+k)\). Thus each stage requires
\(\wtO((n_t+k)B)\) work. There are \(O(B)\le O(k)\) edits in the
phase, so \(n_u+2k\) and \(n_t+2k\) differ by at most a constant
factor. We can therefore allocate fixed per-update work quotas that
meet every deadline.

If the prepared path contains at most \(K_1BL_\Delta^3\) transitions,
use \(\lceil512CK_1L_\Delta^3\rceil\) transitions per Stage~4
update. Opening the edited location costs at most one additional change;
an atomic swap costs two. Early completion waits for the scheduled
boundary.
\end{proof}

\paragraph{Small windows.}
If \(B<1024C\), then \(B\) is bounded by an absolute constant.
Background scheduling is unnecessary: we can refresh and install
the new core during the next record edit. At \(u=t+1\),
compute
\[
 U_u=\Refresh(P_p,U_p,B,P_u),\qquad
 (B_u,H_u,K_u)=\WindowCore(P_u,U_u),
\]
install \(K_u\), and adopt checkpoint time \(p'=u\).
The refresh input contains at most
\(u-p\le\lfloor B/16\rfloor+1\le B\) edits. Lipschitz stability
and the window comparison give
\[
 B_u\le b(P_u)+1\le b(P_p)+1+(u-p)
      \le(C+1/16)B+1.
\]
Both windows are bounded by absolute constants, so the work is
\(\wtO(n_u+k)\). The old serving set and the new core differ in
\(O(BL_\Delta^3+B+B_u)=O(L_\Delta^3)\) facilities. The new checkpoint
has age zero, even if its window enters the large regime. In particular,
when \(B=1\), no free buffer slot is assumed: the whole update is
completed before publication. Check the size and recourse first, then
reserve the work for an uninterrupted installation and projection commit.

\subsection{Maintaining the Distance Data Structures}\label{subsec:trees}
The scheduling argument assumes that the work of a static call can be
split into small operations. We now describe the distance data kept
for the shadow facilities and explain how a frozen snapshot can be
used while new edits continue to arrive. In particular, starting a
new shadow must not require rebuilding all client rows at once.

A row stores the distances from one client or auxiliary location to the
current shadow facilities in a balanced search tree. A persistent
location dictionary maps keys to rows. Initialization uses
\(\wtO(k(n_{\rm init}+k))\) work; freezing a snapshot retains one root in
constant time. Path copying preserves all frozen dictionaries.

Distance orders do not depend on client weights, so a row can be shared
by snapshots that assign different weights to the same location.
Building a row costs \(\wtO(k)\), while deleting an unneeded row costs
one dictionary update. Positive-to-positive weight changes preserve
distance orders. A shadow-center insertion or deletion changes one entry
in every row, including any newly required auxiliary row, and costs
\(\wtO(n+k)\). A restricted-removal call uses a private persistent
version, charging its temporary deletions to its removal budget.
Consequently, \(s\) input edits and \(h\) shadow changes cost
\[
 \wtO\bigl(ks+(n+k)h\bigr)
\]
in addition to the static calls. Augmentation columns satisfy the same
accounting. Background workers pause between individual distance queries
or completed dictionary operations. A short dictionary commit reserves its full
\(\wtO(1)\) allowance and finishes within one record handler.

\paragraph{Reconciling a new shadow with the live input.}
The target shadow \(U_t\) is constructed for the frozen instance
\(P_t\), but the input may have changed by the time it is ready. We
need to reconcile only the locations edited in the meantime.
While Stage~1 constructs \(U_t\), maintain the old shadow's live rows
and a deduplicated dictionary of locations edited since time \(t\).
When Stage~1 finishes, retain its frozen row root \(\mathcal R_t\),
share it to create a separate live root \(\mathcal R\), and freeze the
edited-location dictionary as \(\mathcal D\). Its size is \(O(B)\).
An entry asks us to inspect the location's current state. It does not
replay its earlier insertions, deletions, or weight assignments.

Pin the auxiliary set \(E_t=U_t\cup\{\text{dummy locations}\}\).
The keys of \(\mathcal R_t\) are exactly \(\supp(P_t)\cup E_t\).
These auxiliary rows were already charged to Stage~1, so a boundary
does not build \(\Theta(k)\) fresh rows. Temporary augmentation rows remain
only in private subroutine versions. At record time \(u\), define
\[
 \operatorname{need}_u(x)
   =[w_{P_u}(x)>0]\ \lor\ [x\in E_t].
\]
Inside a call at the current record time, abbreviate
\(\operatorname{need}(x):=\operatorname{need}_u(x)\).
A row records its location, the shadow generation, and distances to
\(U_t\), but no weight or birth identifier. Thus a private row under
construction remains geometrically valid even if the client at that
location is deleted and later reinserted. Costs always use the weight
dictionary of the relevant snapshot.

After the target becomes available, each foreground edit immediately
reconciles its location in \(\mathcal R\), in \(\wtO(k)\) work.
Background work processes each entry of \(\mathcal D\) once and
finishes by the end of Stage~3. A later edit to the same location is
handled in the foreground; it does not restart a pending row or add
another item to \(\mathcal D\).

The order of operations matters here. A row built in the background
may finish after its location has been edited again. We therefore
build the row privately and check the latest membership state before
installing it. The same check protects an already complete foreground
row from being replaced by stale background work.

\begin{algorithm}[H]
\caption{\(\textsc{ReconcileRow}(x)\) for a fixed shadow \(U_t\)}
\label{alg:reconcile-row}
\normalsize
\textbf{State:} current record time \(u\), current weights, latest live
row root \(\mathcal R\), and pinned auxiliary set \(E_t\).
\begin{enumerate}[label=\arabic*.,leftmargin=2.1em,labelsep=0.6em,
                  itemsep=0.3ex,parsep=0pt,topsep=0.5ex]
\item Inspect current \(\operatorname{need}(x)\). If false, remove only
      key \(x\) from the latest live root, complete this item, and return.
\item If a complete row for \(x\) and this shadow generation exists,
      complete this item and return.
\item Build a private row \(R_x\) against \(U_t\); construction may
      pause across record updates.
\item In one short commit, reread current need and the latest entry.
      Remove an unneeded entry, retain an existing complete row, or install
      \(R_x\), as appropriate. Complete this item.
\end{enumerate}
Every commit changes one key in the latest live root. It never replaces
that root by a version captured before a pause or changes the frozen root.
\end{algorithm}

\begin{lemma}[Consistent row catch-up]\label{lem:row-catch-up}
Under any interleaving of foreground edits and catch-up work, exhaustion
of \(\mathcal D\) leaves a complete row for exactly the currently
needed locations. Frozen rows and weights remain unchanged. Catch-up
costs \(\wtO(k|\mathcal D|)\) work, in addition to
\(\wtO(k)\) per foreground record edit, independently of weight
magnitudes and repeated deletion and reinsertion.
\end{lemma}
\begin{proof}
The key point is that every write checks the current live root.
Initially, locations outside \(\mathcal D\) have their correct
snapshot row and membership; every subsequent foreground edit restores
its location before background work resumes. A private row stays valid
across pauses because both its coordinate and \(U_t\) are fixed.
Its final commit rechecks current need and changes only its own key in
the latest root. Completed items remain correct under later foreground
edits. This proves the state claim when all items finish, and path copying
protects the frozen versions.

No item is restarted when the foreground changes its location.
Each pending item builds at most one row and uses constantly many
dictionary accesses. Hence its work is \(\wtO(k)\), with no restart.
The only excess live rows belong to \(\mathcal D\); there are
\(O(n_u+k+B)=O(n_u+k)\) rows during catch-up. The entire phase has
\(O(B)\) foreground edits, so foreground maintenance and catch-up
together cost \(\wtO(kB)\). A zero pending counter and no unfinished
item certify completion without a scan.
\end{proof}

\paragraph{The foreground interface.}
The foreground is responsible for the current record even while a
background stage is running. After committing an edit at \(x\), a handler that has not halted updates the live
distance rows by calling \(\textsc{ForegroundRows}(x)\). It reconciles
\(x\) in the committed
shadow's live root, retaining a row precisely for current support,
committed shadow locations, and dummies. It never changes the checkpoint's
frozen root. During Stage~1 it also records \(x\) in the edited-location
dictionary. Phase setup clears the previous target handle. At Stage~1
completion, including the edit in that handler, the worker freezes
\(\mathcal D\), creates the target live root, and marks that target
available. Thereafter the foreground call reconciles \(x\) in the
target root as well, without changing \(\mathcal D\) or its counter.

At most two live roots are updated, each against a fixed \(k\)-set.
A separate fixed \(\wtO(k)\) foreground allowance therefore finishes
the call in the current handler; background work pauses during its
commits. By the Stage~3 deadline, both the frozen and live target roots
are ready. Adoption switches the committed root to the target and retires
the old live root. A synchronous refresh already describes the current
input and adopts its roots directly.

Only constantly many live versions coexist. To avoid a large memory
reclamation step in a single record handler, use an append-only allocator
over the prescribed finite horizon, charging node creation to row and
path updates and retiring roots without recursive reclamation. Space is
bounded by preprocessing plus the total charged work. This choice avoids
an uncharged reclamation spike; no phase boundary performs a fresh
\(n\times k\) scan.

\subsection{The Complete Record Handler}

For \(k\ge2\), the committed checkpoint is
\((p,P_p,U_p,B,H,K_p)\), with \(H=B-1\) and \(|K_p|=k-H\).
The state also contains the virtual serving set \(Z\), its maintained
proper projection \(\Pi\) onto the live support, persistent edit logs
and row roots, a phase flag, and a permanent failure flag
\(\mathsf{halted}\). An active phase has its own generation, snapshot
time \(t\), immutable start set \(A\), mutable base \(M\), common
buffer \(J=I(t,u)\), and target-availability flag. Initially the phase
is idle; the \(k=1\) routine is given at the end of this section.

\paragraph{Commit rules.}
A record edit cannot be rolled back because a background job fails.
The handler separates mandatory input maintenance from optional
reconstruction and uses the following rules.
\begin{itemize}[leftmargin=1.6em,itemsep=0.4ex]
\item \emph{Input is mandatory.} Pre-edit phase setup consists only of
bounded root retention and scalar assignments. The handler then applies
its record exactly once, increments \(u\), appends its edit-log entry,
and updates live support and \(\Pi\). This commit has a separate
\(\wtO(n_u+k)\) allowance. Optional work cannot interrupt it.
\item \emph{Failure stops only optional work.} A failed guard, exhausted
total job allowance, or missed deadline fixes the current \(Z\),
sets \(\mathsf{halted}\), and cancels background jobs. Return to the
current handler, then to its caller. Halted handlers perform the
mandatory commit but add no buffer location, maintain no shadow rows,
and install no target. Reaching an ordinary work-slice boundary merely
pauses the worker.
\item \emph{Union membership is counted by constituent.} Maintain
\(Z=M\cup J\) from separate base and buffer memberships. Close a
location only when neither membership remains. At phase setup, share
the current \(Z\) as \(A\) and \(M\), and create an empty \(J\);
this changes no virtual facility and requires no scan.
\item \emph{Output changes are indivisible commits.} Before opening a
new buffer location, executing a migration transition, or installing a
synchronous target, check its prospective cardinality and reserve its
full work and recourse allowance, including any required projection
update. Adding a buffer membership at a location already in \(Z\)
does not change the projected source set. No pause or failure check
occurs inside a commit. Persistent roots are retired without recursive
reclamation.
\end{itemize}

The record handler has two branches. A small window is refreshed
synchronously, while a large window starts or advances a background
phase. Both branches first commit the record itself; the remaining
steps may fail without losing that edit.

\begin{algorithm}[H]
\caption{Weighted-record handler: \(\textsc{ProcessRecord}(e)\)}
\label{alg:process-record}\label{alg:worst-case-rebuilding}
\normalsize\raggedright
\textbf{Input:} one record edit \(e\) at location \(x\).
\begin{enumerate}[label=\arabic*.,leftmargin=1.9em,labelsep=0.5em,
                  itemsep=0.45em,parsep=0pt,topsep=0.4em]
\item If not halted, idle, and \(B\ge1024C\), start a phase before the
      edit: retain \(t=u\), \(P_t\), and \(A=Z\); set \(M=A\),
      \(J=\varnothing\), and \(q=\lfloor B/(256C)\rfloor\).
      Clear the target handle and edit dictionary; schedule four stages
      of \(q\) record updates.
\item Apply \(e\) exactly once, increment \(u\), append its edit-log
      entry, and finish the live-support and projection updates as one
      mandatory commit.
\item If halted, return to the caller. Otherwise complete
      \(\textsc{ForegroundRows}(x)\).
\item If \(B<1024C\), compute
      \(U_u=\Refresh(P_p,U_p,B,P_u)\) and
      \((B_u,H_u,K_u)=\WindowCore(P_u,U_u)\) under their caps.
      Check the synchronous guards of \Cref{subsec:failure-rule},
      including the prescribed \(O(L_\Delta^3)\) difference bound.
      Atomically install \(K_u\), adopt its checkpoint at time \(u\)
      and its row roots, set the phase idle, and return. No buffer slot
      is opened in this branch.
\item For a large-window phase, check the ages and the capacity of
      \(M\cup J\cup\{x\}\), and reserve the buffer-opening work.
      Add \(x\) to \(J\) and set \(Z=M\cup J\). If \(x\) was not
      previously in \(Z\), update \(\Pi\) for this source insertion
      in the same atomic commit; otherwise its source set is unchanged.
      Keep deleted and zero-weight locations in \(J\).
\item Run \(\textsc{AdvancePhase}()\) (\Cref{alg:advance-phase}), then return to the caller for
      the next record. Any failed optional step invokes the failure rule
      and returns after the already completed input commit.
\end{enumerate}
\end{algorithm}

\begin{table}[H]
\centering
\normalsize
\begin{tabularx}{\linewidth}{@{}lX@{}}
\toprule
Boundary & Required checks before advancing \\
\midrule
Stage 1 & Refresh is complete within its allowance; the current phase's
          target annotations and frozen row root exist, and its target
          handle is available. \\
Stage 2 & The window is complete,
          \(15B/(16C)\le B_t\le(C+1/16)B\),
          \(H_t=B_t-1\), \(K_t\subseteq U_t\), and
          \(|K_t|=k-H_t\). \\
Stage 3 & The migration plan is ready and has at most
          \(K_1BL_\Delta^3\) transitions; the catch-up counter is zero
          and no row task is unfinished. Both endpoint ages and the
          starting union satisfy their capacity guards. \\
Stage 4 & The plan is exhausted, \(M=K_t\),
          \(u-t\le B_t/16\), and \(|K_t\cup J|\le k\).
          Both target row roots are ready for adoption. \\
\bottomrule
\end{tabularx}
\caption{Deterministic stage guards. The fixed constant \(K_1\) is
chosen from the phase-budget bound. These tests use computed objects,
counts, and times; they do not test optimality or sampled robustness.}
\label{tab:inner-stage-guards}
\end{table}

The phase worker performs only one stage per slice. At stage
boundaries it checks the prepared target, the number of remaining
row tasks, the migration plan, and the available center slots.
The old checkpoint remains in service until the final stage
has installed the new core.

\begin{algorithm}[H]
\caption{One background slice: \(\textsc{AdvancePhase}()\)}
\label{alg:advance-phase}
\normalsize\raggedright
\textbf{State:} an active phase with start time \(t\), stage length
\(q\), base \(M\), and common buffer \(J\).

\medskip
Set \(i=\lceil(u-t)/q\rceil\). Resume only Stage \(i\) using its
fixed work slice; early completion waits for the scheduled boundary.
\begin{enumerate}[label=\textbf{Stage \arabic*.},leftmargin=5.5em,
                  labelsep=0.5em,itemsep=0.45em,parsep=0pt,topsep=0.4em]
\item Compute \(U_t=\Refresh(P_p,U_p,B,P_t)\) using the frozen
      inputs and edit-log segment \((p,t]\). On completion, install
      the target's frozen/live row handles, freeze \(\mathcal D\),
      and mark this phase's target available as specified above.
\item Compute \((B_t,H_t,K_t)=\WindowCore(P_t,U_t)\).
\item Prepare the \(A\)-to-\(K_t\) migration and finish all
      current-live reconciliation items with \Cref{alg:reconcile-row}.
\item Execute the next prescribed \(O(L_\Delta^3)\) migration
      transitions on \(M\). Check and reserve each complete transition
      before changing \(Z=M\cup J\); keep every buffer membership.
\end{enumerate}
\medskip
If \(u=t+iq\), apply the corresponding checks in
\Cref{tab:inner-stage-guards}; failure freezes \(Z\).
At the successful Stage~4 boundary, adopt
\((t,P_t,U_t,B_t,H_t,K_t)\) and its prepared row roots, retain
\(Z=K_t\cup J\), and set the phase idle. The checkpoint time is
\(t\), not the current time \(u\); do not clear the reconstruction-time
buffer. The next stage or phase starts only in the next record handler.
\end{algorithm}

\subsection{Failure Handling and the Weighted Guarantee}
\label{subsec:failure-rule}

Step~3 initializes the annotations and distance rows in
\(\wtO(k(n_{\rm init}+k))\) work (\Cref{lem:annotation-initialization}).
The dummy fallback and its proper projection are installed before any
randomized preprocessing, so the failure rule below also applies during
initialization. On success, the first reference has cost at most
\((3/2)\beta F_k\le\alpha F_k\), and its computed core starts at age
zero. This one-time preprocessing cost is separate from the cost of
each subsequent record edit.

\paragraph{Deterministic quotas and guards.}
The worst-case guarantees require a fixed limit on optional work,
including work that uses randomness. Each randomized static invocation
receives a failure allowance and fixed
sample and iteration caps. Each stage has a fixed work quota. Before
using a large-window target, check
\[
 15B/(16C)\le B_t\le(C+1/16)B,\qquad |K_t|=k-H_t,
\]
and require a prepared path of at most \(K_1BL_\Delta^3\) transitions.
At the Stage~3 boundary require the catch-up counter to be zero and no
row task to be unfinished. Before migration and adoption, check the age
and capacity inequalities. For a synchronous update, check
\(B_u\le(C+1/16)B+1\), cardinality, and the actual difference from
the current virtual set before installation. These tests use only
timestamps, cardinalities, and computed lists; none requires knowing
the optimum. On the success event, the preceding proofs
ensure that they pass and every quota suffices.

\paragraph{Failure is local to optional work.}
The mandatory processing of a record edit must finish even if a
background computation fails. We therefore separate that work from
reconstruction and migration. A failed guard, exhausted total job allowance,
or missed deadline permanently
freezes the current virtual facility set and cancels reconstruction and
migration. The record handler has already completed its mandatory input,
support, and proper-projection commit. That commit has a separate
deterministic allowance and cannot be abandoned; bounded pre-edit
snapshot retention invokes no randomized or budgeted worker.

After failure, return from the current handler after its mandatory
commit. Every later call still applies its input edit and maintains
the fixed virtual set's proper projection. Halted handlers add no
buffered facility, advance no live row task, and install no target.
Thus feasibility, properness, and resource bounds hold on every
execution, while approximation may fail. When this solver processes a
batch emitted by the summary in Step~6, the caller continues with every
remaining record in order before publication; \Cref{lem:batch-drain}
formalizes that composition rule. Each failure
that affects approximation is contained in the union of failed static
calls: when all calls succeed, all guards pass and every deadline is met.

\paragraph{Adaptive probability guarantee.}
An adaptive adversary can see previously published centers, but it
cannot change a static call's input once that input has been fixed.
Every such invocation uses fresh randomness on a frozen input.
Conditional on the complete history before that invocation, its failure
bound applies. Later updates may depend on published centers, but cannot
change that input. Allocate total failure probability at most \(\delta\)
among the calls, for example with an allowance proportional to
\(\delta/(i+1)^2\) for call \(i\). A conditional union bound gives
simultaneous success over the finite adaptive sequence. Confidence factors
are logarithmic in the horizon and initialization work.

\paragraph{Proper output.}
The schedule may temporarily maintain virtual facilities, which need
not be occupied client locations. Only their proper projection is
published. Apply \Cref{lem:projection} to the virtual set and current support.
The mandatory input commit updates the target-support side of the
projection. Opening a new buffer location subsequently updates its
virtual-source side, as a separate atomic event; an existing buffer
membership requires no source update.
One record edit changes the target support at one location and changes
the virtual set in \(O(L_\Delta^3)\) locations. The proper output
therefore has the same recourse order, requires \(\wtO(n_u+k)\)
work per edit, and costs at most \(6C_0F_k(P_u)\). Saturated projection
covers support smaller than \(k\), including empty support; dummies
are never targets.

\paragraph{One center.}
The preceding construction uses distinct anchors and is intended for
\(k\ge2\). The case \(k=1\) has a direct solution. On every nonempty weighted input,
sample \(\lceil\log_2(1/\xi)\rceil\) locations independently in
proportion to weight and evaluate each candidate's exact cost. A weighted
random location has expected cost at most \(2F_1(P)\), so with
probability at least \(1/2\) it costs at most \(4F_1(P)\).
Return the cheapest sample. Failure probability is at most \(\xi\),
work is \(\wtO(n+1)\), and recourse is at most two. Zero optimum
makes every sample optimal; empty input returns the empty set.
Initialization uses the same routine, with the preceding allocation of
failure allowances.

\begin{theorem}[Weighted worst-case inner algorithm]\label{thm:inner}\label{overview:theorem-inner}
For the weighted model in \Cref{sec:prelim}, the construction maintains
a proper constant-factor \(k\)-median solution under insertion,
deletion, or arbitrary weight change at one location. Each record edit
takes \(\wtO(n+k)\) time and \(O(L_\Delta^3)\) recourse in the
worst case, on every random execution. With any prescribed total failure
budget, approximation holds simultaneously over a finite adaptive
sequence. Initialization costs \(\wtO(k(n_{\rm init}+k))\); suppressed
logarithms include the positive weight range.
\end{theorem}
\begin{proof}
The initialization procedure constructs the first annotated checkpoint
within the claimed work bound. For a large stability window, the
four-stage schedule finishes reconstruction and migration before the
current checkpoint expires. For a small window, the synchronous update
uses the computable-core guarantee to preserve the same approximation
factor. In either case, the refresh theorem supplies an approximate
target, and the migration lemma keeps the number of facilities within
\(k\) throughout the transition.

The work budgets and migration quotas give the stated worst-case
bounds for every record edit. Deterministic guards ensure these bounds
even when a randomized call fails, and the conditional union bound
gives simultaneous approximation over adaptive updates. Finally,
projection makes each published solution proper, padding handles
small support, and the separate routine covers \(k=1\).
\end{proof}

%% file: sections/step6_summary.tex
\section{Step 6: Dynamic Weighted Summaries and Composition}
\label{sec:step6}\label{outer:section}

The weighted solver from Step~5 uses \(\widetilde O(n+k)\) time per
record edit, where \(n\) is its current support size. To obtain
\(\widetilde O(k)\) time for updates to the original clients, we
replace that input by a weighted summary on \(\widetilde O(k)\)
locations. This requires more than a small support: each original
update must also produce only \(O(1)\) weighted-record edits, and
the summary must preserve the cost of a solution at the same update
boundary at which the solution is published.

We maintain a partition of the clients into clusters, each represented
by a live client location with weight equal to the cluster's size.
A background controller rebuilds this partition as the input changes.
Rather than replacing all its records at once, a separate process moves
between partitions one cluster at a time. It activates each new cluster
with its full current weight before retiring the old records. During
this transition, a client is represented once or twice. We first prove
the resulting support and edit bounds, then bound the transportation
cost from clients to their representatives. Finally, we compose the
summary with the inner solver by processing each emitted batch in full
before publishing centers.

Throughout this section, \(P\) is the original unit-weight client multiset
and \(Q\) is the aggregate of active weighted records. The index \(t\)
counts original client updates, while \(u\) counts record edits passed to
the inner solver. One original update may emit a constant-size ordered
batch of such edits; the two indices need not agree.

The controller follows Grilnberger and Skarlatos~\cite[Sections~5.1--5.4]{GS26}.
The fixed-assignment count representation and the static weighted reduction
are related to \cite[Claim~3.7 and Appendix~C]{BCLP23}. We give the weighted
record, publication, and transportation arguments explicitly: neither a
small representative set nor a bound on changes to its representatives
alone bounds changes to its weights.

\subsection{The Summary Model and Static Layers}

An original update inserts or deletes one unit-weight client. We give
every insertion a fresh \emph{birth identifier}, even if that location
was occupied before. This distinguishes a new client from one that
has already been assigned to a frozen cluster. All sets in this
section consist of birth identifiers; distances use their associated
metric locations. For a set \(Y\) of births, \(\supp(Y)\) denotes the
distinct locations they occupy.

Let \(N\) bound the number of live clients, enlarged if necessary so
that \(N\ge\max\{2k,2\}\). Fix an update horizon \(T\) and failure
allowance \(\delta\). The distance oracle also accepts locations
whose clients have since been deleted.

For \(p\in\{1,2\}\), write
\[
 \Phi_{P,p}(C)=\sum_{x\in P}d(x,C)^p,
 \qquad
 \Phi_{Q,p}(C)=\sum_{q\in\supp(Q)}w_Q(q)d(q,C)^p,
\]
and let \(\OPT_{k,p}(P)\) denote the proper optimum on the original
instance. We omit \(p\) when it equals one. The same summary construction
will serve both objectives. A weighted-record edit changes one coordinate
of \(w_Q\), possibly from or to zero, by an arbitrary amount.

Fix \(\beta_{\mathrm{o}}=1/4\), \(\gamma_{\mathrm{o}}=3/4\), and
sufficiently small positive absolute constants
\(\varepsilon_{\mathrm{o}},\lambda_{\mathrm{o}}\le10^{-3}\). Set
\[
 \theta=1-\beta_{\mathrm{o}}+5\varepsilon_{\mathrm{o}}<1,
 \qquad
 \eta_{\mathrm{o}}=3\varepsilon_{\mathrm{o}}+2\lambda_{\mathrm{o}}<1/8.
\]
Let \(L_{\mathrm{out}}=\Theta(1+\log N)\) be large enough to bound the
number of levels in every construction below. Define
\begin{equation}\label{outer:parameters}
 \begin{aligned}
 s_{\mathrm{samp}}
 &=\left\lceil\log\frac{8(N+1)(T+2)L_{\mathrm{out}}^2}{\delta}\right\rceil,\\
 M_{\mathrm{term}}&=\lceil A_0k s_{\mathrm{samp}}L_{\mathrm{out}}\rceil.
 \end{aligned}
\end{equation}
Choose the absolute constant \(A_0\) so that
\(M_{\mathrm{term}}\ge2k\),
\(M_{\mathrm{term}}\ge100L_{\mathrm{out}}/\varepsilon_{\mathrm{o}}\), and
\(M_{\mathrm{term}}\ge100/\lambda_{\mathrm{o}}\). The notation \(\widetilde O\) in
this section hides the resulting logarithmic factors and dictionary
costs. The empty instance has an empty summary.

\paragraph{The state of a layer.}
The controller builds successive layers, each covering a constant
fraction of its input. A sampling layer freezes an execution set
\(U\) of size \(a>M_{\mathrm{term}}\).
Let \(P^{\mathrm{gen}}\) be the complete original instance at that
freeze; thus \(U\subseteq P^{\mathrm{gen}}\). The layer selects at most
\(2k\) sample locations and a radius \(\nu\), and covers the frozen set
\(B^{\mathrm{gen}}\subseteq U\) of births within distance \(\nu\) of a
sample. Assign each covered birth to one such sample, using fixed tie
breaking. These assignments define the \emph{native clusters}. A sample
is retained as a \emph{ghost sample} after deletion; it defines the
assignment but need not be the live representative of its cluster.

The assignments are fixed when the layer is built. In particular,
deleting a client does not send its former cluster members to a
different sample. The covered ownership \(B^{\mathrm{gen}}\) never changes. Its live view is
\(B^{\mathrm{live}}=B^{\mathrm{gen}}\cap P\), and each nonempty cluster
chooses a representative among its live members. An \emph{identity layer}
on an execution set of size at most \(M_{\mathrm{term}}\) consists of singleton native
clusters and has radius zero. Identity origin is permanent: a singleton
never gains members if its layer later becomes nonterminal.

We also need to distinguish the frozen input from the set that would
result if all subsequent updates were applied to it. For any frozen
birth set \(U\), define \(\operatorname{Lazy}(U)\) by
starting from \(U\), inserting every subsequent fresh birth, and deleting
each subsequently deleted birth if present. This live set is distinct
from both \(U\) and \(B^{\mathrm{gen}}\). Public descriptors have an
immutable identity consisting of a worker, generation, and level. Their
frozen sizes always refer to the generating sets of the \emph{published}
descriptors, not to the inputs of their current private workers.

For nonempty \(Y\subseteq P\) and \(\xi\in(0,1)\), define
\begin{equation}\label{outer:quantile-definition}
 \mu_\xi(Y;P)=
 \min_{\substack{C\subseteq\supp(P)\\|C|\le k}}
 \min\{r\ge0:|\{y\in Y:d(y,C)\le r\}|\ge\xi|Y|\}.
\end{equation}

The static task for a sampling layer is now precise: given its frozen
execution set, produce a bounded number of samples, fixed assignments
for a constant fraction of the births, and a covering radius. The next
lemma supplies this task with a deterministic work bound. Taking
exactly \(2k\) births per trial fixes the number of samples and the
work even if the chosen radius is poor; randomness is used only to
control that radius.

\begin{lemma}[Fixed-size sampling]\label{outer:fixed-sampling}
On a frozen execution set \(U\) of size \(a>M_{\mathrm{term}}\), perform \(s_{\mathrm{samp}}\)
independent trials. A trial chooses a uniform subset of exactly \(2k\)
births and computes the smallest radius covering at least
\(\beta_{\mathrm{o}}a\) births of \(U\). Keep a trial of minimum radius
\(\nu\). Every outcome uses at most \(2k\) ghost sample locations,
covers at least \(\beta_{\mathrm{o}}a\) frozen births, and requires
\(O(ak s_{\mathrm{samp}}\operatorname{polylog}N)\) work, including its assignments.
Conditioned on the full history \(\mathcal H\) through the freeze of
\(U\), before fresh randomness for this invocation is exposed,
\begin{equation}\label{outer:sample-quality}
 \Pr\!\left[\nu>2\mu_{\gamma_{\mathrm{o}}}(U;P^{\mathrm{gen}})
       \mid\mathcal H\right]\le e^{-s_{\mathrm{samp}}}.
\end{equation}
\end{lemma}
\begin{proof}
A reservoir pass samples the births, and \(O(ak)\) distance queries and
a selection algorithm determine a trial's covering radius. The chosen
assignments require a further \(O(ak\operatorname{polylog}N)\) work.
All these computations use the frozen input, including births that have
since been deleted.

For the radius bound, put
\(\mu=\mu_{\gamma_{\mathrm{o}}}(U;P^{\mathrm{gen}})\). A witnessing
center set covers at least \(\gamma_{\mathrm{o}}a\) births. Partition
those births among at most \(k\) centers into groups of sizes
\(c_1,\ldots,c_k\). Hitting one such group places a sample within
\(2\mu\) of all its members. The expected mass of unhit groups is at
most
\[
 \sum_{i=1}^k c_i(1-c_i/a)^{2k}
 \le\sum_{i=1}^k c_i e^{-2kc_i/a}
 \le\frac{a}{2e}.
\]
A trial can fail to cover \(\beta_{\mathrm{o}}a\) births within radius
\(2\mu\) only if its unhit mass exceeds
\((\gamma_{\mathrm{o}}-\beta_{\mathrm{o}})a=a/2\). Markov's inequality
therefore bounds the failure probability by \(1/e\). Independence of
the \(s_{\mathrm{samp}}\) trials proves the claim.
\end{proof}

Fixed-size sampling is also discussed in \cite[Section~4.4]{GS26}.
Here its role is to make support size and construction work deterministic;
only the radius comparison can fail.

\subsection{The Whole-Suffix Controller}

Rebuilding one layer may change all later layers. The controller
therefore rebuilds an entire suffix in the background while keeping
the earlier public layers unchanged. There is one private worker for
each public level. A level-\(i\) worker starts from the complement
of the public prefix below \(i\), constructs the remaining layers
in order, and freezes each successor from the latest residual root.
Every worker receives a fixed quota after each original update.
Its computation may pause, but membership changes are committed
against current liveness and the latest dictionary roots.

The controller has two separate responsibilities. Its public layers
must always give a complete partition at a snapshot boundary, and
their generating sizes and ages must remain controlled after a suffix
is replaced. The first follows from residual consistency; the second
requires checking publication, terminal overflow, and restart
explicitly.

\paragraph{Publication and restart.}\label{outer:suffix-controller}
The controller publishes complete suffixes, never partially rebuilt
ones. After forwarding an update to the live data, resume the private workers.
Choose the smallest level whose worker reports completion. Publish that
worker's complete suffix, retain the public prefix, restart the affected
workers from their current residual roots, and retire workers beyond the
new terminal level. A private restart never changes a retained public
generation or a transition snapshot. At most one completed suffix is
selected in a handler; a worker made immediately complete by a restart is
eligible for selection on the next original update.

When several workers finish in the same handler, only the smallest
completed level is published. Completed workers at larger levels are
discarded, together with their generation-specific tasks; workers at
smaller levels keep their inputs because their public prefixes have not
changed. The affected workers restart from the new public residuals.
The original edit is forwarded before these decisions are made.

At the last level, the residual is small enough to represent each
remaining birth by a singleton. The terminal identity partition is
read from the latest live residual and
refreshed at every completed update. If its residual has size at most
\(M_{\mathrm{term}}\), this takes
\(O(M_{\mathrm{term}}\operatorname{polylog}N)\) work. If an
insertion changes its size from \(M_{\mathrm{term}}\) to
\(M_{\mathrm{term}}+1\), retain the old identity descriptor of
size \(M_{\mathrm{term}}\) and append a singleton terminal layer for the new
birth. Start the old terminal level's nonterminal worker on these
\(M_{\mathrm{term}}+1\) births. The new terminal worker is immediately complete, without
recursively selecting another suffix in the same handler. If an earlier
suffix is published, its terminal identity state is instead refreshed
from that suffix's current residual. These rules are the terminal
boundary convention throughout the proof.

The terminal construction uses the latest residual root and its
reserved immediate work allowance before a worker reports its suffix
complete. A completed suffix is published in that handler or discarded
in favor of a smaller completed level, except for a worker made complete
by a restart. That worker is considered on the next update, when its
terminal residual is refreshed again. At most one birth can have been
inserted in the meantime, so this deferred case needs at most the
single-overflow rule above.

\paragraph{Residual transactions.}
A worker may pause between deciding which births to cover and
recording their assignments. We maintain a residual that remains
correct during this interval. A worker generation \(g\) has an
immutable covered-ownership set and a mutable live residual.
Once its sampling trial is fixed, let
\(C_g\) be the covered births whose assignment transactions have
committed, including births that are already dead. Initially
\(C_g=\varnothing\), and the residual shares the current
\(\operatorname{Lazy}(U_g)\) root. Maintain
\begin{equation}\label{outer:residual-invariant}
 R_g=\operatorname{Lazy}(U_g)\setminus C_g.
\end{equation}
An original insertion or deletion is forwarded to the lazy set and the
latest residual root. To commit a covered birth \(b\), reserve the
necessary dictionary operations from the worker's current quota, check
its generation and current liveness, and atomically add \(b\) to
\(C_g\) and erase \(b\) from \(R_g\). The transaction also installs its
ownership entry and, if it is live, updates the cluster's member tree,
count, and representative. Dead births are erased idempotently and add
no live mass. A fresh insertion never belongs to the old covered set.
Superseded tasks are discarded, and no paused task can install a root
prepared before its pause. Once the assignments are complete, the
next execution set is frozen simply by sharing \(R_g\); no scan to
form a set difference is needed.

Write \(B_j^{\mathrm{pub,live}}\) for the current live covered set of
public level \(j\), and let
\(H_{<i}=\bigcup_{j<i}B_j^{\mathrm{pub,live}}\).

\begin{lemma}[Residual and public-prefix consistency]
\label{outer:prefix-residual}
Every current generation satisfies \eqref{outer:residual-invariant}.
After its assignment pass,
\(R_g=\operatorname{Lazy}(U_g)\setminus B_g^{\mathrm{gen}}\).
Every active level-\(i\) worker, whose starting execution set is
\(U_i^i\), satisfies
\begin{equation}\label{outer:prefix-complement}
 \operatorname{Lazy}(U_i^i)=P\setminus H_{<i}.
\end{equation}
Consequently, after the structural publication and terminal-identity
operation, the public native clusters partition the live births.
These statements do not require successful sampling-radius estimates.
\end{lemma}
\begin{proof}
We first check that an interrupted assignment cannot cause a birth
to disappear from both the covered clusters and the residual.
An insertion belongs to neither \(C_g\) nor the frozen covered set and
enters both live roots. A deletion removes its birth from both roots,
leaving \(C_g\) unchanged. A commit adds exactly its covered birth to
\(C_g\) and erases it from the latest residual. These operations preserve
\eqref{outer:residual-invariant}; after all assignments,
\(C_g=B_g^{\mathrm{gen}}\).

For consecutive layers of a private generation, sharing the completed
residual gives
\[
 \operatorname{Lazy}(U_{j+1})
 =\operatorname{Lazy}(U_j)\setminus B_j^{\mathrm{gen}}.
\]
The equality persists under future original updates: no fresh birth
belongs to \(B_j^{\mathrm{gen}}\). Iterating it, and reading the terminal
identity partition from the latest residual, partitions the worker's live
input.

It remains to check that publishing a suffix does not change who owns
a birth in the retained prefix. The public-prefix equality holds
initially. An insertion belongs to no
old prefix cluster and enters every worker input. A deletion in the
prefix removes a birth from both \(P\) and \(H_{<i}\); one outside it
is removed from the worker's lazy input. If level \(h\) publishes, its
completed suffix partitions \(P\setminus H_{<h}\), while lower levels
are unchanged. Restarting every affected worker from the corresponding
new residual restores \eqref{outer:prefix-complement}. The terminal
overflow rule partitions its residual into the old singleton births and
the one fresh birth. Induction over updates and publications proves all
claims.
\end{proof}

We next check the three events at which the public structure can
change: suffix publication, terminal overflow, and worker restart.
The distinction between public and private generations matters here.
A public descriptor may survive after the worker that produced it has
been replaced, so its age must be measured from its own freeze time.

\begin{lemma}[Closure under publication, identity overflow, and restart]
\label{outer:boundary-events}
Abbreviate \(\varepsilon=\varepsilon_{\mathrm{o}}\) and
\(\beta=\beta_{\mathrm{o}}\). Suppose the whole-suffix controller has
the following deterministic guarantees: each sampling layer frozen on
\(a>M_{\mathrm{term}}\) births covers at least \(\beta a\); from every nonterminal
internal execution set of size \(a\), its worker completes the remaining
suffix or is superseded within \(\varepsilon a\) updates; and the
controller uses the latest-root and terminal-refresh rules above.
Then, after every structural update:
\begin{enumerate}
\item The public clusters partition \(P\), and each private input is the
complement of its public prefix. Private restarts leave retained public
and snapshot ownership unchanged.
\item Consecutive public generating sizes satisfy
\begin{equation}\label{outer:boundary-geometric}
 a_{j+1}\le\theta a_j.
\end{equation}
Every nonterminal identity-origin descriptor has frozen size exactly
\(M_{\mathrm{term}}\) and retains singleton memberships.
\item Let a sampling-origin descriptor of size \(a\) be frozen at
\(\sigma\), first published at \(\pi\), and first replaced by a suffix
at its own or an earlier level at \(\omega\). Then
\begin{equation}\label{outer:boundary-lifetime}
 \pi-\sigma\le\varepsilon a,
 \quad \omega-\pi\le\varepsilon(1+\varepsilon)a,
 \quad |U\triangle\operatorname{Lazy}(U)|
       \le(2\varepsilon+\varepsilon^2)a\le3\varepsilon a
\end{equation}
while it survives. If the horizon ends first, use any surviving current
time in place of \(\omega\). A terminal identity layer has age zero
after refresh. An identity layer promoted on overflow satisfies
\begin{equation}\label{outer:boundary-identity-age}
 |U\triangle\operatorname{Lazy}(U)|
 \le1+\varepsilon(M_{\mathrm{term}}+1)\le3\varepsilon M_{\mathrm{term}}.
\end{equation}
\item Suffix publication and private restart change no active summary
tokens. They replace roots and descriptors; terminal refresh uses
\(O(M_{\mathrm{term}}\operatorname{polylog}N)\) work. The foreground update and the
separate synchronization process account for all weighted-record edits.
\end{enumerate}
\end{lemma}
\begin{proof}
The partition statement is \Cref{outer:prefix-residual}. It remains to
check sizes and ages across publications, identity overflow, and terminal
refreshes. We first compare consecutive layers within one private
construction, then account for the time elapsed when a new suffix
meets an older public prefix.

Within a private generation, suppose a layer of frozen size \(a\)
freezes its successor after \(d\le\varepsilon a\) updates. Deterministic
coverage and at most one new birth per update imply
\begin{equation}\label{outer:boundary-internal}
 b\le a-|B^{\mathrm{gen}}|+d
   \le(1-\beta+\varepsilon)a.
\end{equation}
Deletions do not increase this upper bound.

At a suffix publication, pairs entirely inside the new suffix are
handled by \eqref{outer:boundary-internal}, while pairs inside the
retained public prefix do not change. Only the pair crossing the
retained-prefix boundary may combine different generations. Consider
such a publication at level \(h\)
with a sampling-origin predecessor. Let \(a\) be that predecessor's
unchanged frozen size and
\(b\) the frozen size of its successor residual in the producing
generation (possibly zero). Let \(\sigma_b\) be the successor's freeze
time and \(\pi_{h-1}\) the predecessor's publication time. The
producing worker publishes that predecessor only after completing its
whole suffix. Its remaining construction after \(\sigma_b\) takes at
most \(\varepsilon a\) original updates, so
\begin{equation}\label{outer:cross-generation-old-time}
 0\le\pi_{h-1}-\sigma_b\le\varepsilon a.
\end{equation}
At \(\pi_{h-1}\), the level-\((h-1)\) worker restarts on the lazy
predecessor input, of size at most \((1+\varepsilon)a\): from the
predecessor's own freeze to its publication there were at most
\(\varepsilon a\) updates. The publication also discards any older
level-\(h\) worker, so a new level-\(h\) input associated with this
retained predecessor cannot have frozen before \(\pi_{h-1}\).
While that public descriptor survives,
the restarted worker cannot have completed and been published, and
no smaller-level publication can have superseded it; either event
would replace the predecessor. Thus, if \(\sigma_h\) is the freeze
time of a new level-\(h\) input published with this predecessor still
present, the restarted worker's completion deadline implies
\begin{equation}\label{outer:cross-generation-new-time}
 0\le\sigma_h-\pi_{h-1}\le\varepsilon(1+\varepsilon)a.
\end{equation}
By public-prefix consistency \eqref{outer:prefix-complement}, the
new level-\(h\) input is the lazy version, at \(\sigma_h\), of the
producing generation's successor residual. Hence its frozen size is
at most \(b\) plus the number of intervening insertions. Combining
the two elapsed-time bounds gives
\begin{equation}\label{outer:boundary-sampling-cross}
 \begin{aligned}
 a_h&\le b+(\sigma_h-\sigma_b)\\
    &\le(1-\beta+\varepsilon)a+\varepsilon a
                       +\varepsilon(1+\varepsilon)a\\
    &=(1-\beta+3\varepsilon+\varepsilon^2)a\\
    &\le(1-\beta+5\varepsilon)a=\theta a.
 \end{aligned}
\end{equation}
New internal pairs satisfy \eqref{outer:boundary-internal}; retained
prefix pairs keep their previous bound. This proves the size inequality
at every sampling-origin boundary.

A terminal identity layer follows a different rule because it does
not cover births by sampling. It can become nonterminal only on an insertion
into its size-\(M_{\mathrm{term}}\) residual. Its old singleton memberships are retained,
and the inserted birth forms the appended terminal layer. The restarted
nonterminal worker completes or is superseded after at most
\begin{equation}\label{outer:boundary-overflow-time}
 d_{\rm id}\le\varepsilon(M_{\mathrm{term}}+1)
\end{equation}
further updates. Every suffix birth while that identity prefix survives
was inserted after its freeze. Thus any first suffix generating set has
size at most \(1+d_{\rm id}\le\theta M_{\mathrm{term}}\), including at later suffix
publications. The same edit count proves
\eqref{outer:boundary-identity-age}. The lower bounds on \(M_{\mathrm{term}}\) absorb
both the overflow insertion and the one-update terminal convention.
In particular, \(1+d_{\rm id}<M_{\mathrm{term}}\): the appended terminal
residual cannot overflow a second time before its old identity prefix
is replaced. A newly published suffix obtains its terminal partition
from the latest residual, as required by the immediate-construction
rule above.

To bound age, follow one descriptor from its freeze to publication
and then until it is replaced. A sampling-origin descriptor completes
between its freeze and publication in at most \(\varepsilon a\) updates. At publication, its
same-level private worker starts from \(\operatorname{Lazy}(U)\), of
size at most \((1+\varepsilon)a\). It completes or is superseded within
another \(\varepsilon(1+\varepsilon)a\) updates, replacing that
public descriptor. This proves \eqref{outer:boundary-lifetime}.
A restart at an earlier level removes the old descriptor; a publication
at a later level does not restart its same-level worker. Hence repeated
private restarts cannot accumulate age against one retained descriptor.
Terminal identity refresh has age zero by construction.

A terminal refresh without overflow also preserves
\eqref{outer:boundary-geometric}. If its predecessor is sampling-origin,
with generating set \(U\), frozen size \(a\), and covered set
\(B^{\mathrm{gen}}\), prefix consistency identifies the current residual
as \(R=\operatorname{Lazy}(U)\setminus B^{\mathrm{gen}}\). Hence
\eqref{outer:boundary-lifetime} gives
\[
 |R|
 \le a-|B^{\mathrm{gen}}|
       +|\operatorname{Lazy}(U)\setminus U|
 \le(1-\beta+3\varepsilon)a
 \le\theta a.
\]
For an identity-origin predecessor, the bound
\(|R|\le1+d_{\rm id}\le\theta M_{\mathrm{term}}\) above applies.
With no predecessor the interlayer bound is vacuous; overflow was
handled separately above.

All these operations concern native partitions. Their tokens are changed
only by the synchronization process defined below. Publications and
restarts therefore emit no weighted-record edits.
\end{proof}

\begin{lemma}[Adapted layer scheduling {\cite{GS26}}]
\label{outer:scheduling}
Suppose a sampling layer on \(a>M_{\mathrm{term}}\) births has deterministic coverage
at least \(\beta_{\mathrm{o}}a\), at most \(2k\) ghost samples, and
work \(O(ak s_{\mathrm{samp}}\operatorname{polylog}N)\). The whole-suffix controller
can be implemented with \(O(L_{\mathrm{out}})\) workers and
\(\widetilde O(k)\) work per original update. Its public native
clusters partition the live instance, and consecutive frozen sizes obey
\begin{equation}\label{outer:geometric}
 a_{j+1}\le\theta a_j.
\end{equation}
Every published descriptor obeys
\begin{equation}\label{outer:upd-age}
 |U_j\triangle\operatorname{Lazy}(U_j)|
 \le3\varepsilon_{\mathrm{o}}a_j.
\end{equation}
From any nonterminal internal execution set of size \(a\), a worker
completes its remaining construction or is superseded within
\(\varepsilon_{\mathrm{o}}a\) original updates. These guarantees hold
on every sampling outcome.
\end{lemma}
\begin{proof}
The quota is chosen so that a layer's work can be spread over a small
fraction of its frozen size. Let \(g(N)\) bound dictionary overhead,
enlarging it by logarithmic confidence factors when needed.
Give each worker a per-update quota
\(\mathsf{quota}_{\mathrm{worker}}
  =A_1k s_{\mathrm{samp}}L_{\mathrm{out}}g(N)/\varepsilon_{\mathrm{o}}\).
For sufficiently large \(A_1\), one nonterminal layer of size \(a>M_{\mathrm{term}}\)
finishes in at most \(\varepsilon_{\mathrm{o}}a/(4L_{\mathrm{out}})\)
updates, including a constant number of boundary operations. The lower
bound on \(M_{\mathrm{term}}\) absorbs rounding. Its successor size is at most
\((1-\beta_{\mathrm{o}}+\varepsilon_{\mathrm{o}})a\), by
\eqref{outer:boundary-internal}. The contraction bounds the remaining
number of layers by \(L_{\mathrm{out}}\); summing their work and boundary
allowances gives a remaining completion time of at most
\(\varepsilon_{\mathrm{o}}a\).

Terminal constructions need a separate allowance because a small
residual must be refreshed immediately at publication. Including the
possible one-birth overflow singleton, they involve
\(O(M_{\mathrm{term}})\) records. Reserve their
\(O(M_{\mathrm{term}}g(N))\) work separately, including the public terminal
refresh; even if all workers reach a terminal stage in the same update,
the total is
\(O(L_{\mathrm{out}}M_{\mathrm{term}}g(N))=\widetilde O(k)\). The remaining worker
quotas also total \(\widetilde O(k)\). Assignment commits add only
\(O(a g(N))\) work per layer, and original updates affect only
\(O(L_{\mathrm{out}}^2)\) live roots.

For completeness, we close the scheduling and boundary arguments by
induction over completed original-update handlers. The induction state
consists of the public partition and geometric size bound, together
with the age and worker-input invariants of
\Cref{outer:prefix-residual,outer:boundary-events}.
These assertions hold after static initialization. Assume they hold
at the start of a handler. Geometric decay bounds the number of
active workers by \(O(L_{\mathrm{out}})\), so the handler can give each
worker its full quota while reserving the terminal and boundary work
separately. For any worker that is not superseded, the per-layer work
bound and the deterministic coverage guarantee yield the completion
deadline above, independently of future publication decisions. A
worker restarted at the end of a handler receives its first quota on
the next update; the lower bound on \(M_{\mathrm{term}}\) absorbs this
one-update convention. These deadlines are exactly the hypotheses of
\Cref{outer:boundary-events}. Applying that lemma to the handler's
publication, overflow, and restart events preserves the public
partition, geometric sizes, and descriptor ages. Its prefix-consistency
claim also supplies the inputs for newly restarted workers. The same
induction state therefore holds after the handler, completing the
argument without assuming the next handler's size bounds in advance.
This scheduling pattern follows \cite[Sections~5.1--5.2]{GS26}.
The argument above verifies worker deadlines and public-layer boundary
events for our latest-root, fixed-ownership implementation. The
weighted-record synchronization and transportation guarantees are proved
separately in \Cref{outer:transition,outer:transport}; they are not
consequences of the cited \(k\)-center controller.
\end{proof}

\begin{corollary}[Deterministic support bound]\label{outer:support}
For an absolute constant \(\rho\), the public native partition contains
at most \(\rho M_{\mathrm{term}}\) clusters on every execution.
\end{corollary}
\begin{proof}
The geometric size decay limits the number of layers, while
fixed-size sampling limits the number of clusters in each layer.
Sampling-origin layers contribute at most \(2kL_{\mathrm{out}}\)
clusters. Identity-origin layers are singleton partitions, start at size
at most \(M_{\mathrm{term}}\), and have geometrically decreasing generating sizes.
Their total contribution is at most \(M_{\mathrm{term}}/(1-\theta)\). Thus
\[
 \#\{\text{public native clusters}\}
 \le2kL_{\mathrm{out}}+\frac{M_{\mathrm{term}}}{1-\theta}\le\rho M_{\mathrm{term}}.
\]
Deleting members or changing representatives creates no native clusters.
\end{proof}

\begin{table}[t]
\centering
\setlength{\tabcolsep}{4pt}
\renewcommand{\arraystretch}{1.12}
\begin{tabularx}{\linewidth}{@{}>{\raggedright\arraybackslash}p{0.28\linewidth}>{\raggedright\arraybackslash}X@{}}
\toprule
Interface & Source and scope \\
\midrule
Whole-suffix controller and native partition
 & \cite[Section~5.2 and Lemma~5.8]{GS26}; latest-root realization in
   \Cref{outer:prefix-residual}. \\
Completion deadlines, public sizes, and ages
 & \cite[Lemmas~5.11--5.17 and 5.21--5.24]{GS26}; deterministic sampling
   work and identity/restart boundaries in \Cref{outer:boundary-events,outer:scheduling}. \\
Retained snapshot records and constant record changes
 & \Cref{outer:records,outer:transition,outer:event-consistency};
   separate counts for each snapshot and full-mass activation. \\
Transportation for sum objectives
 & \Cref{outer:age-quantile,outer:tail-packing,outer:transport};
   a charge that does not require nested execution sets. \\
\bottomrule
\end{tabularx}
\caption{Structural ingredients inherited from the layer controller and
weighted interfaces proved in this section.}
\label{tab:outer-correspondence}
\end{table}

\subsection{Live Records and Full-Mass Synchronization}

The public partition can be rebuilt without immediately changing the
weighted input of the inner solver. We make that separation explicit
through \emph{transition snapshots}. Such a snapshot freezes cluster
ownership, while its counts and representatives continue to reflect
insertions and deletions. By contrast, an inner-solver snapshot freezes
the weighted input itself for a background computation. Keeping
transition records live ensures that a target cluster receives its
current full mass when activated.

\begin{lemma}[Cluster records]\label{outer:records}
Current cluster counts, live representatives, and retained transition
snapshots can be maintained in worst-case \(\widetilde O(k)\) time per
original update. Freezing a snapshot takes \(\widetilde O(k)\) time
without scanning all original clients. In either of two retained
snapshots, the cluster containing a deleted birth can be found in
\(\operatorname{polylog}N\) time.
\end{lemma}
\begin{proof}
The main requirement is to find and update a deleted birth's record
without searching a cluster. Each layer generation stores an ownership dictionary from covered birth
identifiers to native cluster identifiers. Each cluster stores a live
count, a representative birth, and a persistent balanced tree of live
members. The assignment transactions above build these structures within
the layer work bound. If a deletion precedes an assignment commit, the
commit's liveness check excludes it. If the deletion follows a commit,
the completed ownership entry locates its record. No membership can be
lost or counted twice because a worker pauses.

A deletion queries all relevant worker, public, and snapshot ownership
roots, of which there are \(O(L_{\mathrm{out}}^2)\). It updates each
distinct mutable record once and obtains any replacement representative
from that record's live-member tree. A record identity includes the
snapshot or generation identity: two records sharing an initial tree
root must both be updated. Path copying keeps their subsequent live
versions independent. Phase-insertion singletons have the separate
handles specified below.

Freezing a snapshot copies cluster records, not clients.
By \Cref{outer:support}, freezing a target copies only \(O(M_{\mathrm{term}})\)
count/representative records and retains \(O(L_{\mathrm{out}})\)
ownership roots. It shares the persistent member trees. It copies no
client-sized ownership array. Public suffix changes replace descriptor
references; inactive-record maintenance emits no summary edits.
Root retirement uses the append-only persistent allocator of
\Cref{subsec:trees}, so it triggers no recursive deallocation. The
costs of constructing and maintaining records are already included in
\Cref{outer:scheduling}; each snapshot copy costs
\(O(M_{\mathrm{term}}\operatorname{polylog}N)=\widetilde O(k)\).
\end{proof}

\paragraph{Tokens and phases.}
A \emph{token} is a cluster record belonging to a particular snapshot,
together with an active flag. An active token contributes its entire
live cluster mass at its representative location. Tokens from different
snapshots are distinct, even if they initially share a representative
or an ownership root. The summary \(Q\) adds their contributions
when several representatives occupy the same location.

The transition has two subphases. First every new cluster is activated;
only then are the old clusters removed. This order avoids a gap in
coverage, at the cost of temporarily counting some clients twice.

During initialization, freeze the first native partition into a
snapshot \(S^{(0)}\), activate all its nonempty tokens, and let
\(I^{(0)}=\varnothing\). Also initialize the first target snapshot
\(S^{(1)}\) and its empty insertion family before the first original
update. Thereafter, at the start of phase \(\ell\), freeze the current
public partition into an inactive target \(S^{(\ell)}\). The active
old family is \(S^{(\ell-1)}\cup I^{(\ell-1)}\). First activate all
new-target tokens; then retire all old-family tokens. An insertion during
phase \(\ell\) immediately creates its own active singleton in
\(I^{(\ell)}\). At phase completion, precisely
\(S^{(\ell)}\cup I^{(\ell)}\) remains active.

Both subphases traverse frozen lists of token identifiers. A deletion
updates all retained records, active or inactive, but does not remove a
list position. Empty tokens leave tombstones, and changing a
representative does not change list order. The old family remains live
throughout its retirement subphase.

\paragraph{Singleton handles.}
A client inserted after a snapshot freeze cannot be assigned to one
of that snapshot's fixed clusters. Its singleton token keeps it
represented until a later snapshot incorporates the birth.
Retain two dictionaries keyed by birth identifier, one each for
\(I^{(\ell-1)}\) and \(I^{(\ell)}\). A handle identifies its own
singleton token, active flag, mass, and list position. A deletion reads
at most two handles. If the token is active, remove its contribution;
then clear its mass and active flag and retain a tombstone at its list
position. Retiring an already dead singleton does no further work on
\(Q\). At a phase boundary, drop the completed old-family roots, promote
the current family, and create an empty new family. No boundary scans or
renumbers a singleton list.

\paragraph{Original-update order.}\label{outer:update-order}
The order matters when one update also finishes a reconstruction or
crosses a synchronization boundary. Let \(\mathcal Q\) denote the
complete outer state: the public layers, private workers, retained
cluster records, and active tokens. The routine below takes one
unit-weight original update \(\omega\), updates this state, and
returns an ordered list of weighted-record edits transforming the old
aggregate \(Q\) into the new one. Its synchronization quota is the
absolute constant \(q_0\) chosen in \Cref{outer:transition}.

\begin{algorithm}[H]
\caption{Dynamic weighted summary: \(\textsc{SummaryUpdate}(\mathcal Q,\omega)\)}
\label{alg:summary-update}
\normalsize\raggedright
\begin{algorithmsteps}
\item \textbf{Begin.} Save the synchronization phase at handler
  entry and initialize the ordered weighted-edit list \(\mathcal E\).
\item \textbf{Update live records.} Apply the insertion or deletion
  to all relevant live, lazy, and residual sets and retained records.
  A fresh birth creates an active singleton in the entry phase's
  insertion family. A deletion adjusts every affected active-token
  contribution. Append the resulting coordinate edits to \(\mathcal E\).
\item \textbf{Advance workers.} Spend the fixed work quotas.
  Commit covered births using current liveness, current roots, and
  generation checks.
\item \textbf{Publish layers.} If workers have completed, select
  the smallest completed level and publish its complete suffix.
  Apply the \hyperref[outer:suffix-controller]{terminal boundary rule}
  to the current residual. Restart affected workers from current roots
  and retire workers beyond the terminal level. Select at most one
  completed suffix in this handler.
\item \textbf{Synchronize tokens.} Execute \(q_0\) frozen-list or
  phase-boundary steps. Activate new-target tokens before retiring
  old-family tokens, appending each coordinate edit to \(\mathcal E\).
  At a phase boundary, freeze the fully updated public partition as
  the next target and create an empty insertion family.
\item \textbf{Return.} Emit \(\mathcal E\) in execution order.
  The driver processes the entire list before publishing centers
  (\Cref{alg:dynamic-kmedian}).
\end{algorithmsteps}
\end{algorithm}

The publication step uses the current residual after the original edit,
including a terminal identity overflow caused by the current insertion
or encountered when an already-completed worker becomes eligible.
The successor execution set shares the completed worker's latest
committed residual root; publishing it requires no bulk set-difference
scan.

Thus a birth already included in a newly frozen target is not also
inserted into that target's new singleton family. This remains true if
one handler crosses several phase boundaries. Every coordinate change
is logged in execution order as a weighted-record edit; aggregation
preserves nonnegative intermediate weights, and zero coordinates are
removed.

\begin{lemma}[Synchronization invariant]
\label{outer:transition}\label{outer:record-changes}
For a sufficiently large absolute constant
\(q_0=\Theta(\rho/\lambda_{\mathrm{o}}^2)\), performing \(q_0\)
synchronization steps per original update gives the following deterministic
guarantees. Each phase lasts at most \(\lambda_{\mathrm{o}}M_{\mathrm{term}}\) updates.
Every live birth belongs to one or two active tokens, including between
activation and retirement steps. There are \(O(M_{\mathrm{term}})\) active tokens,
and their aggregate has positive integer weights of total at most
\(2|P|\). Every original update emits \(O(1)\) weighted-record edits.
\end{lemma}
\begin{proof}
The first issue is the phase length. Its old and new token lists have
bounded size, but the old list also contains singleton insertions
from the preceding phase.
Let \(D^{\mathrm{sync}}_\ell\) be the duration of phase \(\ell\), with
\(D^{\mathrm{sync}}_0=0\). Its two frozen lists have at most
\(2\rho M_{\mathrm{term}}+D^{\mathrm{sync}}_{\ell-1}\) positions: the preceding phase
created at most one singleton per update. Charge a constant number of
steps for each subphase and phase boundary, so empty lists cannot cause
unbounded looping. Allowing for an incomplete quota at a phase's start,
\begin{equation}\label{outer:phase-duration}
 D^{\mathrm{sync}}_\ell
 \le\frac{2\rho M_{\mathrm{term}}+D^{\mathrm{sync}}_{\ell-1}+O(1)}{q_0}+O(1)
 \le\lambda_{\mathrm{o}}M_{\mathrm{term}}.
\end{equation}
The last inequality follows by induction from the choice of \(q_0\)
and the lower bound on \(M_{\mathrm{term}}\). A handler performs at most \(q_0\)
boundary or list steps, hence also at most constantly many snapshot
copies.

For coverage, consider a birth that is already present at phase
start. The old snapshot and its previous-phase singletons
partition the current births once. During activation, each such birth
keeps its old token and receives at most its one target token. During
retirement, every target token is already active, so removing old tokens
cannot uncover a surviving birth. New births belong only to current-phase
singletons. The fresh-birth convention preserves this argument under
deletion and reinsertion at the same location. At phase completion the
remaining family again partitions live births once, establishing the
next phase's induction base. Therefore multiplicity is always one or two,
and
\[
 \#\{\text{active tokens}\}
 \le2\rho M_{\mathrm{term}}+D^{\mathrm{sync}}_{\ell-1}+D^{\mathrm{sync}}_\ell=O(M_{\mathrm{term}}).
\]
Summing token multiplicities gives total mass at most \(2|P|\).

Finally, count record edits, not merely changes in representative
locations. One activation or retirement changes one weighted coordinate. An
insertion adds its singleton. A deletion belongs to at most two active
tokens. In one affected token, it either subtracts one from the existing
coordinate or removes the old token contribution and moves the remaining
mass to a new live representative. This changes at most two coordinates
per affected token. Any active singleton found by a handle is already
among these two tokens. Updates to inactive records produce no weighted
edits. The fixed synchronization quota therefore gives the asserted
constant bound.
\end{proof}

\begin{lemma}[Event consistency of Algorithm~\ref{alg:summary-update}]
\label{outer:event-consistency}
After every completed original update, the following hold on every
random execution.
\begin{enumerate}
\item The public native clusters partition \(P\); their frozen sizes
decrease by \(\theta\), and their sampling-origin descriptors have age
at most \(3\varepsilon_{\mathrm{o}}a_j\).
\item A retained transition snapshot keeps the geometric sizes recorded
at its freeze. For each of its positive-radius sampling-origin layers,
\begin{equation}\label{outer:event-retained-age}
 |U_j\triangle\operatorname{Lazy}(U_j)|\le\eta_{\mathrm{o}}a_j.
\end{equation}
Its native memberships can lose births but cannot gain them.
\item Each live birth belongs to one or two active tokens, every
nonempty token has a live representative, and \(Q\) is the aggregate
of their full masses. In particular,
\(\supp(Q)\subseteq\supp(P)\), \(\sum_qw_Q(q)\le2|P|\), and
\(|\supp(Q)|=O(M_{\mathrm{term}})\).
\item Outer work is \(\widetilde O(k)\), and the emitted ordered list
contains \(O(1)\) weighted-record edits.
\end{enumerate}
A public partition need only be complete after the structural step and
before a new snapshot freeze. Token coverage also holds between
individual synchronization steps.
\end{lemma}
\begin{proof}
We check the invariant at every publication boundary, since the
inner solver only sees the summary after all record edits of an
original update are applied. The first claim follows from
\Cref{outer:prefix-residual,outer:boundary-events}.
A snapshot is frozen only after the structural step. It copies the
current records and retains immutable ownership and generating sizes;
private restarts and later publications cannot change them. At freeze,
a sampling-origin layer has age at most
\(3\varepsilon_{\mathrm{o}}a_j\). A target is retained for its own
phase and, as the next old family, for at most one more phase. Since
\(a_j>M_{\mathrm{term}}\), \Cref{outer:transition} gives
\[
 |U_j\triangle\operatorname{Lazy}(U_j)|
 \le3\varepsilon_{\mathrm{o}}a_j+2\lambda_{\mathrm{o}}M_{\mathrm{term}}
 \le\eta_{\mathrm{o}}a_j.
\]
Identity-origin clusters stay singleton and need no retained age bound.

Latest-root membership commits and live-record deletion handlers give
current counts and representatives by \Cref{outer:records}. The token
claims follow from \Cref{outer:transition}. If multiple phases finish in
one handler, its earlier insertion remains in its entry-phase family;
new families are empty and their snapshots already include that birth.
The same one-or-two coverage induction therefore continues across every
boundary.

The structural quotas, record updates, and terminal refresh cost
\(\widetilde O(k)\) by \Cref{outer:scheduling,outer:records}.
There are only \(q_0=O(1)\) synchronization steps, so at most constantly
many \(O(M_{\mathrm{term}}\operatorname{polylog}N)\) snapshot copies. Publication,
restart, and inactive-record changes emit no edits to \(Q\); the active
edits are precisely those charged in \Cref{outer:transition}.
\end{proof}

\subsection{Transportation Without Nested Execution Sets}

The summary is useful only if moving a client to the representative
of each of its active clusters has small total cost. The difficulty
is that a retained snapshot may contain layers produced by different
worker generations. Their generating sets shrink geometrically in
size, but their current live sets need not be nested. We will use
this size decay to charge transportation without assuming nesting.

We first specify the common event used in the cost proof. Index every
sampling invocation started within the horizon, including preprocessing,
by \(g\in\mathcal J\), and write \(U_g\), \(P_g^{\mathrm{gen}}\),
and \(\nu_g\) for its frozen input, contemporaneous original instance,
and sampled radius. Fresh random tapes can be assigned at freeze time and
read as the corresponding budgeted operations execute. A superseded
invocation's tape still defines \(\nu_g\), even if the algorithm never
finishes computing it. Let
\[
 \mathcal E_{\mathrm{samp}}
 =\bigcap_{g\in\mathcal J}
   \{\nu_g\le2\mu_{\gamma_{\mathrm{o}}}(U_g;P_g^{\mathrm{gen}})\}.
\]
There are at most \(L_{\mathrm{out}}\) workers, each restarts at most
once per original update, and each generation starts at most
\(L_{\mathrm{out}}\) sampling layers. Including static initialization
and the initial private workers gives a deterministic bound
\(H_{\mathrm{samp}}=(T+2)L_{\mathrm{out}}^2\) on the number of invocations.
The sample for each invocation is chosen only after its input has
been frozen. The conditional bound in \Cref{outer:fixed-sampling},
followed by a union bound over these invocation slots, yields
\[
 \Pr[\mathcal E_{\mathrm{samp}}^{\,c}]
 \le H_{\mathrm{samp}}e^{-s_{\mathrm{samp}}}\le\delta.
\]
Each frozen input is fixed before its fresh samples are exposed. Later
updates may depend on those samples: the deterministic age bounds below
hold for every later update sequence. This proves the required adaptive
rather than merely oblivious probability guarantee.

We next compare a layer's radius, chosen when it was built, with a
covering radius for the current input. Its age bound says that only
a small fraction of its generating births could have changed.

\begin{lemma}[Age and quantiles of a retained layer]
\label{outer:age-quantile}
On \(\mathcal E_{\mathrm{samp}}\), a positive-radius sampling-origin
layer in either retained snapshot, with generating set \(U\),
\(a=|U|\), and current lazy set \(Y=\operatorname{Lazy}(U)\), satisfies
\begin{equation}\label{outer:retained-layer-bounds}
 |U\triangle Y|\le\eta_{\mathrm{o}}a,
 \qquad \nu\le4\mu_{\gamma_{\mathrm{o}}'}(Y;P),
 \qquad
 \gamma_{\mathrm{o}}'=
 \frac{\gamma_{\mathrm{o}}+\eta_{\mathrm{o}}}{1-\eta_{\mathrm{o}}}<1.
\end{equation}
\end{lemma}
\begin{proof}
The age estimate is \eqref{outer:event-retained-age}. Put
\(r=\mu_{\gamma_{\mathrm{o}}'}(Y;P)\) and choose a proper center set
witnessing this quantile. Even after losing the births not shared
with \(U\), its radius-\(r\) balls cover at least
\[
 \gamma_{\mathrm{o}}'|Y|-|Y\setminus U|
 \ge\bigl(\gamma_{\mathrm{o}}'(1-\eta_{\mathrm{o}})
                   -\eta_{\mathrm{o}}\bigr)a
 =\gamma_{\mathrm{o}}a
\]
old births in \(U\). A witnessing center that is absent from
\(\supp(P^{\mathrm{gen}})\) can be replaced by one old covered birth
in its ball. Every other old birth in that ball lies within \(2r\) of
the replacement. Centers covering no old birth can be discarded. Thus
at most \(k\) proper centers of \(P^{\mathrm{gen}}\) cover the required
old births within radius \(2r\), and
\(\mu_{\gamma_{\mathrm{o}}}(U;P^{\mathrm{gen}})\le2r\).
Sampling success then gives \(\nu\le4r\).
\end{proof}

It remains to sum these radius bounds over the layers. A direct
sum could count the same distant client many times, because the
retained lazy sets may overlap without being nested. The next
argument removes the overlap with sufficiently later layers and
then charges disjoint witnesses within each of a constant number
of residue classes.

\begin{lemma}[Geometric tail packing]\label{outer:tail-packing}
Let \(a_{i+1}\le\theta a_i\), and suppose \(Y_i\subseteq P\) with
\((1-\eta_{\mathrm{o}})a_i\le|Y_i|\le(1+\eta_{\mathrm{o}})a_i\).
Set \(\mu_i=\mu_{\gamma_{\mathrm{o}}'}(Y_i;P)\), where
\(\gamma_{\mathrm{o}}'<1\) is fixed. There is an absolute constant
\(K\) such that, for \(p\in\{1,2\}\) and every proper center set
\(C\) of size at most \(k\),
\begin{equation}\label{outer:tail-sum}
 \sum_i a_i\mu_i^p\le K\Phi_{P,p}(C).
\end{equation}
No nesting assumption on the sets \(Y_i\) is needed.
\end{lemma}
\begin{proof}
Let \(c=(1-\gamma_{\mathrm{o}}')(1-\eta_{\mathrm{o}})>0\).
For the fixed comparison set \(C\), at least \(ca_i\) births in
\(Y_i\) have distance at least \(\mu_i\) from \(C\). Indeed, a
larger fraction at distance strictly below \(\mu_i\) would contradict
the definition of the optimal covering quantile. Write this witness set
as \(F_i\).

The later layers of every \(r\)-th subsequence have total size much
smaller than \(a_i\) when \(r\) is a sufficiently large constant.
Choose a positive integer \(r\) such that
\[
 (1+\eta_{\mathrm{o}})\frac{\theta^r}{1-\theta^r}\le c/2,
 \qquad
 D_i=F_i\setminus\bigcup_{q\ge1}Y_{i+qr}.
\]
The removed tail has size at most \(ca_i/2\). Thus at least
half of the witnesses remain:
\(|D_i|\ge ca_i/2\). Within one residue class modulo \(r\), the
sets \(D_i\) are pairwise disjoint: an earlier set excludes the entire
\(Y_i\) containing every later witness. Consequently,
\[
 \sum_{i\equiv h\ ({\rm mod}\ r)}a_i\mu_i^p
 \le\frac2c\sum_{i\equiv h\ ({\rm mod}\ r)}
                 \sum_{x\in D_i}d(x,C)^p
 \le\frac2c\Phi_{P,p}(C).
\]
Sum over the \(r\) residue classes. The same witnesses work for both
powers, with \(K=2r/c\).
\end{proof}

We measure transportation before aggregating tokens at equal
representative locations. Let \(\mathcal G\) be the active token
family and \(q_G\) the live representative of token
\(G\). Its transportation cost is
\begin{equation}\label{outer:transport-definition}
 E_p=\sum_{G\in\mathcal G}\sum_{x\in G}d(x,q_G)^p,
 \qquad p\in\{1,2\},\qquad E=E_1.
\end{equation}

\begin{lemma}[Transportation bound]\label{outer:transport}
\label{overview:lemma-transport}
On \(\mathcal E_{\mathrm{samp}}\), at every completed original update,
\begin{equation}\label{overview:transport-bound}
 E_p\le a_p\OPT_{k,p}(P),
 \qquad a_p=2\cdot8^pK\quad(p\in\{1,2\}),
\end{equation}
where \(K\) is from \Cref{outer:tail-packing}. In particular,
\(a_*=a_1=16K\) is an absolute constant for \(k\)-median.
\end{lemma}
\begin{proof}
It suffices to analyze one snapshot, since at most two snapshots
contribute active tokens. Fix one retained snapshot. Identity-origin clusters and insertion
singletons contribute zero. Zero-radius sampling layers also contribute
zero. The remaining layers form a subsequence of the snapshot's frozen
level order, so their generating sizes still decrease by a factor at most
\(\theta\).

A native cluster at layer \(j\) has diameter at most \(2\nu_j\): every
member was assigned within \(\nu_j\) of the same ghost sample. Any live
representative therefore stays within \(2\nu_j\) of every remaining
member. There are at most \(a_j\) members in that layer.
\Cref{outer:age-quantile,outer:tail-packing} imply that the complete
snapshot, and hence its active subfamily, has transportation cost at most
\[
 2^p\sum_j a_j\nu_j^p
 \le8^p\sum_j a_j
       \mu_{\gamma_{\mathrm{o}}'}(\operatorname{Lazy}(U_j);P)^p
 \le8^pK\OPT_{k,p}(P).
\]
At most two snapshots contribute. Summing their bounds proves the claim;
insertion singletons add nothing. In particular, no age bound on retained
identity-origin layers is needed.
\end{proof}

\subsection{The Summary Interface}

We can now state the summary guarantee. The deterministic part
concerns support, weights, and the number of edits. The
approximation event is needed only to compare transportation
with the optimum. The same argument works for distances and
squared distances.

\begin{theorem}[Dynamic weighted summary]\label{outer:summary}
\label{overview:lemma-summary}
For unit-weight point insertions and deletions, there is a weighted
summary \(Q\) satisfying, at every completed original update,
\begin{equation}\label{overview:outer-interface}
 \begin{gathered}
 \supp(Q)\subseteq\supp(P),\qquad
 |\supp(Q)|=\widetilde O(k),\qquad
 \sum_qw_Q(q)\le2|P|,\\
 \text{maintenance work per update}=\widetilde O(k),\qquad
 \text{weighted-record edits per update}=O(1).
 \end{gathered}
\end{equation}
Its nonzero weights are positive integers. These guarantees hold on
every realization of the random choices, and the update bounds are
worst-case. With probability at least \(1-\delta\)
over the prescribed adaptive horizon, simultaneously at every completed
update, for \(p\in\{1,2\}\) and every ambient center set \(C\),
\begin{subequations}\label{outer:summary-inequalities}
\begin{equation}\label{overview:transport-two-way}
 \begin{aligned}
 \Phi_{P,p}(C)&\le2^{p-1}\bigl(E_p+\Phi_{Q,p}(C)\bigr),\\
 \Phi_{Q,p}(C)&\le2^{p-1}E_p+2^p\Phi_{P,p}(C).
 \end{aligned}
\end{equation}
\begin{equation}
 E_p\le a_p\OPT_{k,p}(P).
\end{equation}
\end{subequations}
The first two inequalities hold on every execution; only the bound on
\(E_p\) uses sampling success. Initialization takes
\(\widetilde O(k|P_0|+k)\) time.
\end{theorem}
\begin{proof}
The resource bounds follow from the controller and token invariants.
\Cref{outer:event-consistency} gives the deterministic resource, weight,
and support guarantees. Static layer construction contracts execution
sizes geometrically, so its total sampling and record work is
\(\widetilde O(k|P_0|+k)\), including the initial snapshots and worker
roots.

For the cost comparison, each live birth is represented once or
twice, always with its full mass. Let \(m_x\in\{1,2\}\) be the
number of active tokens containing a live birth \(x\). The triangle inequality and
\((a+b)^p\le2^{p-1}(a^p+b^p)\) give
\[
 \begin{aligned}
 \Phi_{P,p}(C)
 &\le\sum_xm_xd(x,C)^p\\
 &\le2^{p-1}\sum_{G\in\mathcal G}\sum_{x\in G}
           \bigl(d(x,q_G)^p+d(q_G,C)^p\bigr)\\
 &=2^{p-1}\bigl(E_p+\Phi_{Q,p}(C)\bigr).
 \end{aligned}
\]
To bound the weighted cost from the original cost, apply the same
inequality in the other direction:
\[
 \begin{aligned}
 \Phi_{Q,p}(C)
 &\le2^{p-1}\sum_{G\in\mathcal G}\sum_{x\in G}
            \bigl(d(x,q_G)^p+d(x,C)^p\bigr)\\
 &\le2^{p-1}E_p+2^p\Phi_{P,p}(C).
 \end{aligned}
\]
Aggregating tokens at the same location preserves these identities.
On \(\mathcal E_{\mathrm{samp}}\), \Cref{outer:transport} gives the
remaining inequality. The conditional union bound above proves its
simultaneous probability guarantee against adaptive updates.
\end{proof}

\subsection{Composition and Worst-Case Implementation}
\label{sec:operation-audit}\label{sec:composition}\label{sec:end-to-end}

An original client update may cause several changes to the weighted
summary. The inner solver must process all of them before a new
center set is published, even if a background computation fails in
the meantime. We account for this work at the level of one
original update. The elementary bounds from
\Cref{outer:records,outer:transition} and \Cref{subsec:trees}
are collected below. Here
\(n\) is the current inner support size and
\(L_\Delta=1+\lceil\log_2(2\Delta)\rceil\); logarithmic factors
include dictionary and confidence costs.

\begin{center}
\normalsize
\begin{tabularx}{\linewidth}{@{}Xl@{}}
\toprule
Operation & Worst-case allowance \\
\midrule
Advance structural workers and update live ownership records
  & \(\widetilde O(k)\) per original update \\
Refresh the terminal identity state
  & \(\widetilde O(k)\) per original update \\
Copy a transition snapshot with \(O(M_{\mathrm{term}})\) cluster records
  & \(\widetilde O(k)\) per freeze \\
Execute the fixed synchronization quota
  & \(O(1)\) weighted-record edits \\
Commit one inner record and advance its optional work
  & \(\widetilde O(n+k)\) per record \\
Change inner virtual facilities and their proper projection
  & \(O(L_\Delta^3)\) recourse per record \\
Retain or retire a persistent root
  & \(O(1)\) per root \\
\bottomrule
\end{tabularx}
\end{center}

Two kinds of boundary work need particular care. First, the
controller selects at most one structural suffix publication per
original update. Membership transactions reserve their dictionary
operations, read current liveness and the latest roots, and reject
superseded generations. Second, the synchronization quota allows
only constantly many target freezes, including the steps that
advance past empty lists. The first phase is initialized before
updates, and the \hyperref[outer:update-order]{original-update order}
places a new birth in the entry phase's singleton family before
any later target is frozen. Consequently, neither boundary can
trigger a scan over all clients or an unbounded batch of record edits.

The inner solver's distance-row catch-up is charged to the
four-stage schedule. A phase boundary shares or adopts roots that
have already been prepared; it does not construct a new \(n\times k\)
distance table. A phase contains \(O(B)\le O(k)\) record edits,
so its frozen and current padded sizes \(n+2k\) differ by at most a
constant factor. The per-record allowance can therefore be expressed
using the current support size. Append-only persistent allocation
charges each new node to its creating operation; retiring a root does
not recursively reclaim its descendants.

\begin{proposition}[Worst-case implementation]
\label{prop:operation-ledger}
The combined algorithm uses \(\widetilde O(k)\) RAM operations and
\(O(L_\Delta^3)\) published-center recourse per original update, for
every realization of the random choices.
\end{proposition}
\begin{proof}
The summary has \(\widetilde O(k)\) support, performs
\(\widetilde O(k)\) work, and emits \(O(1)\) record edits on every
execution. A partial batch differs from the starting weighted input
at only constantly many locations, so every prefix also has
\(\widetilde O(k)\) support. Apply the inner resource bounds to each
record and sum over the constant-size list.

For unconditional guarantees, we must also account for failed
background work. The driver drains the list even if an inner guard fails
(\Cref{lem:batch-drain}). In halted mode, subsequent records maintain
only the mandatory input and support projection; no lazy buffer grows.
Before failure, all optional work and output changes obey deterministic
caps. Thus failure cannot add uncharged computation or recourse.
If \(\Pi_0,\ldots,\Pi_r\), with \(r=O(1)\), are the projected
answers before and after the successive records, then
\[
 |\Pi_0\sym\Pi_r|\le\sum_{j=1}^{r}|\Pi_{j-1}\sym\Pi_j|
 =O(L_\Delta^3).
\]
Only \(\Pi_r\) is published. The persistent-allocation convention
precludes an additional reclamation cost at publication or retirement.
\end{proof}

\paragraph{Three levels of execution.}
There are three clocks in the implementation: original client updates,
weighted-record edits, and background work slices.
\Cref{alg:summary-update} handles the first;
\Cref{alg:process-record,alg:advance-phase} in Step~5 handle the other
two. The driver below connects these interfaces. After \(u\)
weighted-record edits, the inner input is
\(Q_u\), denoted by \(P_u\) in the inner-only analysis of
\Cref{sec:inner}. It equals the aggregate of the outer state
\(\mathcal Q\) at the end of each original update, but not necessarily
halfway through its emitted list
\(\mathcal E\). We therefore process the whole list before publishing
the next answer. Historical locations may appear in intermediate
weighted inputs without appearing in the final output.

The driver needs only the inner solver's virtual serving set \(Z\),
its maintained proper projection \(\Pi\) onto the current inner
support, and its permanent failure flag \(\mathsf{halted}\).
Checkpoint, phase, and distance-row state are maintained by the
record handler. We describe the driver for \(k\ge2\); the separate
\(k=1\) routine is specified in \Cref{sec:weighted-inner}.

\paragraph{Initialization.}
We initialize both components before the first original update, so
the batch driver never encounters an unfinished initial state.
Here \(n_0=|P_0|\) is the number of original clients at initialization.
Allocate failure budget \(\delta/2\) to each component. Construct the
initial outer summary and its first synchronization phase before any
original update; set \(Q_0=\mathcal Q_0\). Start with \(k\) dummy
virtual facilities and their proper projection, so the failure rule is
available before randomized inner preprocessing. Compute a static
approximate set, pad it to a distinct \(k\)-set \(W\), and apply
\Cref{lem:annotation-initialization}. On success,
\(\Phi_{Q_0}(U_0)\le(3/2)\beta F_k(Q_0)\le\alpha F_k(Q_0)\).
Construct its window and persistent roots. After the cardinality and
work checks, install \(K_0\), set \(p=u=0\), and enter the idle state.
Initialization uses \(\wtO(k(n_0+k))\) work. A failed check retains
the dummy fallback and sets \(\mathsf{halted}\).

\paragraph{The record-handler contract.}
An emitted input edit cannot be rolled back because a randomized
background job fails. In \Cref{alg:process-record}, pre-edit phase
setup uses only bounded root retention and scalar assignments.
The handler then applies its record exactly once, increments \(u\),
appends its edit-log entry, and updates live support and \(\Pi\) as
one mandatory commit. This commit has its own
\(\wtO(|\supp(Q_u)|+k)\) allowance and precedes all optional work.
If the solver has not halted, the foreground row maintenance of
Step~5 completes before a background slice can run; frozen row roots
are never changed by this maintenance.

A failed deterministic guard, exhausted total job allowance, or
missed deadline fixes \(Z\), sets \(\mathsf{halted}\), and cancels
only optional reconstruction. The current handler returns after its
mandatory commit, and later handlers still commit their records and
maintain the proper projection. They add no buffer location, maintain
no shadow rows, and install no target. An ordinary slice boundary
merely pauses background work. Prospective capacity checks and
indivisible, fully reserved output commits keep \(|Z|\le k\) before
failure; freezing keeps this bound afterward. These are the inner
guarantees used below, rather than an assumption that every random
execution gives a good approximation.

Thus the following driver takes one original insertion or deletion,
obtains its ordered record list from the outer state, and publishes
the final proper projection only after the complete list has been
processed. An empty list leaves the inner input unchanged.

\begin{algorithm}[H]
\caption{Original-update driver, after initialization (\(k\ge2\))}
\label{alg:dynamic-kmedian}
\normalsize\raggedright
\textbf{For each unit-weight original insertion or deletion \(\omega\):}
\begin{enumerate}[label=\arabic*.,leftmargin=1.9em,labelsep=0.5em,
                  itemsep=0.5em,parsep=0pt,topsep=0.4em]
\item Execute \(\mathcal E\gets\textsc{SummaryUpdate}(\mathcal Q,\omega)\)
      in the \hyperref[outer:update-order]{original-update order}.
      Record every emitted coordinate edit in order.
\item For every \(e\in\mathcal E\), call \(\textsc{ProcessRecord}(e)\).
      Complete this loop even if the inner solver halts.
\item Publish the maintained projection \(\Pi\) after the entire list
      has been processed. No intermediate inner answer is published.
\end{enumerate}
\end{algorithm}

\begin{lemma}[Batch completion and properness on every execution]
\label{lem:batch-drain}
For every realization of the random choices, the driver applies every
record of an emitted batch exactly once and publishes only after the
full batch. Its final inner input equals \(\mathcal Q\), and its
published output is a subset of the current original support of size
at most \(k\). Inner failure does not increase the worst-case work or
recourse order of the reduction.
\end{lemma}
\begin{proof}
The key point is that every emitted record is handled even if
an earlier record caused the inner solver to freeze.
Before committing its input, a handler performs only bounded deterministic
phase setup. Every optional call occurs after that commit. Thus a failure
leaves the current edit applied, fixes \(Z\), and returns to the driver;
all remaining records are applied by the frozen branch. Induction on the
position in \(\mathcal E\) shows that the inner instance equals the
result of exactly its processed prefix. At the end it equals the outer
summary, including when the list is empty.

We next check feasibility at publication. Capacity guards and
uninterrupted output commits keep \(|Z|\le k\)
before failure, and freezing preserves that bound afterward. The mandatory
support commit maintains \(\Pi\subseteq\supp(Q_u)\) with
\(|\Pi|\le k\) on every execution. At batch completion,
\Cref{outer:event-consistency} gives
\(\supp(\mathcal Q)\subseteq\supp(P)\). The published answer is
therefore proper for the current original input. No claim of original
properness is needed for unpublished prefixes.

Finally, each batch has \(O(1)\) edits. The support of a prefix differs from
its starting support at only \(O(1)\) coordinates, so it remains
\(\widetilde O(k)\). Each mandatory commit costs
\(\wtO(|\supp(Q_u)|+k)\) work and constant projection recourse.
Optional work and virtual changes have their fixed caps, including the
separately reserved work of an output commit. Summing over the batch
gives \(\widetilde O(k)\) work and \(O(L_\Delta^3)\) recourse.
\end{proof}

\paragraph{The reduction.}
We now connect the summary to the weighted inner solver. The
summary supplies a small weighted input and a constant number of
edits per original update. Its transportation bound lets us compare
the inner solution with the optimum on the original clients.
The reduction applies to any proper weighted solver whose
comparison benchmark is the ambient optimum.

\begin{theorem}[Composition with a weighted solver]\label{outer:reduction}
Suppose a dynamic algorithm on \(m\) weighted locations supports
arbitrary positive integer record weights and adaptive record edits.
On its success event, its proper output costs at most
\(\alpha_{\rm in}F_k(Q)\), where \(\alpha_{\rm in}\) is its
approximation factor.
Let \(T_{\rm in}(m)\) and \(R_{\rm in}(m)\) be its worst-case time
and recourse per record edit, and assume its mandatory input and
proper-projection updates remain valid after any internal failure.
Combined with \Cref{outer:summary}, it gives an algorithm for original
unit-weight updates with bounds
\[
 \begin{aligned}
 \text{update time}&=\widetilde O(k)
                  +O(1)T_{\rm in}(\widetilde O(k)),\\
 \text{recourse}&=O(1)R_{\rm in}(\widetilde O(k)).
 \end{aligned}
\]
On the common success event, the proper output costs at most
\([a_*+\alpha_{\rm in}(a_*+2)]\OPT_k(P)\).
\end{theorem}
\begin{proof}
First consider the order and number of updates. Use the
\hyperref[outer:update-order]{original-update order}. Let
\(c_{\rm sync}\) bound the number of emitted records per original
update, and run the inner solver for a record horizon of at least
\(c_{\rm sync}T\). Process the emitted list in order and publish
only the answer after its last record.

For approximation, take a proper optimum \(C^*\) for \(P\).
It is a legal ambient comparison for \(Q\), even if its centers
do not belong to \(\supp(Q)\).
For the final inner answer \(C\), the transportation inequalities give
\[
 \Phi_P(C)\le E+\alpha_{\rm in}F_k(Q)
 \le E+\alpha_{\rm in}\Phi_Q(C^*)
 \le[a_*+\alpha_{\rm in}(a_*+2)]\OPT_k(P).
\]
The constant batch size gives the time bound. For recourse, compare
the answer before the first record with the answer after the last.
If the inner answers through a batch are \(Z_0,\ldots,Z_r\), then
\[
 |Z_0\triangle Z_r|
 \le\sum_{i=1}^r|Z_{i-1}\triangle Z_i|
 \le O(1)R_{\rm in}(\widetilde O(k)),
\]
which proves the published recourse bound.

Properness needs to be checked only after the whole batch. A
partial batch may still contain a deleted original location while its
remaining tokens are moved. Such a record is valid in the fixed metric.
The driver must drain every record even if optional inner work fails.
At batch completion, \(\supp(Q)\subseteq\supp(P)\), and the inner
proper output is therefore proper for \(P\) as well. Empty intermediate
supports are allowed. For our weighted solver this requirement is
\Cref{lem:batch-drain}. Separate random streams and a conditional union
bound combine the two approximation guarantees against adaptive updates.
\end{proof}

\begin{proof}[Proof of \Cref{thm:main}]
Apply \Cref{outer:reduction} with the weighted algorithm of
\Cref{thm:inner}.
The inner factor is \(\alpha_{\rm in}=6C_0\), its time is
\(\widetilde O(m+k)\), and its recourse is \(O(L_\Delta^3)\).
Since the summary has \(m=\widetilde O(k)\) support, the final
factor is \(a_*+6C_0(a_*+2)=O(1)\), and both worst-case bounds
follow. \Cref{prop:operation-ledger} accounts for the implementation
on every execution. Summary weights are positive integers at most
\(2N\), and padding changes the aspect ratio by only a factor two.

For initialization, the outer layers cost
\(\widetilde O(kn_0+k)\) work.
Their representatives lie in the initial client support. Initializing
the weighted solver, its distance rows, and its proper projection
therefore fits \(\widetilde O(k(n_0+k))\), including empty input.
Allocate total failure probability \(\delta/2\) to each component,
including preprocessing. Fresh randomness on frozen inputs and the
deterministic record-horizon bound give simultaneous success probability
at least \(1-\delta\).
\end{proof}

\paragraph{Exact cardinality when requested.}
The algorithm normally maintains at most \(k\) proper centers.
If exactly \(\min\{k,|\supp(P)|\}\) are desired, we can fill
unused slots without increasing the objective. To do so,
augment the proper answer with the first available live locations
outside it in a fixed identifier order. If an original update changes the
unaugmented answer by \(r\) center insertions or deletions, an
order-statistic dictionary maintains the fillers with \(O(r+1)\)
changes and \(O((r+1)\operatorname{polylog}N)\) work, without distance
queries. Indeed, each elementary center change affects the required
prefix by only \(O(1)\) locations, and an original update changes the
support by at most one location. Added centers cannot increase cost.
Thus the time and recourse bounds are unchanged, including through
empty support.

\begin{remark}\label{outer:weighted-domain}
The outer construction counts unit client births in its sampling,
coverage, and age budgets. It does not support arbitrary changes of
original client mass. The stronger weighted-update model is used by
the inner solver for the summary records it receives.
\end{remark}

%% file: sections/kmeans_and_conclusion.tex
\section{Extension to Metric \texorpdfstring{\(k\)}{k}-Means}\label{sec:kmeans}

We now consider metric \(k\)-means, where the objective sums squared
distances to the nearest center. The underlying distance \(d\) is
still a metric, but \(d^2\) need not satisfy the triangle inequality.
We therefore continue to define balls, nearest neighbors, and
separation using \(d\), and adjust the cost inequalities where
squaring is involved. For a weighted instance \(P\), write
\[
 \Psi_P(S)=\sum_{x\in\supp(P)}w_P(x)d(x,S)^2,\qquad
 F_j^{(2)}(P)=\min_{\substack{S\subseteq X\\|S|\le j}}\Psi_P(S),
\]
and restrict facilities to \(\supp(P)\) to define
\(\OPT_{j,2}(P)\).
For a weighted client subset \(B\), put
\(\cost_2(z,B)=\sum_{x\in B}w_P(x)d(x,z)^2\) and
\(F_1^{(2)}(B)=\min_{z\in X}\cost_2(z,B)\).
The empty-input and zero-budget conventions are unchanged. We assume
attainment of the squared optima, as in \Cref{sec:prelim}.

The inner algorithm compares its proper output with an ambient
optimum, which may place facilities outside the client support.
These benchmarks differ by only a constant factor: projecting
each ambient facility to a nearest positive-support location
at most doubles each assigned distance. Consequently,
\begin{equation}\label{km:proper-comparison}
 F_j^{(2)}(P)\le\OPT_{j,2}(P)\le4F_j^{(2)}(P)
 \qquad(j\ge1).
\end{equation}
The extension reuses the six-step construction, but its cost
inequalities must be checked with care: squared distances do not
form a metric. We first adapt the robust certificates and batch
repair from Step~3. We then verify the static, window, and refresh
interfaces used in Steps~1--3, and transfer the migration and
scheduling arguments of Steps~4--5. Finally, we apply the summary
of Step~6 with squared transportation. The update rules and the
deterministic operation counts are unchanged; the approximation
constants are replaced by the ones proved below.

\subsection{Squared Robust Certificates}\label{km:robustification}

The robust-center construction relies on the geometry of the metric,
so its anchors, witness scales, and separation tests still use \(d\).
What changes is the local cost test. We choose its squared-cost
threshold so that proxy displacement remains small enough for the
separation and overlap arguments from the median case.

\begin{definition}[Squared robust sequence]\label{km:robust-sequence}
A sequence \((p_0,\ldots,p_\ell)\) is squared robust on \(P\) if the
following rule holds for every \(i=\ell,\ldots,1\). Put
\(R_i=10^i\) and \(B_i=\Ball_{R_i}^P(p_i)\).
If \(w(B_i)=0\), or if its average squared cost at \(p_i\) is at
least \(R_i^2/25\), set \(p_{i-1}=p_i\). Otherwise choose
\(p_{i-1}\in B_i\cup\{p_i\}\) with
\begin{equation}\label{km:local-choice}
 \cost_2(p_{i-1},B_i)
 \le\min\{\cost_2(p_i,B_i),8F_1^{(2)}(B_i)\}.
\end{equation}
Height zero imposes no input-dependent condition. Retain the entire
sequence and its generation data.
\end{definition}

The procedure \(\MakeRobustProc_2\) can implement this choice by
sampling candidates according to weight and evaluating their squared
costs. It also keeps the previous point as a candidate, so the
cost cannot increase even if sampling fails. For a positive-mass
set \(B\), a weighted sample
\(Y\) satisfies
\[
 \mathbb E[\cost_2(Y,B)]\le4F_1^{(2)}(B).
\]
If the optimum is zero, every sample is optimal. Otherwise Markov's
inequality makes a sample an \(8\)-approximation with probability at
least \(1/2\). Use \(O(\log((\ell+2)/\varepsilon))\) independent
samples at each level. Conditional on the higher-level choices, the
current ball is fixed; a union bound gives failure probability at most
\(\varepsilon\) over the sequence. Work is
\(O((n+1)(\ell+1)\log((\ell+2)/\varepsilon))\).
Comparison with \(p_i\) guarantees local cost nonincrease on every
outcome. The deterministic certificate rules omit only the sampled
\(8F_1^{(2)}\) comparison; full squared robust validity includes it.

\begin{lemma}[Squared sequence bounds]\label{km:sequence}
Every generated sequence, on every sampling outcome, satisfies
\[
 d(p_{i-1},p_i)\le\tfrac25\,10^i,\qquad
 d(p_0,p_i)\le\tfrac12\,10^i\quad(1\le i\le \ell).
\]
Its ambient defining balls are nested. Set
\(B_0=\Ball_1^P(p_0)\). For \(B_i\subseteq C\subseteq P\),
including \(i=0\),
\begin{equation}\label{km:sequence-cost}
 \cost_2(p_0,C)\le\tfrac94\cost_2(p_i,C).
\end{equation}
\end{lemma}
\begin{proof}
We first bound how far one step can move. Only a low-average step
of positive mass can change the point. Applying the triangle
inequality to the ball, then Cauchy--Schwarz and local cost
nonincrease, gives
\[
 d(p_{i-1},p_i)
 \le\sqrt{\cost_2(p_i,B_i)/w(B_i)}
    +\sqrt{\cost_2(p_{i-1},B_i)/w(B_i)}
 <\tfrac25\,10^i.
\]
The geometric sum gives the displacement from any level to the
final proxy. It also yields
\(10^{i-1}+d(p_{i-1},p_i)<10^i\), so the defining balls
are nested. For \(x\notin B_j\), endpoint displacement implies
\(d(x,p_0)\le\tfrac32d(x,p_j)\).
Apply successive local cost nonincrease on the nested balls:
\[
 \begin{aligned}
 \cost_2(p_i,C)
 &\ge\cost_2(p_0,B_1)
    +\sum_{j=1}^{i-1}\cost_2(p_j,B_{j+1}\setminus B_j)
    +\cost_2(p_i,C\setminus B_i)\\
 &\ge\cost_2(p_0,B_1)+\tfrac49\cost_2(p_0,C\setminus B_1)
 \ge\tfrac49\cost_2(p_0,C).
 \end{aligned}
\]
Each annulus is compared directly with the final proxy \(p_0\).
The factor therefore does not multiply with the number of levels.
For \(i=0\), the claim is immediate.
\end{proof}

As in the median construction, the geometric conditions must hold
even on a failed sampling outcome. A deterministic squared state
is Invariant~\ref{block:state} with
these cost-nonincreasing certificates. A full squared-robust state
additionally requires validity of every stored sampled comparison on
the current input. The displacement lemma preserves proxy separation,
strong-reset history, and historical top-ball overlap on every outcome.
Below, a squared-robust state means the full state.

\begin{lemma}[Squared replacement and restart]
\label{km:replacement}\label{km:restart}
Use the same separation constant \(\Gamma\ge400000\).
\begin{enumerate}[label=(\roman*)]
\item If \(v\) belongs to a squared-robust state \(V\), and
  \((v,z)\in V\times Z\) is well separated, then the comparison
  Voronoi cell \(C_z\) satisfies
  \(\cost_2(v,C_z)\le18\cost_2(z,C_z)\).
\item If \(R=10^\ell\), \(\ell\ge1\), and \(d(a,w)\le R/20\), a
  successful call \(v'=\MakeRobustProc_2(Q,a,\ell)\) satisfies
  \(\cost_2(v',C)\le18\cost_2(w,C)\) for every
  \(\Ball_R^Q(a)\subseteq C\subseteq Q\).
\end{enumerate}
\end{lemma}
\begin{proof}
For (i), we compare the robust proxy with a nearby comparison
center on the latter's Voronoi cell. The case \(v=z\) is immediate.
Otherwise put
\(d_0=d(v,z)>0\) and choose \(20d_0\le R=10^i<200d_0\).
The proxy-radius bound \eqref{block:proxy-radius} places this level
in the stored certificate \((p_0,\ldots,p_\ell)\), with \(v=p_0\).
The geometry in \Cref{block:replacement} places all defining balls
through this level inside \(C_z\).

In the positive-mass high-average case, \(p_{i-1}=p_i\) and
\(d(z,p_i)\le R/10\). The reverse triangle inequality for the
weighted Euclidean norm of the distance vector gives
\[
 \sqrt{\cost_2(z,B_i)}
 \ge\sqrt{\cost_2(p_i,B_i)}-\sqrt{w(B_i)}R/10
 \ge\tfrac12\sqrt{\cost_2(p_i,B_i)}.
\]
Outside \(B_i\), \(d(x,p_i)\le2d(x,z)\).
Equation~\eqref{km:sequence-cost} therefore gives factor \(9\).
In the other case, let \(b=p_{i-1}\). Robustness, or zero mass,
gives inside cost at most \(8\cost_2(z,B_i)\).
Also \(d(z,b)\le R/10\). For \(x\notin B_i\), the step bound gives
\(d(x,b)\ge R/2\). The reverse triangle inequality then implies
\(d(x,z)\ge d(x,b)-R/10\ge(4/5)d(x,b)\), so
\(d(x,b)\le(5/4)d(x,z)\) there. Apply the lower
sequence cost bound to obtain \((9/4)\cdot8=18\).

For (ii), the old comparison center is close to the new anchor.
This makes the top-ball comparison possible even though the
certificate is built on the updated input. Let
\(B=\Ball_R^Q(a)\). Outside \(B\),
\(d(x,w)\ge(19/20)d(x,a)\). If the ball has zero mass, all lower
balls do as well, so the output is \(a\). In the high-average case,
the same norm inequality gives
\(\sqrt{\cost_2(w,B)}\ge(3/4)\sqrt{\cost_2(a,B)}\).
Combining inside and outside comparisons with
\eqref{km:sequence-cost} gives factor \(4\).
For a low-average top step, let \(b=p_{\ell-1}\). Its inside cost is
at most \(8\cost_2(w,B)\). Outside the ball,
\(d(x,w)\ge19R/20\) and \(d(w,b)\le11R/20\), so
\(d(x,b)\le(30/19)d(x,w)\). Since \((30/19)^2<8\), the lower
sequence again gives factor \(18\).
\end{proof}

We can now run the same four-phase batch repair. The combinatorial
arguments controlling anchor relocations and overlapping historical
certificates depend only on the displacement bounds, while the
cost comparison uses the squared replacement lemma.

\begin{theorem}[Squared batch robustification]\label{km:batch}
Assume a deterministic squared state for \((V,P)\), with inputs
\(Q,W,s,h\) as in \Cref{block:theorem}. The four-phase procedure
of \Cref{block:algorithm-subsection}, using \(\MakeRobustProc_2\),
preserves that deterministic state on every outcome. With
\(L=1+\lceil\log_{10}\Delta_d\rceil\), its bounds remain
\[
 \text{work}=\widetilde O((n+k)(h+s+1)),\qquad
 \text{recourse}=O((h+s)L^3).
\]
If inherited retained certificates are squared robust-valid and all
new sampled choices satisfy \eqref{km:local-choice}, the output is
fully squared robust and
\begin{equation}\label{km:batch-cost}
 \Psi_Q(V')\le18\Psi_Q(W).
\end{equation}
\end{theorem}
\begin{proof}
The unconditional displacement bounds preserve distinct proxies,
strong resets, historical overlap, and the no-feedback property of the
four phases. The same defining balls detect input contamination.
Consequently the deterministic state and all operation counts are
unchanged. Full robust validity additionally uses the inherited and
new sampling assumptions in the statement.

For the cost bound, keep the fixed Voronoi cells of the proposed centers
\(w_i\in W\), which are identical for \(d\) and \(d^2\).
A new or marked label starts at \(w_i\); localization and
\eqref{km:sequence-cost} give factor \(9/4\), or factor one at
height zero. An unmarked label rebuilt at a larger level has
\(R_i'\ge10R_i\) and
\(d(a_i,w_i)\le R_i/2\le R_i'/20\), so restart gives factor
\(18\). Unchanged proxies retain their cell costs. Summing over the
fixed disjoint cells applies this factor only once, regardless of
the number of repair steps.
\end{proof}

\subsection{Static Interfaces and the Weighted Solver}
\label{km:static}\label{km:stability}\label{km:refresh-section}\label{km:scheduling}

The worst-case schedule uses only a few properties of its static
routines: a removal bound, an augmentation bound, and a convex
benchmark for stability windows. We establish their squared-cost
versions below. Lazy protection, padding, and the ordering of
distances are unchanged. LP arguments use the finite facility-domain
convention of \Cref{sec:prelim}, and \(F_j^{(2)}\) denotes the
ambient optimum throughout. Static work bounds assume the stated
cached distances.

\begin{lemma}[Squared static primitives]
\label{km:lp}\label{km:removal}\label{km:augmentation}
The weighted squared objective has these interfaces.
\begin{enumerate}[label=(\roman*)]
\item On a nonempty finite facility domain, its assignment LP has
  integrality gap at most \(9\). Its value is convex in the real
  opening budget.
\item For \(q\ge1\), \(r\ge0\), and \(|X_0|=q+r\),
  \(\Remove_2(P,X_0,r)\) returns a \(q\)-subset \(Y\) of \(X_0\)
  satisfying
  \begin{equation}\label{km:removal-bound}
   \Psi_P(Y)\le a_R\Psi_P(X_0)+b_RF_q^{(2)}(P),
   \qquad a_R=8,\quad b_R=4608,
  \end{equation}
  in \(\widetilde O((n+|X_0|)(r+1))\) work, given ordered
  client-to-candidate distance trees.
\item Given distances to \(U\) and an integer \(a\ge0\),
  \(\Develop_2(P,U,a)\) adds at most \(a\) support facilities and
  gives an absolute \(\beta_2\)-approximation to the best such
  augmentation in \(\widetilde O((n+|U|)(a+1))\) work.
  A static solution with at most \(q\ge1\) facilities and cost at
  most \(\beta_2F_q^{(2)}(P)\) is also available in
  \(\widetilde O((n+1)(q+1))\) work.
\end{enumerate}
Randomized approximation guarantees use assigned failure allowances
and fixed work caps. Cardinality holds on every outcome; weight-range
factors exclude zero-demand candidates.
\end{lemma}
\begin{proof}
\emph{LP gap.} We use a constant integrality gap to transfer
the convex LP comparison to integral center budgets. The gap in (i) is
\cite[Theorem~1.3]{ANSW20} for squared distances in general metrics
with a finite candidate domain. Rational weights are handled by demand
replication while keeping that domain fixed; continuity of the finite
integral and LP optimum values extends the value bound to real weights.
This argument does not expand the algorithmic input. Convexity follows
by combining feasible solutions, independently of the cost coefficients.

\emph{Removal.} Apply \Cref{lem:restricted-local-search} with
\(p=2\), candidate set \(X_0\), and budget \(q\). Its comparison
with the ambient optimum gives
\[
 \sqrt{\Psi_P(Y)}\le
 2\left(\sqrt{\Psi_P(X_0)}+12\sqrt{F_q^{(2)}(P)}\right).
\]
Using \((a+b)^2\le2a^2+2b^2\), we obtain
\[
 \Psi_P(Y)\le8\Psi_P(X_0)+1152F_q^{(2)}(P),
\]
which implies the stated bound with \(a_R=8\) and \(b_R=4608\).
The same lemma supplies the fixed work cap and permits candidates
outside the positive-demand support.

\emph{Augmentation.} We must add only a prescribed number of
facilities while retaining the entire starting set \(U\).
Handle \(a=0\), zero cost, and
\(|\supp(P)\setminus U|\le a\) as in the median procedure.
For empty \(U\), use the static algorithm. Otherwise contract \(U\)
to \(u_*\) in the underlying metric:
\[
 d'(x,u_*)=d(x,U),\qquad
 d'(x,y)=\min\{d(x,y),d(x,U)+d(y,U)\}.
\]
With \(u_*\) open, distances to \(\{u_*\}\cup H\) equal the
original distances to \(U\cup H\), so squared costs agree as well.
Apply the weighted static algorithm of~\cite[Section~1.1]{MP04},
which extends to squared metric costs. Let \(V=\Psi_P(U)>0\) and
let \(\delta_*\) be the smallest positive distance to \(U\) among
positive clients. In addition to transferred demand, give \(u_*\)
weight \(W_*=2\beta_2V/\delta_*^2\), choosing \(\beta_2\) to
cover the static factor. Omitting \(u_*\) costs at least
\(2\beta_2V\), whereas an optimum containing it costs at most
\(V\). A successful \((a+1)\)-facility output must include it,
leaving at most \(a\) new support facilities after lifting.
The added weight lies between \(2\beta_2w_{\min}\) and
\(2\beta_2nw_{\max}\Delta_d^2\), so it changes only logarithmic
range factors. Cached distances make each contracted query constant-time.

The boundary cases preserve cardinality on every execution.
Return \(X_0\) for zero removal budget. If its cost is zero and the
support has at most \(q\) locations, retain the support and fill from
\(X_0\). Empty support is immediate. If a capped augmentation omits
the mandatory center or violates its budget, report failure and return
\(U\). Static initialization absorbs the factor-four
proper-to-ambient comparison into \(\beta_2\); requests larger than
the support are handled by opening that support.
\end{proof}

The remaining static ingredient controls how many facilities are
needed to augment a robust solution. As in the median case,
well-separated pairs allow us to delete facilities, but the
squared objective changes the loss in the comparison.

\begin{lemma}[Squared separated-pairs deletion]\label{km:separated-delete}
Let \(\Gamma\ge400000\), \(|U|=k\), and \(|V|=k+s\).
If there are \(k-m\) well-separated pairs and \(m\ge s\), some
\(\widetilde U\subseteq U\) satisfies
\[
 |\widetilde U|\le k-\left\lfloor\frac{m-s}{4}\right\rfloor,
 \qquad
 \Psi_P(\widetilde U)\le D_{\rm sep}
        \bigl(\Psi_P(U)+\Psi_P(V)\bigr),
 \quad D_{\rm sep}=81\Gamma^2.
\]
\end{lemma}
\begin{proof}
The idea is to open facilities participating in the difficult
pairs fully and open the others fractionally. The integrality
gap then converts this cheaper fractional solution into a
smaller integral set. Use the fractional construction of
\cite[Lemma~15.6 and Section~20.5]{BCFJJL26}.
Project \(V\) to \(U\). Put in \(U_I\) all endpoints of
well-separated pairs and all facilities with at least two preimages;
write \(U_F=U\setminus U_I\). Counting preimages gives
\[
 k+s\ge k-m+2(|U_I|-(k-m)),\qquad
 |U_I|\le k-(m-s)/2.
\]
Open \(U_I\) fully and \(U_F\) halfway, using budget at most
\(k-(m-s)/4\).

We next assign demand to these openings. For a client \(x\),
let \(v=\pi_V(x)\) and \(u=\pi_U(v)\).
If \(u\in U_I\), assign all demand to \(u\). Otherwise assign
half to \(u\). For the other half, use a distinct \(u'\) with
\(d(u,u')\le\Gamma d(u,v)\) when one exists. If none exists,
nonseparation supplies \(v'\ne v\) with
\(d(v,v')\le\Gamma d(u,v)\); use \(u'=\pi_U(v')\).
This facility differs from \(u\), since a receiving facility in
\(U_F\) has exactly one preimage. These assignments respect all
fractional openings.

The distance inequalities control the cost of these fractional
assignments. Put \(a=d(x,U)\), \(b=d(x,V)\). We have
\(d(x,u)\le a+2b\), \(d(u,v)\le a+b\), and in the last case
\(d(x,u')\le d(x,u)+2(\Gamma+1)d(u,v)\).
Thus the assigned squared cost is at most
\[
 \begin{aligned}
 \tfrac32d(x,u)^2+4(\Gamma+1)^2d(u,v)^2
 &\le[12+8(\Gamma+1)^2](a^2+b^2)\\
 &\le9\Gamma^2(a^2+b^2),
 \end{aligned}
\]
where the last inequality already holds for \(\Gamma\ge40\).
Multiply by weights, sum, and apply gap \(9\) on domain \(U\) at
integer budget \(k-\lfloor(m-s)/4\rfloor\). The argument uses only
metric inequalities in \(d\).
\end{proof}

\paragraph{Window constants.}
The stability-window argument uses the same two-sided slack
measure as before, with constants adjusted for the squared LP
gap. For \(g=9\), \Cref{lem:gap-interpolation} gives factors
\(37/4\) and \(12\) whenever \(H\le R/(36M)\).
Their product is \(111<128\). Define
\begin{equation}\label{km:slack}
 b_{128}(P)=\max\{r\in\{0,\ldots,k-1\}:
       F_{k-r}^{(2)}(P)\le128F_{k+r}^{(2)}(P)\},
\end{equation}
and let \(r_M^{(2)}\) be the analogous one-sided slack, comparing
\(F_{k-r}^{(2)}\) with \(MF_k^{(2)}\). The same lazy and
interpolation arguments yield
\[
 |b_{128}(P)-b_{128}(Q)|\le1\ \text{for one edit},\qquad
 b_{128}\le r_M^{(2)}<36M(b_{128}+1)\quad(M\ge128).
\]
Choose constants in the following order:
\begin{align}
 \alpha_2&\ge18(72a_R\beta_2+b_R),\label{km:alpha}\\
 \eta_2&\ge\max\{128,D_{\rm sep}(\alpha_2+4)\},\nonumber\\
 A_{{\rm test},2}&=a_R+b_R\eta_2,\qquad
 M_{0,2}=A_{{\rm test},2}\alpha_2,\nonumber\\
 D_2&=\lceil36M_{0,2}\rceil,\qquad C_2=2D_2,\qquad
 C_{0,2}=12\left(a_R\alpha_2+\tfrac{37}{4}b_R\right).
 \label{km:window-constants}
\end{align}
The algorithm still obtains the window from computed removal
costs, without querying any optimum. Run the same geometric
search with squared removal and threshold
\(A_{{\rm test},2}\Psi_P(U)\). From the last accepted \(R\), or
zero if none, set \(H=\lfloor R/D_2\rfloor\), \(B=H+1\), and
\(K=\Remove_2(P,U,H)\), taking \(K=U\) when \(H=0\).

\begin{proposition}[Squared windows and refresh]\label{km:window}\label{km:refresh}
For \(k\ge2\) and a squared-robust \(k\)-set \(U\) with
\(\Psi_P(U)\le\alpha_2F_k^{(2)}(P)\), successful window computation
gives \(K\subseteq U\), \(|K|=k-H\), \(H=B-1\), and
\begin{align}
 B&\le b_{128}(P)+1\le C_2B,\qquad
 r_{\eta_2}^{(2)}(P)<2D_2B,\label{km:window-scale}\\
 \Psi_P(K)&\le C_{0,2}F_{k+H}^{(2)}(P).\label{km:window-protection}
\end{align}
After \(s\le B\) record edits, successful refresh returns a
squared-robust \(k\)-set \(U'\) with
\(\Psi_Q(U')\le\alpha_2F_k^{(2)}(Q)\) and
\(|U'\triangle U|=O(BL_\Delta^3)\).
Given the checkpoint's cached distance structure, both procedures have
prescribed \(\widetilde O((n+k)B)\) work bounds. The deterministic
squared state and resource bounds do not require sampling success;
approximation and full validity use the common event, including
inherited certificates.
\end{proposition}
\begin{proof}
\emph{Window.} The accepted-prefix argument works because
the removal guarantee accepts every budget up to the
one-sided slack. Every test up to \(r_{\eta_2}^{(2)}\) succeeds by
\eqref{km:removal-bound}, so the accepted-prefix argument gives
\(r_{\eta_2}^{(2)}\le2R\). Acceptance gives
\(F_{k-R}^{(2)}\le M_{0,2}F_k^{(2)}\).
Gap interpolation implies \(H\le b_{128}\) and
\[
 \Psi_P(K)\le
 \left(a_R\alpha_2+\tfrac{37}{4}b_R\right)F_k^{(2)}
 \le C_{0,2}F_{k+H}^{(2)}.
\]
The inequalities \(R<D_2(H+1)\) and \(b_{128}+1\le2R+1\)
prove the window scale, also for \(R=H=0\). Tested budgets sum to
\(O(R+1)=O(B)\).

\emph{Refresh.} We augment the old solution before processing
the changed coordinates, then remove the excess facilities and
repair the certificates. Project a \((k+s)\)-optimum to support, losing
factor four, and apply squared separated-pairs deletion. If there are
\(k-m\) separated pairs, the choice of \(\eta_2\) gives
\(m\le4r_{\eta_2}^{(2)}+s+3\). Retaining the \(m+s\) unpaired
comparison facilities and replacing each paired facility by its robust
partner yields a support set \(F\) with
\[
 |F|\le4r_{\eta_2}^{(2)}+2s+3,\qquad
 \Psi_P(U\cup F)\le72F_{k+s}^{(2)}(P).
\]
Unused dummy facilities can be discarded. Use augmentation budget
\(a=\lceil8D_2B+2s+3\rceil\), open all edited coordinates, and
remove back to \(k\). As in \Cref{thm:refresh}, the proposed set
\(W\) differs from \(U\) in \(O(B)\) facilities and satisfies
\[
 \Psi_Q(W)\le(72a_R\beta_2+b_R)F_k^{(2)}(Q).
\]
Squared batch repair gives
\(\Psi_Q(U')\le18\Psi_Q(W)\le\alpha_2F_k^{(2)}(Q)\)
and the recourse bound. The edited-location set includes every changed
coordinate, regardless of its final weight.

\emph{Work and constants.} The primitives and new distance columns
cost \(\widetilde O((n+k)B)\). Cross entries between \(O(B)\) new
columns and edited rows cost \(O(B^2)\le O(kB)\); unchanged rows
are shared. The refresh factor depends on \(\beta_2,a_R,b_R\),
but not on \(\eta_2\), so the displayed order of constants is
noncircular.
\end{proof}

\paragraph{Scheduling and projection.}
Only the cost estimate during center migration depends on the
objective. The window lengths, phase deadlines, and row
maintenance follow the same schedule as for \(k\)-median.
For snapshots separated by \(s\le\xi B\) edits, \(0\le\xi<1\),
\begin{equation}\label{km:window-comparison}
 (1-\xi)B/C_2\le B'\le(C_2+\xi)B.
\end{equation}
The four-stage schedule in \Cref{sec:step5} therefore has the same
lifetimes and budgets, with \(C_2,C_{0,2}\). Let \(A\) be the
old migration base, \(K_t\) the target core, \(S\) an intermediate
base, and \(J\) their common edit buffer, as in \Cref{sec:step4}.
The objective-dependent migration estimate is
\begin{equation}\label{km:migration}
 \Psi_P(S\cup J)\le6\Psi_P(A\cup J)+3\Psi_P(K_t\cup J),
\end{equation}
which follows from the pointwise migration bound and
\((2a+b)^2\le6a^2+3b^2\). Intermediate virtual outputs cost at most
\(9C_{0,2}F_k^{(2)}\), and proper projection loses factor four.

\begin{theorem}[Weighted squared-cost solver]\label{km:inner}
Arbitrary weighted-record edits admit
\(\widetilde O(n+k)\) worst-case time and
\(O(L_\Delta^3)\) worst-case recourse, with proper output of at most
\(k\) facilities. For any finite adaptive horizon, with total failure
probability at most \(\delta\), simultaneously after every edit,
\[
 \Psi_{P_t}(S_t)\le36C_{0,2}F_k^{(2)}(P_t).
\]
Initialization uses \(\widetilde O(k(n_{\rm init}+k))\) work, where
\(n_{\rm init}\) is the initial weighted support size. Feasibility
and resource bounds hold on every execution, with the same confidence
and weight-range conventions as \Cref{thm:inner}.
\end{theorem}
\begin{proof}
The squared window and refresh proposition supplies the protection,
comparable-window, per-call work, and recourse interfaces used by the
median schedule. Equation~\eqref{km:migration} supplies the cost bound
during migration. Persistent rows and current-live reconciliation are
unchanged; evaluating a squared distance takes constant arithmetic work.
For initialization, use the fixed-anchor construction of
\Cref{lem:annotation-initialization} with \(\MakeRobustProc_2\).
The same displacement bounds establish the deterministic state, and
\eqref{km:sequence-cost} on the fixed initial Voronoi cells gives cost
at most \((9/4)\beta_2F_k^{(2)}\le\alpha_2F_k^{(2)}\).

The deterministic safeguards also carry over. Keep the fixed
static caps, stage guards, and frozen-set failure rule.
The deterministic squared state suffices for work accounting, and full
sampled validity supplies approximation. Fresh randomness on frozen
inputs gives the same conditional union bound. For \(k=1\), weighted
sampling directly gives an \(8\)-approximate proper center in
\(\widetilde O(n+1)\) work with the assigned confidence and recourse
at most two; empty support returns the empty set.
\end{proof}

\subsection{Summary Transportation and the Final Bound}\label{km:outer}

The outer summary depends only on distances and client
memberships, not on whether connection costs are squared.
In particular, its full-mass transitions and constant edit bound
are unchanged. We measure the remaining error by squared
transportation:
\begin{equation}\label{km:transport-definition}
 E_2=\sum_{G\in\mathcal G}\sum_{x\in G}d(x,q_G)^2.
\end{equation}

\begin{lemma}[Squared transportation]\label{km:transport}
On the simultaneous sampling-quality event, the summary satisfies
\begin{align}
 E_2&\le a_2\OPT_{k,2}(P),\qquad a_2=128K,
       \label{km:transport-bound}\\
 \Psi_P(C)&\le2E_2+2\Psi_Q(C),\label{km:transport-forward}\\
 \Psi_Q(C)&\le2E_2+4\Psi_P(C)\label{km:transport-backward}
\end{align}
for every ambient facility set \(C\), where \(K\) is the absolute
constant in the tail-packing lemma. The last two inequalities hold
deterministically, even when sampling fails. Structural and resource
bounds are unchanged.
\end{lemma}
\begin{proof}
The tail-packing proof does not use the first power of distance.
Applying \Cref{outer:tail-packing,outer:transport} with exponent two
gives the bound on \(E_2\). Its comparison is with a proper
optimum because the quantiles use current client locations.
For the two cost inequalities, apply
\((a+b)^2\le2a^2+2b^2\) to each token-client pair.
Every client appears in one or two active tokens, so the
same counting argument as in \Cref{outer:summary} applies.
\end{proof}

Combining this summary bound with the weighted squared-cost solver
gives the claimed dynamic result.

\begin{theorem}[Fully dynamic metric \(k\)-means]\label{thm:kmeans-extension}
Under the fixed-universe model and unit-weight update assumptions of
\Cref{thm:main}, a randomized algorithm maintains a proper set \(S_t\)
of at most \(k\) centers. With probability at least \(1-\delta\),
simultaneously throughout any adaptive sequence of at most \(T\) updates,
\[
 \Psi_{P_t}(S_t)\le A_2\OPT_{k,2}(P_t)
\]
for an absolute constant \(A_2\). Update time is
\(\widetilde O(k)\), recourse is \(O((1+\log\Delta)^3)\), and
initialization costs \(\widetilde O(k(n_0+k))\).
The resource and feasibility guarantees hold on every execution.
Stable fillers permit exactly \(\min\{k,|\supp(P_t)|\}\) centers.
\end{theorem}
\begin{proof}
Run the weighted solver of \Cref{km:inner} on the summary, and write
\(\alpha_{\rm in,2}=36C_{0,2}\). A proper optimum \(C^*\) for
\(P\) is also a valid ambient comparison for \(Q\).
On the common success event,
\[
 \begin{aligned}
 F_k^{(2)}(Q)&\le\Psi_Q(C^*)
              \le2E_2+4\OPT_{k,2}(P),\\
 \Psi_P(S)&\le2E_2+2\alpha_{\rm in,2}F_k^{(2)}(Q)\\
 &\le\bigl[(2+4\alpha_{\rm in,2})a_2
             +8\alpha_{\rm in,2}\bigr]\OPT_{k,2}(P).
 \end{aligned}
\]
The bracket defines the absolute constant \(A_2\).
The implementation argument is the same as for \(k\)-median:
each original update produces constantly many weighted-record
edits, the summary support is contained in the original support
after the whole batch, and all positive record weights lie in
\([1,2N]\). Process every batch before publication and retain
the same failure guards and filler rule. Splitting the failure
allowance between the two components gives simultaneous
approximation for adaptive updates. The fixed work and
recourse caps continue to hold on every execution.
\end{proof}

\begin{corollary}[Comparison with Euclidean centroids]\label{km:euclidean}
Suppose the fixed admissible client universe is Euclidean and satisfies
the same global aspect-ratio assumption, for example a finite set of
Euclidean locations. If \(\OPT_k^{\rm Euc}(P)\) allows arbitrary
centroids in the surrounding Euclidean space, the output has cost at
most \(2A_2\OPT_k^{\rm Euc}(P_t)\). The update model and
distance-query convention are unchanged.
\end{corollary}
\begin{proof}
For a nonempty cluster \(B\) with weighted mean \(\bar x\), a
client \(Y\) sampled in proportion to its weight satisfies
\[
 \mathbb E\!\left[\sum_{x\in B}w_P(x)\|x-Y\|_2^2\right]
 =2\sum_{x\in B}w_P(x)\|x-\bar x\|_2^2.
\]
Thus each optimal Euclidean cluster contains a client whose
squared connection cost is at most twice that of its centroid.
Choosing one such client per cluster shows that
\(\OPT_{k,2}\le2\OPT_k^{\rm Euc}\).
Apply \Cref{thm:kmeans-extension}. The argument changes only the
benchmark; the algorithm still opens centers at client locations.
\end{proof}

This proves \Cref{cor:kmeans-main}. The original update sequence still
consists of unit-weight insertions and deletions. Arbitrary weight
changes arise only in the records passed to the inner solver.

\section*{Concluding Remarks}\label{sec:conclusion}

The six-step construction separates the two roles of a stability
window. It certifies how long a core can serve the changing input,
and it supplies the time in which a replacement can be constructed
and installed. Per-invocation repair bounds make that work predictable;
the weighted summary makes the input to the solver small without
introducing a large batch of record changes. The composition uses
these guarantees at the same completed-update boundaries.

We have not tried to optimize the approximation constants or the
parameters used to choose stability windows. The
\(O((1+\log\Delta)^3)\) recourse bound includes the costs of
certificate contamination, chains of anchor relocations, and
changes across witness scales. A sharper analysis of these
interactions may improve the dependence on \(\Delta\).

The outer summary currently assumes unit-weight client updates,
although the inner solver permits arbitrary changes to weighted
records. Allowing arbitrary original weights would require
sampling and stability bounds that account for mass rather than
the number of births. Our persistent data structures also use
space proportional to the total allocated work over the prescribed
horizon; a bound in terms of the current input would need a
different reclamation scheme. Finally, the algorithm maintains
the center set and its connection-cost guarantee, not an explicit
nearest-center assignment for every client. Reporting all such
assignments after each update would incur the corresponding
output-size cost.

%% file: sections/restricted_local_search.tex
\section{Restricted Randomized Local Search}
\label{app:restricted-local-search}

We use the randomized local-search procedure of~\cite{BCGLP24},
also used in~\cite[Lemma~11.1]{BCF25}. We give the exchange and
convergence arguments for the weighted instances and ambient
facility domain used here. Several comparison facilities may
have the same nearest candidate. We group them by this candidate
so that each possible entering facility is counted only once.

For \(p\in\{1,2\}\), write
\[
 J_p(S)=\sum_{x\in\supp(P)}w_P(x)d(x,S)^p,
 \qquad L_p(S)=J_p(S)^{1/p}.
\]
Let \(F_q^{(p)}(P)\) be the ambient optimum for \(J_p\), so
\(F_q^{(1)}=F_q\) and \(F_q^{(2)}\) agrees with the notation
in \Cref{sec:kmeans}. All nearest-facility ties are broken
consistently. Candidate facilities need not carry demand.

\begin{lemma}[Restricted local search with a fixed work bound]
\label{lem:restricted-local-search}
Let \(q\ge1\), \(r\ge0\), and \(X_0\subseteq X\) satisfy
\(|X_0|=q+r\). Fix a failure allowance \(0<\xi<1\).
Given the ordered client-to-candidate distance trees used by
\Cref{lem:remove}, a randomized procedure returns a
\(q\)-subset \(Y\subseteq X_0\) such that, with probability
at least \(1-\xi\),
\begin{equation}\label{eq:restricted-local-search-bound}
 L_p(Y)\le
 2\left(L_p(X_0)+6p\bigl(F_q^{(p)}(P)\bigr)^{1/p}\right).
\end{equation}
Its work is \(\widetilde O((n+|X_0|)(r+1))\), including
logarithmic factors in the positive-weight range, the distance
range, and \(1/\xi\). Its cardinality and work bounds hold on
every execution. The probability bound also holds conditional
on any history fixing the input before the call's fresh random
choices.
\end{lemma}
\begin{proof}
Return \(X_0\) if \(r=0\), and return any \(q\)-subset if
the support is empty. If \(J_p(X_0)=0\) and \(n\le q\),
retain all support locations and fill the remaining slots from
\(X_0\). Zero cost implies \(\supp(P)\subseteq X_0\), so
this is feasible. We henceforth consider the remaining cases.

\paragraph{The procedure.}
Start with any \(q\)-subset \(S_0\subseteq X_0\). In each
iteration, sample \(z\) uniformly from \(X_0\setminus S\)
and choose
\[
 s\in\arg\min_{a\in S\cup\{z\}}
 J_p\bigl((S\cup\{z\})\setminus\{a\}\bigr).
\]
Replace \(S\) by \((S\cup\{z\})\setminus\{s\}\).
Choosing \(s=z\) leaves the solution unchanged. Thus every
iteration keeps exactly \(q\) facilities and never increases
the cost. The procedure compares \(J_p\) values; roots are
used only in the analysis. For \(z\in X_0\setminus S\), let
\[
 \delta_S(z)=
 \min_{a\in S\cup\{z\}}
 J_p\bigl((S\cup\{z\})\setminus\{a\}\bigr)-J_p(S)
 \le0.
\]
For analysis only, set \(\delta_S(z)=0\) when \(z\in S\).
The procedure uses the fixed iteration count specified below.

\paragraph{Grouping comparison facilities.}
Fix an ambient optimal set \(O\) for budget \(q\). If
\(|O|<q\), add distinct locations from \(X_0\) until
\(|O|=q\). This is possible since \(|X_0|\ge q\), and
optimality implies that the cost remains \(F_q^{(p)}(P)\).
The algorithm does not construct \(O\).

Put \(Z=\pi_{X_0}(O)\), taking this image as a set, and define
\[
 G_z=\{o\in O:\pi_{X_0}(o)=z\}\quad(z\in Z),
 \qquad
 H_s=\{o\in O:\pi_S(o)=s\}\quad(s\in S).
\]
Call \(s\) empty if \(H_s=\varnothing\), pure if its
nonempty set \(H_s\) lies in one group \(G_z\), and mixed
otherwise. Let the numbers of empty, pure, and mixed facilities
be \(e,u,m\), respectively. Then
\[
 q=e+u+m,\qquad
 q=\sum_{s\in S}|H_s|\ge u+2m,
 \qquad m\le e.
\]
For every group having a pure facility, select one such facility
as its closing facility \(s_z\). Distinct groups select distinct
pure facilities. Every remaining group has a member captured by
a mixed facility, and all its members are captured by mixed
facilities. Hence the number of these remaining groups is at most
\[
 \sum_{s\text{ mixed}}|H_s|
 \le q-u=e+m\le2e.
\]
Assign them empty closing facilities, using each empty facility
at most twice. We have therefore chosen \(s_z\) so that
\(H_{s_z}\subseteq G_z\), and no old facility is selected by
more than two groups.

\paragraph{Bounding the exchanges.}
For a client \(x\), abbreviate
\[
 o_x=\pi_O(x),\qquad s_x=\pi_S(x),\qquad
 a_x=d\bigl(x,\pi_{X_0}(o_x)\bigr),\qquad
 b_x=d\bigl(x,\pi_S(o_x)\bigr),\qquad c_x=d(x,S).
\]
In particular, \(b_x\ge c_x\).
For \(z\in Z\setminus S\), open \(z\) and close \(s_z\).
Assign each client with \(o_x\in G_z\) to \(z\).
If \(o_x\notin G_z\) and \(s_x=s_z\), assign \(x\)
to \(\pi_S(o_x)\); leave all other assignments unchanged.
The second target remains open, since
\(H_{s_z}\subseteq G_z\). This gives
\begin{equation}\label{eq:restricted-group-exchange}
 \delta_S(z)\le
 \sum_{x:o_x\in G_z}w_P(x)(a_x^p-c_x^p)
 +\sum_{\substack{x:s_x=s_z\\o_x\notin G_z}}
       w_P(x)(b_x^p-c_x^p).
\end{equation}
For \(z\in Z\cap S\), the first sum is nonnegative because
\(a_x=d(x,z)\ge c_x\) for its clients, and the second sum
is also nonnegative. Thus \eqref{eq:restricted-group-exchange}
continues to hold with \(\delta_S(z)=0\).

Every client belongs to one group, and each closing facility
is used at most twice. Since the second summands are nonnegative
and all other candidate increments are nonpositive, summing over
the distinct groups yields
\begin{equation}\label{eq:restricted-group-sum}
 \begin{aligned}
 \sum_{z\in X_0\setminus S}\delta_S(z)
 &\le\sum_{z\in Z}\delta_S(z)\\
 &\le\sum_x w_P(x)(a_x^p-c_x^p)
       +2\sum_x w_P(x)(b_x^p-c_x^p).
 \end{aligned}
\end{equation}
The nearest-candidate property and the triangle inequality give
\[
 a_x\le d(x,X_0)+2d(x,O),\qquad
 b_x\le d(x,S)+2d(x,O).
\]
Write \(A=L_p(X_0)\), \(B=L_p(O)\), and \(C=L_p(S)\).
Weighted Minkowski applied to \eqref{eq:restricted-group-sum}
now gives
\begin{equation}\label{eq:restricted-drift}
 \sum_{z\in X_0\setminus S}\delta_S(z)
 \le (A+2B)^p+2(C+2B)^p-3C^p.
\end{equation}

\paragraph{A fixed number of iterations.}
Put \(T=A+6pB\). If \(p=1\) and \(C>2T\), the
right-hand side of \eqref{eq:restricted-drift} is
\(T-C<-C/2\). If \(p=2\) and \(C>2T\), then
\(A<C/2\) and \(B<C/24\), so that
\[
 (A+2B)^2+2(C+2B)^2-3C^2
 \le\left(\frac7{12}\right)^2C^2
       +2\left(\frac{13}{12}\right)^2C^2-3C^2
 =-\frac5{16}C^2.
\]
In either case, since there are exactly \(r\) inactive
candidates, one iteration satisfies
\begin{equation}\label{eq:restricted-conditional-drift}
 \mathbb E[J_p(S_{t+1})\mid S_t]
 \le\left(1-\frac1{4r}\right)J_p(S_t)
 \quad\text{whenever }L_p(S_t)>2T.
\end{equation}

The input and \(O\) are fixed during the call, so \(T\) is
constant. Cost never increases, and therefore the condition
\(L_p(S_t)\le2T\) is absorbing. Define
\[
 V_t=J_p(S_t)\,\one\{L_p(S_t)>2T\}.
\]
Conditional on the preceding random choices,
\eqref{eq:restricted-conditional-drift} gives
\[
 \mathbb E[V_{t+1}\mid\mathcal H_t]
 \le\left(1-\frac1{4r}\right)V_t.
\]
In the remaining cases \(T>0\): otherwise both
\(J_p(X_0)\) and \(F_q^{(p)}(P)\) vanish, which forces
\(\supp(P)\subseteq X_0\) and \(n\le q\), a case
already handled directly.

Let \(R_w=w_{\max}/w_{\min}\), using only positive demand
weights. Every positive connection distance lies in
\([1,\Delta_d]\). Hence \(T^p\ge w_{\min}\) and
\(J_p(S_0)\le n w_{\max}\Delta_d^p\). It follows that
\[
 \Pr[L_p(S_K)>2T]
 \le\frac{\mathbb E[V_K]}{(2T)^p}
 \le\exp\!\left(-\frac K{4r}\right)
       nR_w\Delta_d^p.
\]
Choose the smallest nonnegative integer \(\ell\) with
\(2^\ell\ge2nR_w\Delta_d^p/\xi\), and set
\begin{equation}\label{eq:restricted-iteration-cap}
 K=4r\ell.
\end{equation}
Then \(\exp(-K/(4r))\le2^{-\ell}\), so the failure
probability is at most \(\xi/2\). Repeated doubling computes
\(\ell\) using only arithmetic and comparisons, in
\(O(1+\log(nR_w\Delta_d^p/\xi))\) work. Neither \(O\),
\(T\), nor the optimum is needed to set this cap or to execute
the procedure.

\paragraph{Work per iteration.}
We spell out the best-swap implementation of
\cite[Section~7.3]{BCGLP24} to account for \(q>n\) and
zero-demand candidates. The supplied distance rows are ordered
search trees on \(X_0\). Retain their original roots and use
private persistent roots during the call. To initialize the
active rows for \(S_0\), delete the \(r\) inactive candidates
from every row, using \(O(nr\log(|X_0|+2))\) work. We do
not rebuild a tree containing all \(|X_0|\) candidates.

After temporarily inserting the sampled \(z\), every active
row contains \(q+1\ge2\) facilities. Read its first two
facilities \(a_1(x),a_2(x)\). For each
\(a\in S\cup\{z\}\), compute
\[
 \lambda_a=
 \sum_{x:a_1(x)=a}w_P(x)
 \bigl(d(x,a_2(x))^p-d(x,a_1(x))^p\bigr).
\]
Deleting \(a\) from \(S\cup\{z\}\) increases its cost
by exactly \(\lambda_a\). Initialize \(q+1\) accumulators,
scan the clients once, and choose a facility of minimum
\(\lambda_a\). The choice \(a=z\) is included. Insertion
and deletion in all rows, together with this scan, take
\(\widetilde O(n+q)\) work. This remains valid for \(q=1\).
An auxiliary cumulative-weight tree gives each inactive
candidate unit weight. It supports uniform sampling and the
membership changes in \(O(\log(|X_0|+2))\) work after
\(\widetilde O(|X_0|)\) initialization.

Querying the first entry of each initial row computes
\(J_p(X_0)\), and a client scan computes the positive-weight
range. Original row
roots are unchanged throughout. Combining initialization with
\eqref{eq:restricted-iteration-cap} proves the claimed
\(\widetilde O((n+|X_0|)(r+1))\) work bound on every
execution. Finally, the argument holds for every fixed weighted
input. Conditioning on the history at its freeze time therefore
preserves the failure bound when the call uses fresh random
choices, even if its work is interleaved with later updates.
\end{proof}